\pdfoutput=1
\documentclass{report}
\usepackage{suthesis-2e}
\RequirePackage{latexsym}
\RequirePackage{amssymb,amsfonts,amsmath,amsthm}
\RequirePackage[dvips]{graphicx,epsfig}
\RequirePackage[T1]{fontenc}
\RequirePackage{helvet}
\RequirePackage{palatino}
\RequirePackage{times}
\RequirePackage{fancyhdr}
\RequirePackage{sectsty}
\usepackage[ruled]{algorithm2e}
\usepackage{appendix}
\usepackage{amsbsy}
\usepackage{amsfonts}
\usepackage{amsmath}
\usepackage{amsthm}
\usepackage{amssymb}
\usepackage{helvet}
\usepackage{array}
\usepackage{bm}
\usepackage{caption}
\usepackage[color]{changebar}
\usepackage{datetime}
\usepackage{epsfig}
\usepackage[outdir=./]{epstopdf}
\usepackage{enumerate}
\usepackage{enumitem}
\usepackage{fancybox,amssymb}
\usepackage{fancyhdr}
\usepackage{floatflt}
\usepackage[OT1]{fontenc}
\usepackage{graphicx}
\usepackage[dvips]{graphicx}
\usepackage{latexsym}
\usepackage{lastpage}
\usepackage{lineno}
\usepackage{multirow}
\usepackage[sort&compress,numbers]{natbib}
\usepackage[nice]{nicefrac}
\usepackage{overpic}
\usepackage{palatino}
\usepackage{pstricks}
\usepackage{ragged2e}
\usepackage{setspace}
\usepackage{stfloats}
\usepackage{subfigure,amsmath}
\usepackage{threeparttable}
\usepackage{type1cm}
\usepackage{type1ec}
\usepackage{textcomp}
\usepackage{textfit}
\usepackage{upgreek}
\usepackage{dirtytalk}
\usepackage[utf8]{inputenc}
\usepackage{tikzscale}
\usepackage{tikz}
\usepackage{hyperref}
\hypersetup{
    colorlinks,
    citecolor=black,
    filecolor=black,
    linkcolor=black,
    urlcolor=black
}
\usetikzlibrary{shapes.geometric, arrows}
\tikzstyle{every node}=[]
\tikzstyle{arrow} = [thick,->,>=stealth]
\usepackage{xcolor}
\definecolor{red}{RGB}{231, 76, 60}
\definecolor{orange}{RGB}{243, 156, 18}
\definecolor{yellow}{RGB}{241, 196, 15}
\definecolor{green}{RGB}{39, 174, 96}
\definecolor{blue}{RGB}{46, 134, 193}
\definecolor{purple}{RGB}{155, 89, 182}
\definecolor{pink}{RGB}{236, 64, 122}
\definecolor{turquoise}{RGB}{0, 172, 193}

\dept{Aeronautics and Astronautics}

\begin{document}
\title{Probabilistic Analysis of Aircraft Using Multi-Fidelity Aerodynamics Databases}
\author{Jayant Mukhopadhaya}
 
\beforepreface


\figurespagefalse
\tablespagefalse
\afterpreface

\chapter{Introduction} \label{intro}
Engineering design is a predictive discipline involving the estimation of an object's future real-world behavior.
The object, at the  time, may only exist as a thought in the brain, as a sketch on paper, or more recently, as a file on a computer.
The design process starts with a definition of requirements that the object in question must fulfill.
It ends with the manufacturing of prototypes that, hopefully, confirm those requirements' fulfillment. 
Depending on the object's complexity, the interim can be as short as a few hours or as long as multiple years involving thousands of hours of engineering work..

Those hours are spent predicting the real-world behavior of an object that doesn't physically exist yet.
Numerous analysis techniques progressing from basic back-of-the-envelope calculations, through computational numerical simulations, to prototyping and experimental testing of subsystems are employed in this endeavor.
Due to the intricacies of real-world physics, almost none of these techniques are perfect.
Each method has some uncertainty associated with its predictions that must be taken into account by the engineers employing them.
Quantifying these uncertainties can be highly specific to the method in question.

The word \textit{fidelity} is used to refer to how closely a method can mimic real-world behavior.
High-fidelity methods are better at predicting real-world behavior, whereas low-fidelity methods employ simplifications that introduce uncertainties into the analyses.
As an example, consider estimating the weight of an object.
A low-fidelity method would be to pick up the object and estimate the weight based on how heavy it feels.
Factors such as personal bias, left vs. right-hand usage, and muscle soreness, would contribute to the estimate's uncertainty.
A high-fidelity method would be to use a weighing scale that is accurate up to one milligram to measure the weight.
Direct measurement introduces significantly fewer sources of uncertainty, and the weight would be accurate up to $0.5$ milligrams. 

Fidelity comes at a cost.
This is to be expected. 
If high-fidelity analyses were less expensive than low-fidelity ones, there would be no reason to use low-fidelity analyses. 
Continuing with the weight estimation example, the low-fidelity method's only cost is the time taken to pick up and estimate the object's weight.
The high-fidelity method incurs the additional cost of the weighing scale.
Cost minimization is often a priority, and a mix of low- and high-fidelity analyses are needed.
Greater emphasis is placed on lower-fidelity methods in earlier stages of the design process, where rough estimations are sufficient to make design decisions.
This emphasis transfers to higher-fidelity methods as designs progress, and more certainty in performance metrics is required before the significant investment of creating a prototype is made.

This dissertation focuses on combining results from analyses of varying fidelity to create a superior prediction of an engineered object's real-world behavior.
These predictions incorporate the analyses' uncertainties to create a probabilistic representation rather than a single, deterministic result.
A method to quantify the uncertainties due to modeling simplifications, particularly computational fluid dynamics (CFD) simulations, is implemented and validated.
Finally, statistical analysis that allows for the explicit calculation of the likelihood of meeting/failing a particular design requirement is presented.
The engineered object of choice to showcase the work's real-world impact is an aircraft.
It represents one of the most complex, engineered objects, and it requires years of development to design.
Standard industrial analysis techniques are employed.
The design requirements are real-world flight certification tests created by the Federal Aviation Administration (FAA) that are in use today.

\section{Motivation}\label{intro_motivation}

Aircraft design is a complex, non-linear, multi-disciplinary problem that requires many years and thousands of engineering hours to solve.
An aircraft comprises numerous subsystems that work together to make flight possible.
The sheer size, complexity, and number of parts make aircraft manufacturing a very long and expensive process.
It is imperative that all aspects of the aircraft's real-world performance are thoroughly investigated before the manufacturing step is taken in the design process.
While historical experience in designing prior aircraft is useful, more precise analyses are required to push the boundaries of performance. 

The rapid improvement of computational capabilities in the recent past has increased computer simulations' use to predict various aspects of the aircraft's real-world behavior in a virtual setting.
Coupled with advances in understanding the underlying physics of these phenomena, this has led to the development of computer simulations of varying sophistication and computational cost that can describe the relevant quantities of interest (QoI) at different levels of fidelity.
These simulations allow for the critical assessment of engineering designs significantly earlier in the design process than previously possible with purely experimental design campaigns.
These analyses have been used for aerodynamic shape optimizations \cite{jameson1988aerodynamic,anderson_aerodynamic_1999,chen2016aerodynamic}, structural optimizations \cite{bindolino2010multilevel,kirsch2012structural,zhu2016topology}, and more recently, combined aero-structural optimizations \cite{gray2019openmdao,brooks2018benchmark}.

While these simulations can use simplifying assumptions that introduce uncertainties in their analyses, they have significantly improved the ability to predict the satisfaction of performance-based design requirements, such as range, passenger capacity, and weight, early in the design process.
However, performance-based metrics are not the only design requirements on an aircraft. 
Governing bodies, such as the Federal Aviation Administration (FAA), set stringent flight certification requirements that test for an aircraft's air-worthiness \cite{romanowski_flight_2018}.
These are fundamentally different from performance-based requirements as the outcome is binary; either the aircraft passes or fails the test.
Consequently, the ramifications of not meeting the certification requirements are worse than not meeting performance-based requirements. 
\textit{Flight certification} suggests that these tests can only be performed with a full-scale prototype.
Nevertheless, learning from the trend of increased reliance on computational analyses, virtual representations of the aircraft design can be put through simulated air-worthiness testing to estimate the likelihood of passing or failing a requirement. 

The current methodology involves building a virtual representation of the aircraft using aerodynamic databases that contain the force and moment coefficients experienced by the aircraft across its expected flight envelope. 
These databases are created using a single information source and at specific milestones during the design process. 
They do not contain any information about the uncertainties introduced due to the particular information source used.
Simulating a flight certification maneuver using these databases yields a single deterministic outcome.
Such a result belies the uncertainty present in the database and wrongly assumes a $100\%$ certainty in the force and moment coefficient data that populates the database. 
With rigorous handling and propagation of these uncertainties, the deterministic result can be converted to a more realistic likelihood of success or failure.

This work aims to perform statistical analyses on simulations of flight certification maneuvers for commercial aircraft. 
This is achieved by tackling the problem on multiple fronts.
A method to quantify the uncertainties arising from modeling assumptions in the widely used Reynolds-Averaged Navier-Stokes (RANS) computational fluid dynamics (CFD) simulations is presented, implemented, and validated. 
These simulation results are combined with data from other information sources into aerodynamic databases that utilize multi-fidelity models to create a stochastic representation of the aircraft's potential performance.
These models are sampled to create hundreds of individual representations of the databases that have small variations due to the uncertainties in the underlying data. 
These samples are used to propagate the uncertainties through the flight simulation of a current air-worthiness test maneuver.
Statistical analysis of these maneuver simulation results yields the likelihood of an aircraft passing or failing the given test maneuver.

This enables the consideration of flight certification metrics earlier in the aircraft design process.
It also allows for assessing the adequacy of the planned control systems.
Quantifying the risks involved with failing a certification maneuver equips the engineers with the necessary information required to mitigate them.

\section{Uncertainty Quantification} \label{intro_uq}

Computer simulations and real-world experiments alike carry some inaccuracies in their predictions. 
The field of uncertainty quantification (UQ) does exactly as the name suggests; it provides methodologies to efficiently characterize, propagate, and, in some cases, minimize uncertainties that plague analyses.
UQ has been adopted for a wide range of applications, including modeling climate change \cite{katz2013uncertainty,qian2016uncertainty}, understanding uncertainties in numerical simulations \cite{najm2009uncertainty,garcia2014quantifying,schefzik2013uncertainty}, and even predicting the economic effects of COVID-19 \cite{baker2020covid}.
UQ benefits are realized when the process is taken one step further, and the quantified uncertainties are used to make better decisions \cite{kochenderfer2015decision}. 
Information on the underlying uncertainties of analysis techniques can be used to make better medical decisions \cite{begoli2019need}, create reliable and robust designs \cite{reliability,robust,multif}, and create safer autonomous driving algorithms \cite{feng2018towards,brechtel2014probabilistic,xu2014motion}.
This work aims to propagate uncertainties in design analyses through flight simulations to calculate the likelihood that an aircraft will succeed or fail in meeting flight certification requirements.

Uncertainties are often divided into two categories: aleatoric and epistemic.
Aleatoric uncertainties arise due to natural variation in the parameters that describe a situation.
Epistemic uncertainties arise due to incomplete knowledge of the situation. 
To exemplify the differences between the two categories, consider a ball being thrown multiple times using the same amount of force. 
Slight changes in the wind direction and speed will result in slight differences in the distance that the ball travels. 
The uncertainty in the distance traveled due to this natural variation in the wind is aleatoric.
It can be quantified by throwing the ball thousands of times, recording each distance, and calculating statistical information, such as the mean and standard deviation, about the result.
These uncertainties are easier to quantify through Monte Carlo simulations \cite{geraci_multifidelity_2017,menhorn_multifideliy_nodate} and  polynomial chaos expansions \cite{oladyshkin2012data,ng_multifidelity_2014}.
However, they are irreducible. 
Now consider that to remove the effects of natural variation, the same ball throw is simulated using simple calculations employing Newton's laws of motion.
The calculated distance is the same every time, but simplifications in the equations, such as ignoring wind resistance, introduce uncertainty in the result.
These simplifications reflect the lack of knowledge that defines epistemic uncertainties.
Contrary to aleatoric ones, epistemic uncertainties are reducible (through better modeling/measurement) but are difficult to calculate and propagate as they are extremely problem-dependent \cite{FERSON2004355}.

Analyses of varying fidelity are used in this work.
There are uncertainties associated with each. 
For some techniques, the uncertainties are provided by subject matter experts (SME) who rely on historical data and their experience in using the analyses.
Explicit quantification of epistemic uncertainties is performed for one commonly used analysis technique: Reynolds-averaged Navier-Stokes (RANS) computational fluid dynamics (CFD) simulations.
Specifically, the uncertainties introduced due to turbulence models and numerical error due to insufficient discretization are quantified.

\subsection{Turbulence Modeling Uncertainty}
The Navier-Stokes equations are a set of non-linear partial differential equations that describe the motion and behavior of fluids.
CFD is the field of study that involves using computers and numerical analysis techniques to solve the non-linear Navier-Stokes equations to simulate the flow of a fluid over a domain of interest.
This is made more difficult due to turbulence, which is a state of fluid flow characterized by chaotic, small-scale fluctuations in the density and velocity of the fluid.
The range of length and time scales that need to be resolved through spatial and temporal discretization makes it computationally intractable to solve exactly, i.e., without simplifying models.
Most flows of engineering interest are plagued by turbulence.
The difficulty in solving these flows has paved the way for developing a hierarchy of solution techniques that trade computational cost for prediction accuracy.

The most widely used industry method is the Reynolds-averaged Navier-Stokes (RANS) simulation.
Steady RANS simulations are very computationally efficient and can be used for expensive undertakings such as iterative aerodynamic shape optimization \cite{lyu2015aerodynamic,kenway2014multipoint,chen2016aerodynamic} and aircraft database generation \cite{wendorff_combining_2016}.
The computational efficiency comes at the cost of modeling inaccuracies.
The simulations assume that the flow is steady (no time-dependent variation in the flow) and require simplifying turbulence models that aggregate the effects of the turbulent eddies that would be present in the flow.

The inadequacy of turbulence models in predicting certain flow features has been well documented \cite{slotnick_cfd_nodate}.
These models have various parameters and constants calibrated using canonical flow conditions, such as the flow over a flat plate. 
Cheung et al. in \cite{cheung2011bayesian} treat these parameters as random variables and use high fidelity data from direct numerical simulations (DNS) to learn posterior distributions for these parameters. 
Similarly, Dow et al. in \cite{dow2011quantification} use DNS data to solve an inverse RANS problem to determine the eddy viscosity field that would result in a flow field closest to the DNS data. 
Both methods lean on computationally expensive DNS results, which are limited to simple geometries such as flat plates \cite{hoyas_reynolds_2008}, and channels \cite{laval_marquillie_dns_channel,marquillie_instability_2011}.

This work focuses on the eigenspace perturbation methodology \cite{emory2013modeling,iaccarino_eig_pert}, a physics-based UQ method that does not require any high-fidelity information and can provide uncertainty estimates by running 6 RANS simulations.
This has been applied to wind-engineering problems \cite{gorle2015quantifying} and to perform design optimizations under uncertainty \cite{mishra2020design}.
Chapter \ref{chap:rans_uq} details this methodology, provides various validation cases, and explores the relationship between uncertainty introduced by turbulence models and the numerical errors introduced due to insufficient discretization. 

\subsection{Numerical Discretization Error}

Turbulence models are not the only source of uncertainty in RANS CFD simulations.
Solving continuous equations on a discrete domain creates numerical discretization error.
In RANS CFD simulations, the continuous RANS equations that define fluid flow are solved on a discrete domain known as the mesh or grid.
Numerical discretization error is classified as an epistemic uncertainty as it can be reduced by increasing the number of discrete points in the domain.
As the discretization increases and the grid spacing tends towards $0$, the numerical error approaches $0$. 
This is the basis for the "Grid Convergence Study" method to quantify numerical discretization error \cite{american_society_of_mechanical_engineers_standard_2009}.
This is an important tool used in the verification and validation (V\&V) of CFD codes and turbulence models \cite{rumsey2010description}.

Section \ref{sec:num_vs_turb_error} delves into the details of quantifying numerical discretization error. 
It goes on to explore its relationship with the turbulence modeling uncertainty estimate by applying both uncertainty quantification techniques to two benchmark simulations: a 2D NACA0012 airfoil and a 3D ONERA M6 wing.

\section{Multi-Fidelity Analysis} \label{intro_mf}

During the course of the typical aerospace design process, different kinds of performance analysis tools are used at different stages.
Low-fidelity computer simulations accept lower accuracy for faster computations.
They are useful at the very early stages of the design process when the aircraft's geometry is not well defined and is subject to significant change.
For example, AVL \cite{drela2008athena} solves the incompressible potential flow equations, taking mere seconds.
It is used to rapidly explore a large multi-dimensional design space, changing variables like the number of engines or wing placement \cite{botero2016suave,botero2019generative}.
These are often replaced by higher-fidelity techniques, such as RANS CFD simulations, as the design progresses, and more details of the design are finalized.
Experimental data, normally obtained through a costly wind-tunnel test, typically provides the most accurate representation of the phenomena analyzed.
It is obtained late in the design process as expensive prototypes of subsystems need to be manufactured.

Instead of discarding the low-fidelity simulation data when higher-fidelity data is available, there are methods to combine data from multiple fidelity levels to better represent the quantities of interest (QOI).
Multi-fidelity extensions to popular surrogate modeling techniques have been developed.
Polynomial chaos expansion (PCE) \cite{oladyshkin2012data,blatman2011adaptive}, which is a popular technique for sensitivity  analysis \cite{sudret2008global,crestaux2009polynomial}, can be used to combine multi-fidelity information by learning an additive correction on the low-fidelity data \cite{ng2012multifidelity, palar2018global}.
Gaussian processes (GP) \cite{krige1951statistical,matheron1963principles,rasmussen_gaussian_2006}, popular for their direct estimation of the error in its modeling, combines multiple information sources by learning additive and multiplicative corrections on low-fidelity data \cite{kennedy_predicting_2000}.
This correction of the lower-fidelity data $f_{i-1}$ to a higher-fidelity $f_i$ is represented as
\begin{equation}
    f_i(\mathbf{x}) = \rho_i(\mathbf{x}) f_{i-1}(\mathbf{x}) + \delta_i(\mathbf{x}),
\end{equation}
where $\rho_i(\mathbf{x})$ is the multiplicative term, and $\delta_i(\mathbf{x})$ is the additive term.
These are sometimes referred to as bridge functions \cite{han_improving_2013}.
One disadvantage of using GP vs. PCE is that the GP assumes a Gaussian distribution across the uncertainty intervals. 
PCE can handle different probability distributions, but the superior multi-fidelity handling and the direct error estimation of the GP regression make multi-fidelity GP the tool of choice.

Multi-fidelity GP regression has been used extensively in aerospace applications \cite{lam_surrogate_nodate,menhorn_multifideliy_nodate}, modeling oil reservoir production and pressures \cite{kennedy_predicting_2000}, hydrodynamic simulations \cite{le_gratiet_recursive_2014}, and biological tissue growth \cite{lee2020propagation}.
Numerous extensions to the framework have been proposed.
Ghoreishi et al. introduced the ability to have non-hierarchical information sources by relating each fidelity level to the highest fidelity, as opposed to each other \cite{ghoreishi_gaussian_2018}. 
This is unnecessary for this work as there is a well-defined hierarchy based on the physics that is modeled. 
Perdikaris et al. \cite{perdikaris_nonlinear_2017} significantly improved prediction capability by using deep Gaussian processes \cite{damianou2013deep}.
The high computational cost and the loss of a Gaussian posterior precludes its use in this effort. 
Han et al. \cite{han_improving_2013} proposed using gradient information to improve predictive capability.
Unfortunately, gradient information is only available for RANS CFD simulations through adjoint simulations \cite{jameson1988aerodynamic}, but not for the other information sources. 

This work does employ important improvements to the multi-fidelity GP process that were proposed by Gratiet \cite{gratiet_multi-fidelity_nodate}.
The new equations significantly reduced the GP's computational cost by splitting the data into individual information sources.
This results in the inversion of smaller matrices which is more computationally efficient. 
He also extends the multiplicative term to be a function of $\mathbf{x}$ instead of a constant.
Section \ref{sec:mf_modeling} presents these equations in detail and extends them for use with noisy data for design sets that are not nested. 

\section{Aerodynamic Databases} \label{intro_databases}

Aerodynamic databases are a representation of the aircraft's behavior in-flight.
It contains all the forces and moments that are experienced by the aircraft as a function of the aircraft's configuration (control surface deflections), orientation (angle of attack, angle of sideslip), and operating conditions (dynamic pressure, Mach number, altitude).
Calculating these forces and moments at various points in its operating envelope is paramount to predicting real-world performance.
Most aerodynamic analyses, be it computational or experimental, are geared towards creating a database that catalogs these values as a function of the aircraft's orientation and operating conditions.

The industry standard is to have a lookup-table populated by data from aerodynamic analyses that are performed during the design process.
They get updated as the design progresses, and the results from the higher-fidelity analysis techniques replace the lower-fidelity data.
The forces and moments are described as multi-dimensional functions depending on up to 5 input variables: angle of attack, sideslip angle, Mach number, dynamic pressure, and altitude.
Often only a subset of the 5 input variables is used.
Discrete analyses in this multi-dimensional domain provide data points that are used to interpolate values between analysis locations.
These databases are deterministic and have no characterization of the uncertainties present in the analysis techniques. 

Engelund et al. \cite{engelund2001aerodynamic} created databases for the Hyper-X scramjet using wind tunnel data to analyze and predict the expected behavior before flight testing was performed.
Keshmiri et al. \cite{keshmiri2005development} used a mixture of CFD analyses and wind tunnel experiments to create aerodynamic databases for the generic hypersonic vehicle.
Instead of having a lookup table, the coefficients were described using analytic polynomial functions of arbitrary order. 
Databases for the Mars Pathfinder aerodynamics created using CFD data by Gnoffo et al. \cite{gnoffo1999prediction} were validated using flight measurements and were found to be accurate within reason.

Each of these previous works has created deterministic expressions of the databases.
There is no quantification of the uncertainties affecting the analyses used.
Previous work by Wendorff et al. \cite{wendorff_combining_2016} introduces the concept of probabilistic aerodynamic databases that uses multi-fidelity data and its associated uncertainties in a Gaussian Process regression framework to create a non-deterministic representation of the database.
Using a combination of sensitivity and uncertainty analysis, an adaptive sampling technique was developed to find the best location to perform the next analysis to minimize the objective function's uncertainty at minimum analysis cost.
This was extended to include physics-based uncertainty quantification for the RANS CFD simulations \cite{mukhopadhaya2020multi}.

The current work follows from this and extends the databases from exclusively describing longitudinal dynamics to include lateral-directional information. 
The databases are concerned with the force coefficients of lift, drag, and side-force, $C_L, C_D,$ and $C_{SF}$, respectively, and the moment coefficients of pitch, roll, and yaw, $C_m, C_l,$ and $C_n$, respectively. 
These coefficients are treated as functions of the angle of attack ($\alpha$) and angle of sideslip ($\beta$).
The databases are divided into two parts, the aerodynamics databases that describe the bare airframe aerodynamics, and the controls databases that describe the effect of controls surface deflections. 
Simple linear combinations of the two are used to determine the aircraft's final forces and moments.
The generic coefficient $C_i$ is calculated as a function of $\alpha, \beta,$ and the control surface deflections $\delta_j$ as
\begin{equation}
    C_i\left (\alpha, \beta, \delta_1, ..., \delta_N \right ) = C_{i0} \left (\alpha, \beta \right ) + \sum_{j}^{N} C_{i_{\delta_j}}  \left (\alpha, \beta, \delta_j \right ),
\end{equation}
where $i0$ refers to the bare airframe coefficient and $C_{i_{\delta_j}}$ refers to the incremental effect the control surface $\delta_j$ on coefficient $C_i$:
\begin{equation}
    C_{i_{\delta_j}}  \left (\alpha, \beta, \delta_j \right ) = C_i \left (\alpha, \beta, \delta_j \right ) - C_{i0} \left (\alpha, \beta \right ).
\end{equation}
Deflections of the ailerons, rudder, elevator, flaps, and spoilers are used.

Chapter \ref{chap:aero_db} delves into the details of creating multi-fidelity aerodynamics and controls databases using multi-fidelity GP regression.
This creates probabilistic databases that can be sampled to create hundreds of individual databases with slight variations due to the uncertainties present in the underlying data. 
AVL simulations, RANS CFD simulations, and wind and water tunnel experiments provide three different information sources of increasing fidelity.
The required data is generated, and databases are created for the generic T-tail transport (GTT) aircraft \cite{cunningham_generic_2018}.

\section{Certification by Analysis} \label{intro_cba}

To fly a new aircraft design, it needs to go through rigorous air-worthiness testing to ensure it is safe and can perform predictably in a variety of different situations.
These tests are defined and carried out by the Federal Aviation Administration (FAA) \cite{romanowski_flight_2018} in the US.
They occur at the very end of the design process, once a functional full-scale aircraft is built. 
Failing a certification test at this stage would require a redesign that would be incredibly expensive. 
To prevent this from occurring, safety factors are employed to account for potential uncertainty and error in the design analyses.
The quantification of uncertainties in analyses can replace arbitrary safety factors with the explicitly calculated design margins required to account for the errors.
This is the cornerstone of reliability-based design processes \cite{reliability}.

Certification by analysis (CbA) purports that the passing of certification requirements can be done using purely simulation-based analyses. 
It is the logical conclusion to the current trend of increased reliance on computer simulations for design analysis.
CbA has garnered interest from the aerospace community with the maturing of simulation procedures.
To achieve this, simulations would have to be as accurate, if not more accurate, than what is possible with flight testing.  
Many required improvements to simulation capabilities \cite{slotnick_cfd_nodate} will take years to develop.
The effects of uncertainties on flight performance predictions and predicted performance in flight testing could be quantified in the interim.
This provides aircraft designers with a method to estimate the likelihood that a design will pass or fail a certification test.

\subsection{Uncertainty Propagation}

A common method for uncertainty propagation is to perform a Monte Carlo analysis.
This is a brute-force method of characterizing the effect of input uncertainties on an output quantity of interest (QoI) \cite{janssen2013monte,thompson1992monte}.
It involves running multiple deterministic calculations where input variables are randomly sampled from their respective probability distributions.
The results are analyzed to determine the effect of the variation in the input variables on the posterior probability distribution of an output quantity of interest (QoI). 
In this work's context, the deterministic calculations are the flight simulations, the input values are the aerodynamics and controls databases, and the QoI is the success/failure of the certification test. 

A significant advantage of using GP regression to represent the probabilistic aerodynamic databases is the ability to take deterministic samples of the database.
Each sample is a potential candidate aircraft that could explain the data used to create the databases. 
They represent aircraft that have a slightly different performance from each other. 
These differences respect the uncertainty associated with the underlying data.
Since multi-fidelity GP regression is used to model the databases, the input probability distributions are Gaussian. 
The flight simulation does not represent a linear transfer function.
Therefore the posterior probability distribution of the QoI is arbitrary. 
As such, cumulative density functions are used to describe the posterior probabilities. 

\subsection{Aircraft Maneuver Simulation}

Aerodynamic databases contain all the information needed to perform high-fidelity flight simulations.
With multi-fidelity aerodynamics and controls databases, a virtual representation of the GTT aircraft can fly through real-world flight certification maneuvers. 
With the help of industry experts at The Boeing Company, a maneuver was picked from the FAA's \textit{Flight Test Guide for Certification of Transport Category Airplanes} \cite{romanowski_flight_2018}.
Within Chapter \textit{5.3 Directional and Lateral Control} of the guide, the \textit{Lateral Control: Roll Capability \S 25.147(d)} maneuver was chosen.   
The testing procedure is paraphrased as: 
\begin{enumerate}
    \item The airplane starts in a trimmed state for steady straight flight at maximum takeoff speed.
    \item Establish a steady $30^\circ$ banked turn.
    \item Roll the airplane to a $30^\circ$ bank angle in the other direction.
    \item Aircraft must have sufficient roll authority to perform the $60^\circ$ change in bank angle in no more than $11$ seconds. 
    \item The aircraft must do this with one engine inoperative, specifically the one that makes this maneuver more difficult.

\end{enumerate}

This part of the work leans heavily on The Boeing Company's expertise in flight simulation and control law mixing.
Due to proprietary and patent restrictions, the exact implementation of the flight simulation code is unavailable, but enough information is provided to outline the simulations' overarching methodology and workflow.
This maneuver is simulated using a 5 degree of freedom (DOF) flight simulator that The Boeing Company uses in their design processes.
Details of the maneuver simulation process are discussed in Section \ref{sec:sim_procedure}.

\section{Aircraft Configurations}

The methodologies presented in this thesis were developed and demonstrated using two aircraft configurations: the NASA Common Research Model (CRM) and the Generic T-tail Transport (GTT). 

\subsection{NASA Common Research Model}
The NASA CRM is a well-investigated full-configuration aircraft \cite{rivers_further_2012,rivers_experimental_2010} that was developed to create a baseline geometry upon which numerous experimental and computational studies could be performed and compared.
It was created for the 4th Drag Prediction Workshop \cite{morrison20094th} and was used for the subsequent 5th and 6th editions of the workshop as well \cite{levy2013summary,morrison20166th,roy2017summary,tinoco2017summary}.
The wealth of experimental and computational data lends itself well to showcase the performance of the uncertainty quantification and multi-fidelity data fusion techniques developed in this thesis.

\begin{figure}
\center
\subfigure[\label{subfig:nasa_crm_wt_img} NASA Common Research Model]
  {\includegraphics[width=0.48\textwidth]{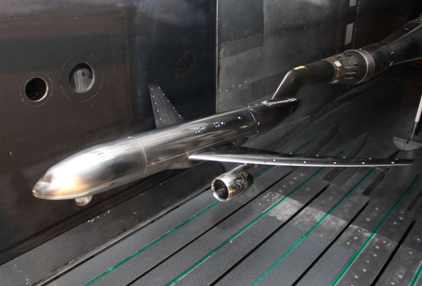}}
\subfigure[\label{subfig:gtt_wt_img} Generic T-tail Transport]
  {\includegraphics[width=0.48\textwidth, trim=0 92 0 0, clip]{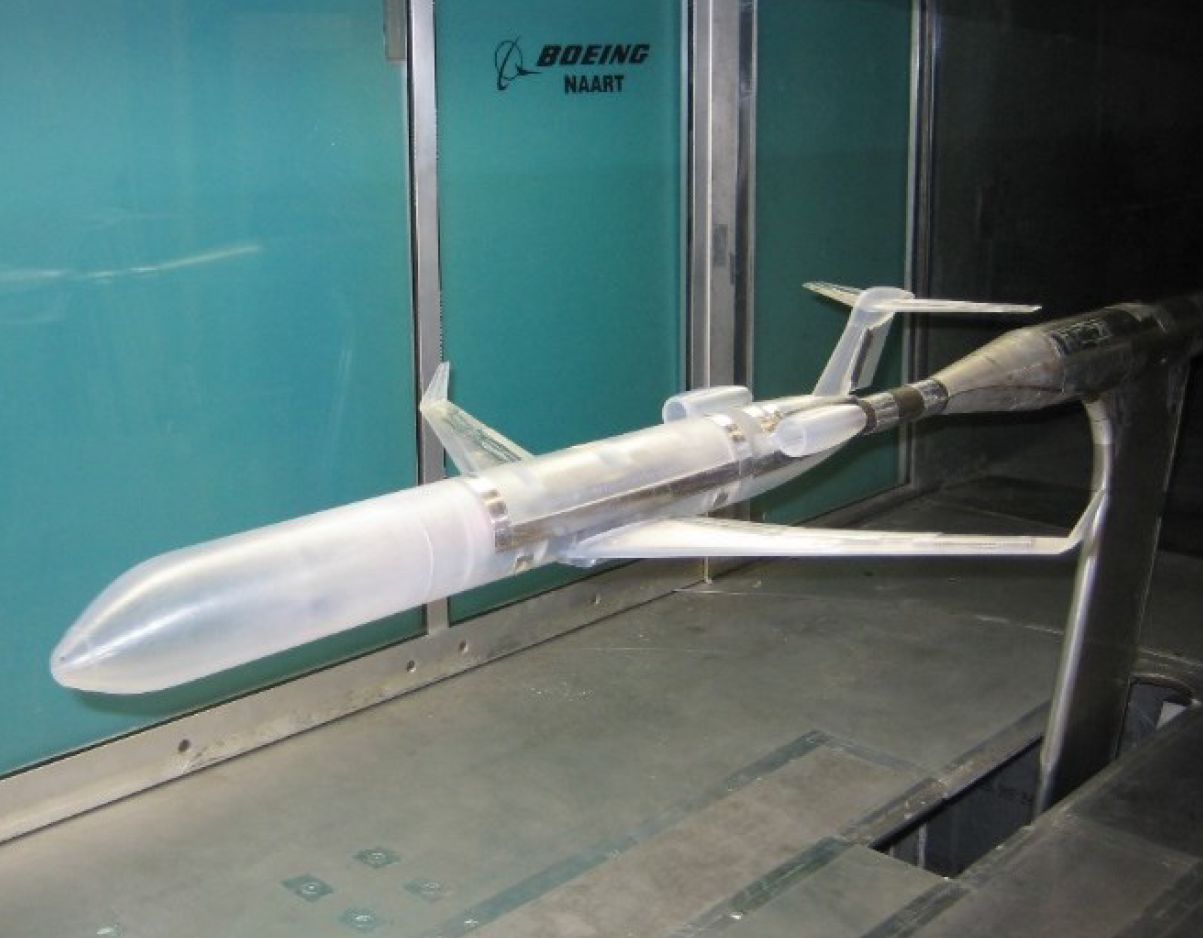}}
\caption{Wind tunnel models for the two aircraft configurations that are investigated in this work.}\label{fig:aircraft_images}
\end{figure}

The model, designed by The Boeing Company, is based on the Boeing 777 aircraft with a modified wing. 
It is a conventional tube-wing configuration designed for a cruise Mach number of 0.85.
The NASA CRM was built to be modular such that additional components could be attached to the baseline geometry. 
Consequently, different configurations of the aircraft were tested in the wind tunnel:
\begin{enumerate}
    \item Baseline wing + fuselage model,
    \item Wing + fuselage + pylon and nacelle,
    \item Wing + fuselage + horizontal tail mounted at either $-2^\circ, 0^\circ,$ or $+2^\circ$.
\end{enumerate}
Figure \ref{subfig:nasa_crm_wt_img} shows one such configuration in a wind tunnel. 
The configuration with the wing + fuselage + horizontal tail mounted at $-2^\circ$ is used for this work.


\subsection{Generic T-tail Transport}

The GTT aircraft was developed in another collaborative effort between NASA and The Boeing Company. 
This effort aimed to provide experimental and computational data to inform a flight simulator that provides stall recovery training to pilots \cite{cunningham_generic_2018}.
The aircraft configuration derives from a remotely piloted airplane configuration designed by NASA and built by Area-I \cite{kuehme_flight_2014}.
It is a $16\%$ scaled version of the Bombardier CRJ700, which is currently in-service.
The configuration resembles a generic short to medium range single-aisle regional jet with a T-tail and twin aft engines.
Tables \ref{tab:gtt_ref_dim} and \ref{tab:gtt_mass_prop} lists the aerodynamic reference dimensions and the mass properties of the GTT aircraft, respectively.

\begin{table}
\parbox{0.5\textwidth}{
\centering
    \renewcommand{\arraystretch}{1.2}
    \captionsetup{justification=centering}
    \begin{tabular}{|c|c|}
        \hline
        Mean Aerodynamic Chord & $3.374~\mathrm{m}$ \\ \hline
        Wing Span & $23.159~\mathrm{m}$ \\ \hline
        Wing Area & $70.079~\mathrm{m^2}$ \\ \hline
    \end{tabular}
    \caption{Aerodynamic reference dimensions for the GTT aircraft.} 
    \label{tab:gtt_ref_dim}
}
\parbox{0.5\textwidth}{
\centering
    \renewcommand{\arraystretch}{1.2}
    \captionsetup{justification=centering}
    \begin{tabular}{|c|c|}
        \hline
        Weight & $25,332~\mathrm{kg}$ \\ \hline
        $I_{xx}$ & $238,419~\mathrm{kg~m^2}$ \\ \hline
        $I_{yy}$ & $1,510,624~\mathrm{kg~m^2}$ \\ \hline
        $I_{zz}$ & $1,717,539~\mathrm{kg~m^2}$ \\ \hline
    \end{tabular}
    \caption{Mass properties of the GTT aircraft.}
    \label{tab:gtt_mass_prop}
}
\end{table}

The primary motivation for using the GTT configuration for this work is the wealth of experimental wind tunnel data that is available \cite{cunningham_preliminary_2018}. 
Three experimental campaigns in different facilities explored various aspects of the aircraft performance characteristics:
\begin{enumerate}
    \item NASA Langley Research Center 12-Foot Low-Speed Tunnel (12-Foot LST): Focus on stability derivatives and $\alpha$ sweeps,
    \item Boeing North American Aviation Research Tunnel (NAART): Focus on control derivatives and $\alpha$ and $\beta$ sweeps,
    \item Boeing Flow Visualization Water Tunnel (FVWT): Focus on stability derivatives at various values of $\alpha$.
\end{enumerate}
Figure \ref{subfig:gtt_wt_img} shows the GTT model mounted in the NAART facility.
This work only uses data from the NAART and the FVWT experimental campaigns.
The data from the 12-Foot LST is a subset of the data from the other two sources. 
Table \ref{tab:gtt_scale_factors} lists the scales of the models used for the various wind tunnel campaigns and the CFD campaign performed using SU2 for this work. 

\begin{table}
\centering
    \renewcommand{\arraystretch}{1.2}
    \captionsetup{justification=centering}
    \begin{tabular}{|c|c|}
        \hline
        Data Source & Scaling factor \\ \hline
        NASA 12-Foot LST & $0.057$ \\ \hline
        Boeing NAART & $0.020$ \\ \hline
        Boeing FVWT & $0.019$ \\ \hline
        SU2 & $0.158$\\ \hline
    \end{tabular}
    \caption{Scale factors for the different sources of data.}
    \label{tab:gtt_scale_factors}
\end{table}

The NAART and 12-Foot LST models tested individual control surface parts to define the control characteristics of the aircraft. 
Table \ref{tab:gtt_cs_range} lists the control surfaces and their associated deflection ranges.
In addition to these, the entire horizontal stabilizer could be rotated to change the its incidence angle. 
However, since most of the wind tunnel tests mounted the horizontal stabilizer at an angle of $-6^\circ$, only that particular incidence angle is used.

\begin{table}
    \renewcommand{\arraystretch}{1.2}
    \centering
    \begin{tabular}{ |c|c| } 
        \hline
        Control Surface & Deflection Range \\ \hline
         Ailerons &  $-25^\circ \leq \delta_a \leq 25^\circ$ \\ \hline
         Elevator &  $-30^\circ \leq \delta_e \leq 30^\circ$ \\ \hline
         Rudder & $0^\circ \leq \delta_r \leq 35^\circ$ \\ \hline
         Flaps & $0^\circ \leq \delta_f \leq 60^\circ$ \\ \hline
         Spoilers & $0^\circ \leq \delta_s \leq 60^\circ$ \\ \hline
    \end{tabular}
    \caption{Control surface deflection ranges for the GTT aircraft.}
    \label{tab:gtt_cs_range}
\end{table}

\section{Contributions} \label{intro_contributions}

This thesis establishes a framework to perform virtual flight testing of an aircraft early in the design process.
While the focus is on aircraft design, the principles of multi-fidelity modeling and uncertainty quantification, and certification testing can be applied to any design problem that requires significant analysis. 

Starting with UQ, Chapter \ref{chap:rans_uq} reviews the eigenspace perturbation methodology to quantify epistemic uncertainties introduced by turbulence models into RANS CFD simulations.
The methodology is implemented in an open-source CFD solver, SU2, to enable its widespread use in the research community. 
It is validated on a bevy of test cases that range from commonly used benchmark flow conditions to those of specific aerospace interest. 
The relationship between turbulence modeling uncertainty and numerical discretization error, another common source of uncertainty in CFD simulations, is investigated.
The UQ methodology is applied to two aircraft, the NASA Common Research Model and the Generic T-tail Transport, to create aerodynamic databases with physics-informed uncertainties. 

Since CFD is not the only analysis technique used in the aircraft design process, Chapter \ref{chap:mf_gp} presents multi-fidelity Gaussian processes (GP) that are used to combine data from different information sources. 
Existing equations for multi-fidelity GP regression are extended to use noisy data when design sets are not nested.
Multi-fidelity aerodynamic databases are modeled using these equations.
AVL simulations are the lowest fidelity level, RANS CFD simulations are the middle-fidelity, and wind tunnel experimental data are the highest fidelity.
The benefits of multi-fidelity data fusion are elucidated using one-dimensional aerodynamic databases created for the NASA CRM aircraft. 
This is extended in Chapter \ref{chap:aero_db}, where the first comprehensive, multi-fidelity, multi-dimensional aerodynamics and controls databases representing a full-configuration aircraft's lateral and longitudinal dynamics are presented. 
These are created for the Generic T-tail Transport (GTT) aircraft.

With all aspects of an aircraft's performance defined, Chapter \ref{chap:cba} delves into the simulation of an FAA flight certification maneuver used to test commercial jets' air-worthiness.
The Monte Carlo method is used to propagate the uncertainties in the design analyses through the flight simulation.
The effect of the input uncertainty on the aircraft maneuver is analyzed, and a probability of succeeding/failing the certification test is explicitly quantified. 
The virtual flight testing is also performed using databases based on lower-fidelity data, representing what would be available at earlier stages in the design process.
By quantifying the likelihood of success/failure of a flight certification maneuver, potential problems in the aircraft design can be identified, and the risk of failure can be mitigated. 

These contributions are discussed in further detail in Chapter \ref{chap:conclusions}.
Potential avenues for future research are presented as well. 

\chapter{Quantifying Uncertainties in RANS CFD Simulations} \label{chap:rans_uq}

The Navier-Stokes equations are a set of non-linear partial differential equations that describe the motion and behavior of fluids.
Broadly, there are two classes of forces acting on a fluid: inertial and viscous.
Inertial forces refer to those exerted by the fluid's momentum or its lack thereof.
Consider the force exerted by water coming out of a tap versus water coming out of a fire hose.
The water flowing out of the fire hose has significantly larger momentum by virtue of its flow velocity.
This results in higher inertial forces being exerted by the water flowing out of the fire hose.
Viscous forces are a result of the fluid's resistance to deformation.
Viscosity is a measure of this resistance.
It is a property specific to the fluid in question.
Honey is more viscous than water.
Pouring both fluids out of a container demonstrates the effect of these viscous forces.
Honey is more resistant to this deformation and flows out more slowly than water.

The ratio of the inertial and viscous forces on a fluid determines the fluid's behavior in motion.
Reynolds number $\left ( Re = \frac{\rho u L}{\mu}\right )$ is a measure of this ratio. When the inertial forces acting upon a fluid are small compared to the viscous forces, i.e., the Reynolds number is low, the fluid flow is  \textit{laminar}.
This flow is smooth, without significant cross-currents or time dependence.
Laminar flow usually occurs at lower fluid velocity, is relatively easy to predict computationally, and does not require any simplifying models.

Fluid flow turns \textit{turbulent} when the ratio of inertial to viscous forces is high, i.e., at high Reynolds numbers.
Turbulence refers to the chaotic, time-dependent, varied-scale fluctuations in flow variables that characterize turbulent flows.
Its unsteady nature, and the range of length and time scales of the flow eddies, make it computationally intractable to predict precisely.
Simplifying assumptions are required to make the computations feasible and productive in an engineering situation. 

Note the ambiguous use of ``low'' and ``high'' to describe the Reynolds numbers for laminar and turbulent flow.
This ambiguity is intentional.
The point of transition between the two regimes is highly dependent on the fluid and the geometry of the flow itself.
It is an active area of research \cite{arnal2000laminar}.
This thesis focuses on turbulent flows. 

\section{Turbulent Flows}

In the realm of engineering, there is often a need to predict fluid flow behavior in and around engineered products.
These situations can range from simulating the flow around an airplane to simulating the flow through laptop cooling systems.
Computational Fluid Dynamics (CFD) is the field of study that involves using computers and complex numerical analysis techniques to solve the non-linear Navier-Stokes equations on the flow domain of interest.
As computational capabilities have increased over the past few decades, so has the reliance on the predictive capability of CFD simulations. 

CFD simulations are made more difficult due to turbulence.
The range of length and time scales that need resolving through spatial and temporal discretization makes it computationally intractable to solve precisely, i.e., without simplifying models.
Turbulence plagues most flows of engineering interest.
The difficulty in solving these flows has paved the way for developing a hierarchy of solution techniques that trade computational cost for prediction accuracy. 

\begin{figure}
    \center
    \includegraphics[width=0.75\textwidth]{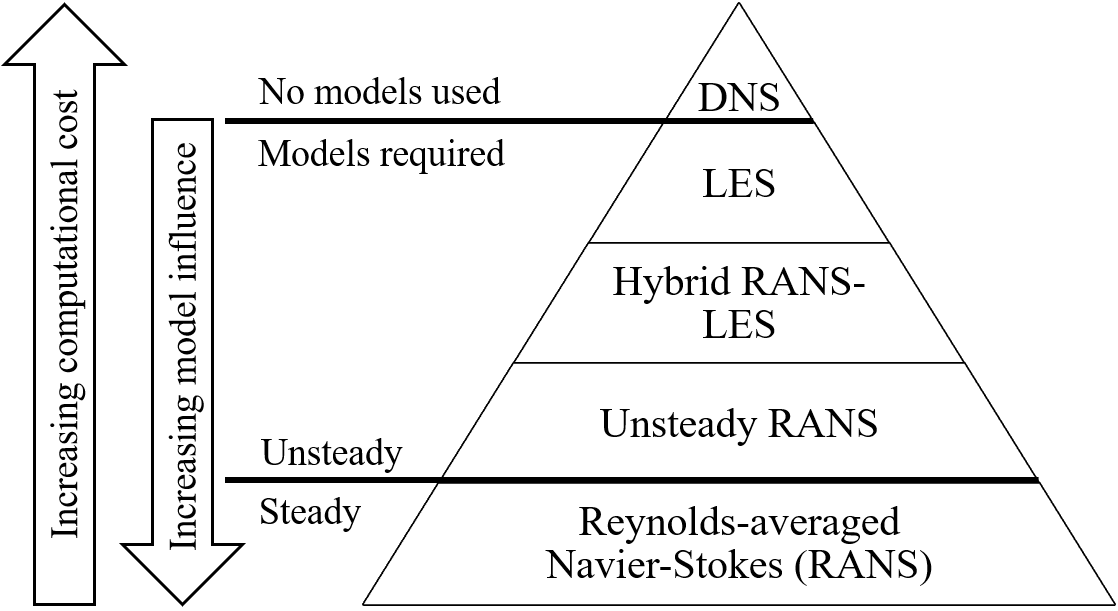}
    \caption{Hierarchy of CFD solution techniques in order of increasing computational cost and decreasing model influence. \label{fig:cfd_types}}
\end{figure}

Figure \ref{fig:cfd_types} presents this hierarchy in the form of a pyramid.
Sitting at the top of this pyramid are Direct Numerical Simulations or DNS.
These calculations do not employ any mathematical models and resolve all spatial and temporal time scales to give an exact reproduction of the fluid behavior.
These time-varying (unsteady) simulations are computationally expensive and memory intensive.
At the time of writing, DNS calculations are only possible at low Reynolds numbers and for simple geometries such as flat plates \cite{hoyas_reynolds_2008}, and channels \cite{laval_marquillie_dns_channel,marquillie_instability_2011}.
These limitations disqualify the use of DNS, in its current state, for practical engineering applications. 

Below DNS, in Figure \ref{fig:cfd_types}, lie Large Eddy Simulations (LES) and Hybrid RANS-LES simulations.
These unsteady simulations resolve some, but not all, of the scales of turbulence.
Unresolved time and length scales are modeled using simplifying assumptions \cite{pope_2000}.
These solution methodologies allow for the fine-tuning of the desired influence of simplifying models through the chosen level of spatial and temporal discretization.
They are in the process of being adopted by industry for specific use cases such as performance predictions for high-lift configuration aircraft \cite{rumsey2019overview}, or engine simulations. 

The pyramid's base represents the most widely used method in the industry, Reynolds-averaged Navier-Stokes (RANS) simulations.
These simulations suffer the most from modeling inaccuracies.
They assume that the flow is steady (no time-dependent variation in the flow features) and require simplifying turbulence models that aggregate the effects of the turbulent eddies that would be present in the flow.
These simplifications significantly reduce the computational cost but inhibit the flow features that the simulations can capture.
While its unsteady counterpart can resolve some of the time variations of the flow, turbulence modeling is still required, affecting prediction accuracy.
Steady RANS simulations are very computationally efficient and can be used for expensive undertakings such as iterative aerodynamic shape optimization \cite{lyu2015aerodynamic}, and aircraft database generation \cite{wendorff_combining_2016}.
This chapter will focus on quantifying the uncertainties injected by turbulence models employed in RANS simulations. 

\section{Uncertainty and Error in RANS Simulations}

In order to use steady RANS simulations to design engineering products, it is essential to understand its strengths and shortcomings.
The extensive use of mathematical models to accelerate these simulations makes it feasible to run multiple function evaluations at different test conditions.
Concurrently, the inadequacy of the models in predicting real-world behavior of fluids limits its use to flow conditions where the models are valid.
Over the years, these restrictions have been found through thorough the validation of CFD simulations and turbulence models \cite{oberkampf_verication_2002} against experimental data. 

Precise language is required to have a productive discussion of uncertainties and errors in these simulations.
For this purpose, definitions and guidelines established by the American Institute of Aeronautics and Astronautics (AIAA) in \cite{computational_fluid_dynamics_committee_guide_1998} are followed.
In a non-academic setting, \textit{error} and \textit{uncertainty} are often used interchangeably.
Ambiguity is avoided in the current discussion by defining uncertainty as: 
\say{A potential deficiency in any phase or activity of the modeling process that is due to lack of knowledge} \cite{computational_fluid_dynamics_committee_guide_1998}. 

In contrast, errors are defined as: 
\say{A recognizable deficiency in any phase or activity of modeling and simulation that is not due to lack of knowledge} \cite{computational_fluid_dynamics_committee_guide_1998}.
Two key differences surface from these definitions: 

\begin{enumerate}
    \item \textbf{Potential vs. recognizable deficiency -} the effects of the uncertainty in some model may or may not cause a knowable deficiency in the resulting prediction. On the other hand, deficiencies introduced due to modeling errors are identifiable upon examination. 
    
    \item \textbf{Lack of knowledge -} uncertainties arise due to inadequate modeling of the real-world physics in the simulation process.  Errors exist even with complete knowledge and often have established practices and methods to reduce them.  
\end{enumerate}

To solidify these definitions with examples, consider the uncertainty introduced by turbulence models in contrast to discretization errors caused by insufficient grid resolution.
Turbulence models are required to solve the closure problem with the RANS equations.
There are numerous models and numerous variations built upon those models. 
Their hyper-parameters are often calibrated to work well in certain flow conditions.
This means that the deficiencies in the employed model may or may not present themselves, depending on the flow regime.
There is a fundamental lack of knowledge that necessitates the development and use of so many different turbulence models.
For these reasons, the deficiencies introduced by turbulence models are considered \textit{uncertainties}.

Solving non-linear PDEs, such as the RANS equations, requires discretizing the continuous spatial and temporal domains into discrete ones.
The discrete spatial domain is known as the grid or mesh.
An insufficient number of grid points introduces discretization errors.
As the number of grid points increases, i.e., the size of the grid spacing tends to zero, the discrete representation approaches the original continuous domain.
Correspondingly, the discretization error approaches zero.
Thus, discretization error results from insufficient grid quality and limited computational resources rather than a lack of knowledge.
It can be quantified using grid convergence studies \cite{american_society_of_mechanical_engineers_standard_2009}.
For these reasons, deficiencies due to insufficient discretization are considered \textit{errors}.
A similar argument can be made for temporal discretization.

Numerous sources of errors and uncertainties can plague a CFD simulation. 
Table \ref{tab:errors_uncert} lists a sampling of these sources.
Errors due to iterative convergence, programming, and usage are easily mitigated.
Computer round-off error is inevitable yet inconsequential when compared to discretization error.
This chapter focuses on turbulence modeling uncertainties and their relationship with discretization errors.
Section \ref{sec:equips_rans_uq} introduces the eigenspace perturbation methodology, which provides a framework to estimate the uncertainties introduced by turbulence models.
This methodology is implemented in the open-source CFD software, SU2 \cite{su2_aiaajournal}.
Section \ref{sec:VandV_rans_uq} presents the validation of this methodology on a wide variety of test cases. 
Section \ref{sec:crm_rans_uq} goes on to present the methodology's performance on the NASA CRM aircraft.  

\begin{table}
    \centering
    \def\arraystretch{1.2}
    \begin{tabular}{c|c}
         \textbf{Errors} &  \textbf{Uncertainties} \\ \hline
         Discretization & Turbulence modeling \\
         Iterative convergence & Simulation conditions \\
         Computer round-off & Complexity of physics \\
         Programming & \\
         Usage & \\
    \end{tabular}
    \caption{Sources of errors and uncertainties in CFD simulations. This is not an exhaustive list.}
    \label{tab:errors_uncert}
\end{table}

\section{Eigenspace Perturbation Methodology} \label{sec:equips_rans_uq}

The balance of computational cost and fidelity provided by Reynolds Averaged Navier-Stokes (RANS) simulations makes it the tool of choice in most engineering CFD applications.
Reynolds averaging starts with the decomposition of the chaotic velocity field ($ u_i$) into its mean ($\overline{u_i}$) and fluctuating ($u_i'$) velocity components such that $u_i = \overline{u_i} + u_i'$. Here $i$ denotes a coordinate direction, $i=1, 2, 3$.
This decomposition allows for the ensemble-averaging of the Navier-Stokes equations \cite{pope_2000}, leading to a closure problem due to the resulting non-linear $\overline{u_i'u_j'}$ term.
This term is also known as the Reynolds stress tensor, $R_{ij}$ and requires modeling. 

Turbulence models are concerned with predicting the behavior of the Reynolds stress tensor throughout a flow field of interest, without any a priori high-fidelity data, in a computationally tractable manner.
To this end, these models often make simplifying assumptions about the tensor that can inject significant uncertainties into their predictions.
For example, a popular assumption employed in numerous turbulence models is the Boussinesq approximation (also known as the linear eddy viscosity hypothesis).
It assumes that $R_{ij}$ can be defined by a combination of the mean rate of strain ($S_{ij}$), an eddy viscosity ($\nu_t$), and the turbulent kinetic energy ($k$):
 \begin{equation}\label{equ:rst}
     R_{ij} = \nu_t S_{ij} - \frac{2}{3} k \delta_{ij},
 \end{equation}
where $S_{ij} = \left ( \frac{\partial \overline{u_i}}{\partial x_j} + \frac{\partial \overline{u_j}}{\partial x_i} \right )$, $x_i$ is a coordinate direction, and $\delta_{ij}$ is the Kronecker delta.
This linear eddy viscosity model purports a simplified proportional relationship between the Reynolds stress tensor and the mean rate of strain tensor.
It assumes that the turbulent fluid is an isotropic medium where Reynolds stresses are always aligned with the mean strain rate.
Although reasonable for simple flows without adverse pressure gradients, this assumption can severely limit the flow features predicted by the turbulence model, thus introducing uncertainty into the QoIs predicted by RANS simulations.
The shortcomings of the Boussinesq assumption are particularly evident in areas of the operating envelope of an aircraft where separated flows and shock-boundary layer interactions exist. 

\begin{figure}
    \center
    \includegraphics[width=0.95\textwidth]{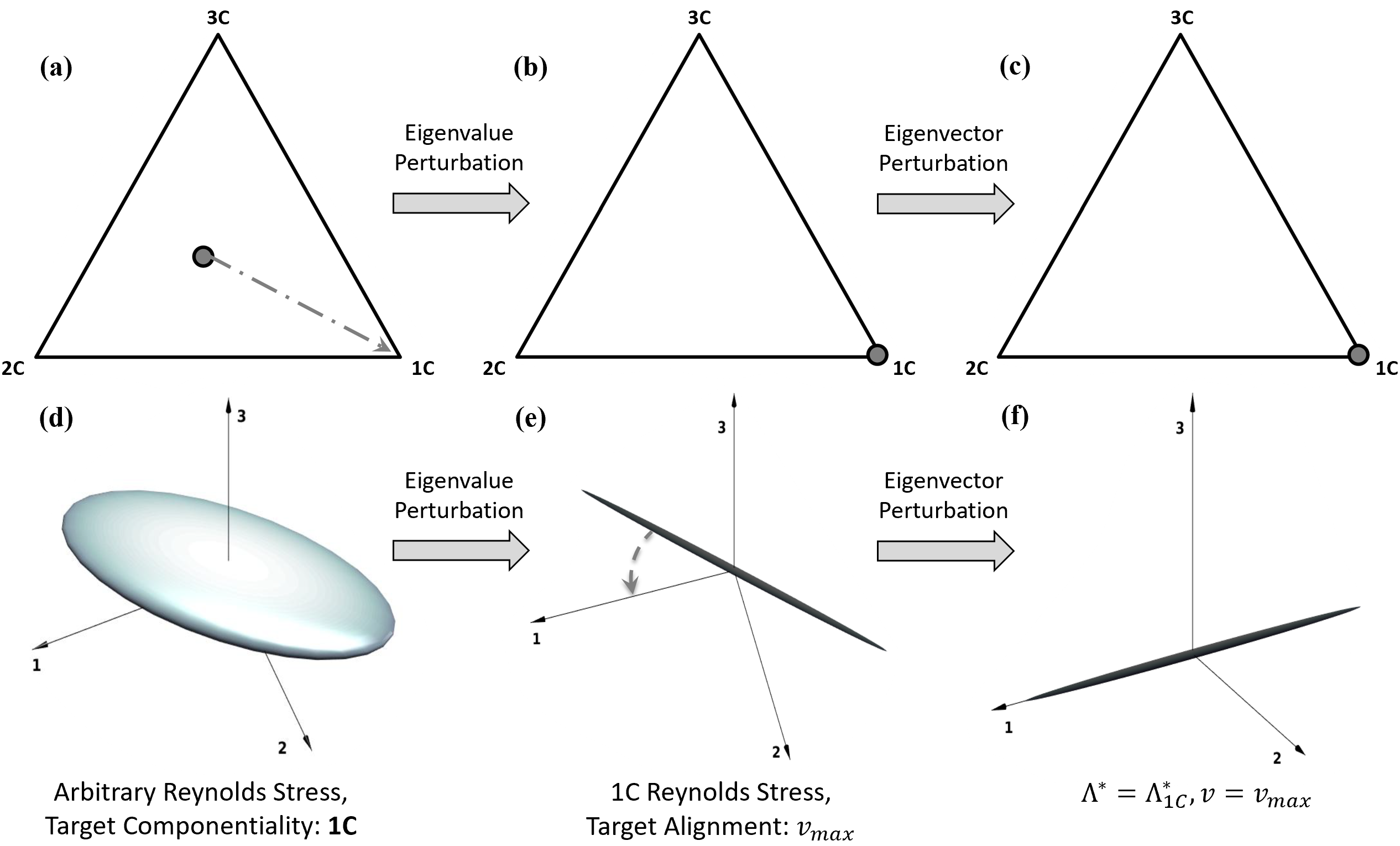}
    \caption{Schematic outline of Eigenspace perturbations from an arbitrary state of the Reynolds stress. \label{fig:pert_vis}}
\end{figure}

To better understand the effects of such an assumption, the eigenspace perturbation methodology was developed by Emory et al. \cite{emory2013modeling} and Mishra et al. \cite{iaccarino_eig_pert}. 
It aims to quantify model-form uncertainties arising from the use of turbulence models in RANS simulations by introducing perturbations in the eigenvalues and eigenvectors of the Reynolds stress anisotropy tensors $\left ( b_{ij} \right )$ that the models predict. 
This methodology does not rely on any higher-fidelity data.
The stress tensor is decomposed into its anisotropic and deviatoric components as
 
\begin{equation}\label{equ:rst_decomp}
    R_{ij}=2k(b_{ij}+\frac{\delta_{ij}}{3}).
\end{equation}
Here, $k~(=\frac{R_{ii}}{2})$ is the turbulent kinetic energy and $b_{ij}$ is the Reynolds stress anisotropy tensor.
The anisotropy tensor can be further decomposed into its eigenvalues and eigenvectors and represented as

\begin{equation}\label{equ:eigendecomposition}
b_{ij}=Q \Lambda Q^T,
\end{equation}
where $\Lambda$ is a diagonal matrix that contains the eigenvalues $\lambda_i \in \mathbb{R}$ \cite{Gerolymos2016AlgebraicPA}, and $Q$ is a matrix where the $i$-th column represents the eigenvector corresponding to $\lambda_i$.
The matrices $Q$ and $\Lambda$ are ordered such that $\lambda_{1}\geq\lambda_{2}\geq\lambda_{3}$.
To understand the eigenspace perturbations, it helps to visualize the Reynolds stress tensor as an ellipsoid where the eigenvectors define the ellipsoid axes, the eigenvalues define the relative lengths along each of the axes, and the turbulent kinetic energy defines the size of the ellipsoid.
Figure \ref{fig:pert_vis} \textbf{(d)} shows an example of such an ellipsoid, represented in a coordinate system defined by the eigenvectors of $S_{ij}$. 

The perturbations exercise the limits of the physical realizability constraints placed on the Reynolds stress tensor \cite{schumann1977realizability,speziale1994realizability,2014realizability}.
One way to graphically represent the constraints placed on the anisotropic part of the stress tensor is to use the eigenvalues ($\Lambda$) to project the stress tensor onto an anisotropy-invariant map.
The map often takes the form of a triangle.
The triangle's vertices represent the one-, two- and three-component limiting states of turbulence (referred to as the $1C$, $2C$, and $3C$ states).
All of the physically realizable states of the stress tensor lie within the triangle.
One such representation of an anisotropy-invariant map is the barycentric map \cite{banerjee2007presentation}.
This mapping allows the writing of the projection as a convex combination of the limiting states of turbulence: 

\begin{equation}\label{equ:barycentric_mapping}
    \textbf{x} = \textbf{x}_{1C} (\lambda_1 - \lambda_2) + \textbf{x}_{2C} (2\lambda_2 - 2\lambda_3) + \textbf{x}_{3C} (3\lambda_3 + 1), 
\end{equation}
where $\textbf{x}_{1C}$, $\textbf{x}_{2C}$, and $\textbf{x}_{3C}$ are the coordinates of the vertices that represent the one-, two-, and three-component limiting states of turbulence.
For example, the stress ellipsoid shown in Figure \ref{fig:pert_vis} \textbf{(d)} would be mapped on to the barycentric map as shown in Figure \ref{fig:pert_vis} \textbf{(a)}.

The eigenvalue perturbations are performed such that the limiting states of turbulence are simulated.
This involves perturbing the state of the tensor, sequentially, to each of the vertices of the triangle.
The eigenvalues resulting from these perturbations are: 

\begin{equation}\label{equ:eigenvalue_pert}
    \Lambda^*_{1C} = 
    \begin{bmatrix}
    2/3 & 0    & 0 \\
    0   & -1/3 & 0 \\
    0   & 0    & -1/3
    \end{bmatrix},~
    \Lambda^*_{2C} = 
    \begin{bmatrix}
    1/6 & 0   & 0 \\
    0   & 1/6 & 0 \\
    0   & 0   & -1/3
    \end{bmatrix},~
    \Lambda^*_{3C} = 
    \begin{bmatrix}
    0 & 0 & 0 \\
    0 & 0 & 0 \\
    0 & 0 & 0
    \end{bmatrix}.
\end{equation}
where $*$ represents the perturbed state.
The eigenvector perturbation involves changing the alignment such that the perturbed state is
\begin{equation}\label{equ:eigenvector_pert}
    Q^* = Qv
\end{equation}
where $v$ is either:
\begin{equation}\label{equ:vmin_vmax}
    v_{max} = 
    \begin{bmatrix}
    1 & 0 & 0 \\
    0 & 1 & 0 \\
    0 & 0 & 1
    \end{bmatrix},~
    v_{min} = 
    \begin{bmatrix}
    0 & 0 & 1 \\
    0 & 1 & 0 \\
    1 & 0 & 0
    \end{bmatrix}.
\end{equation}
Combining the eigendecomposition from Equation \eqref{equ:eigendecomposition}, the eigenvalue perturbations from Equation \eqref{equ:eigenvalue_pert}, and the eigenvector perturbations from Equations (\ref{equ:eigenvector_pert}, \ref{equ:vmin_vmax}), the perturbed anisotropy tensor can be built as
\begin{equation}
    b^*_{ij}=Q^* \Lambda^* Q^{*T}.
\end{equation}

In the case shown in Figure \ref{fig:pert_vis} \textbf{(b)}, the eigenvalues are perturbed to the 1C limiting state.
This state corresponds to changing the shape of the ellipsoid to be infinitely long in one direction, as shown in Figure \ref{fig:pert_vis} \textbf{(e)}.
Consequently, the eigenvector alignment is changed to maximize turbulent kinetic energy production ($v_{max}$).
This is represented in Figure \ref{fig:pert_vis} \textbf{(c)} and \textbf{(f)}.
The change in alignment of the eigenvector can not be represented in the barycentric map, which is why Figures \ref{fig:pert_vis} \textbf{(c)} and \textbf{(b)} are identical.

\begin{figure}
    \centering
    \begin{subfigure}[$2C$ state with $v_{max}$ eigenvector alignment.] {
        \includegraphics[trim=200 40 280 30, clip, width=0.24\textwidth]{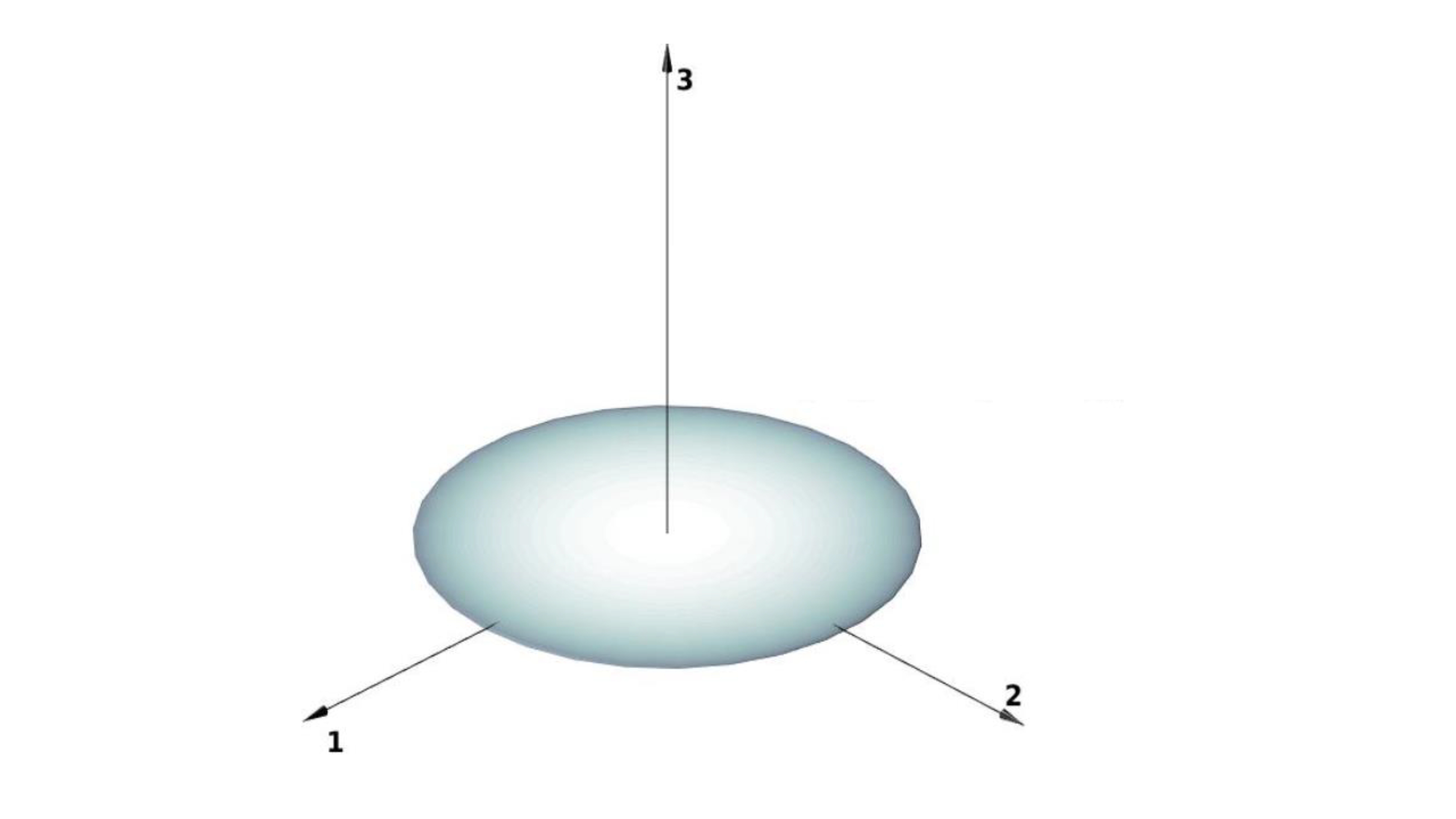}
        \label{fig:2c_max}
    }
    \end{subfigure}
    \hspace{10pt}
    \begin{subfigure}[$2C$ state with $v_{min}$ eigenvector alignment.]{
        \includegraphics[trim=230 20 230 30, clip, width=0.24\textwidth]{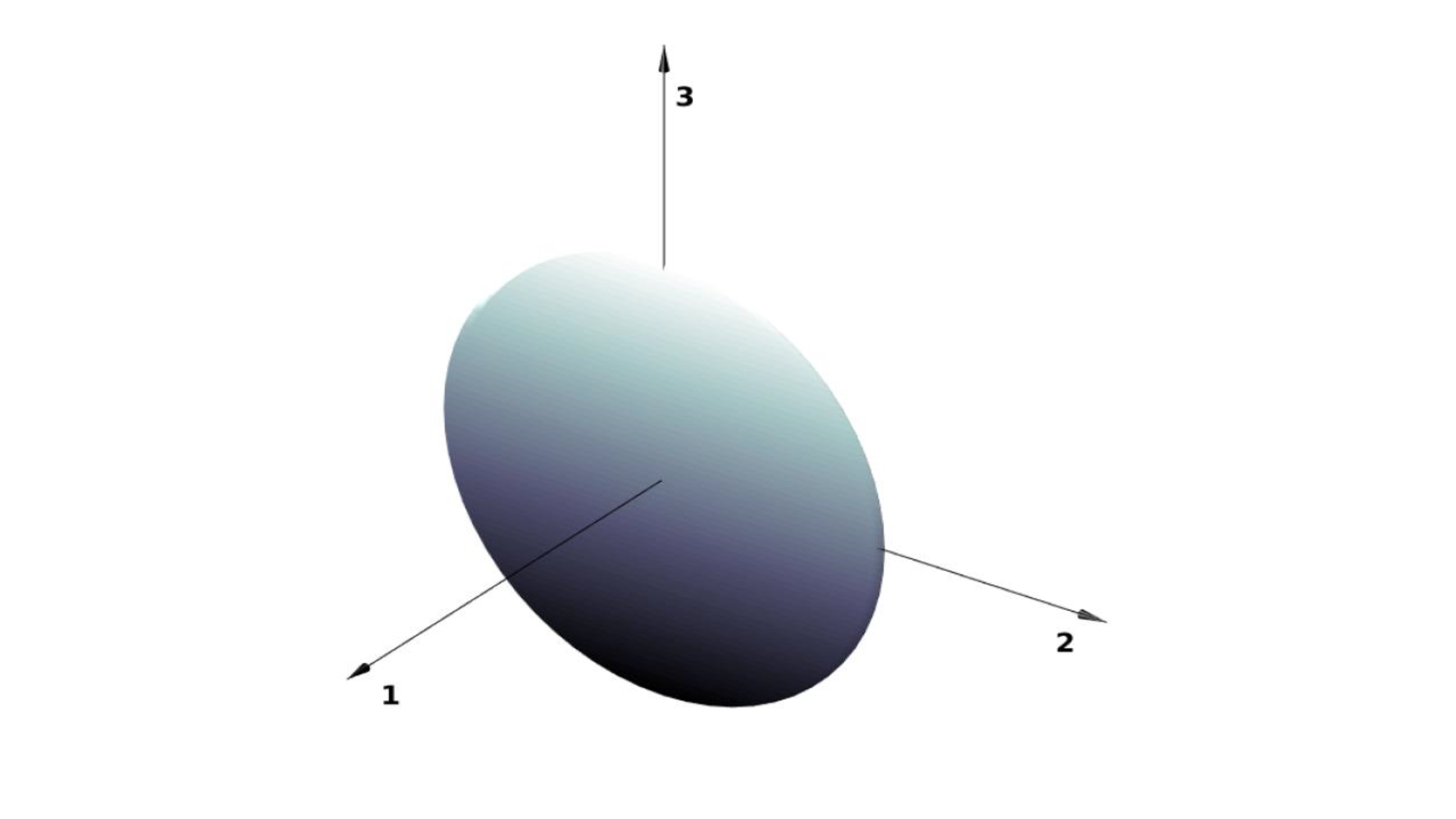} 
        \label{fig:2c_min}
    }
    \end{subfigure}
    \hspace{10pt}
    \begin{subfigure}[$3C$ isotropic turbulence state.]{
        \includegraphics[trim=210 40 280 50, clip, width=0.24\textwidth]{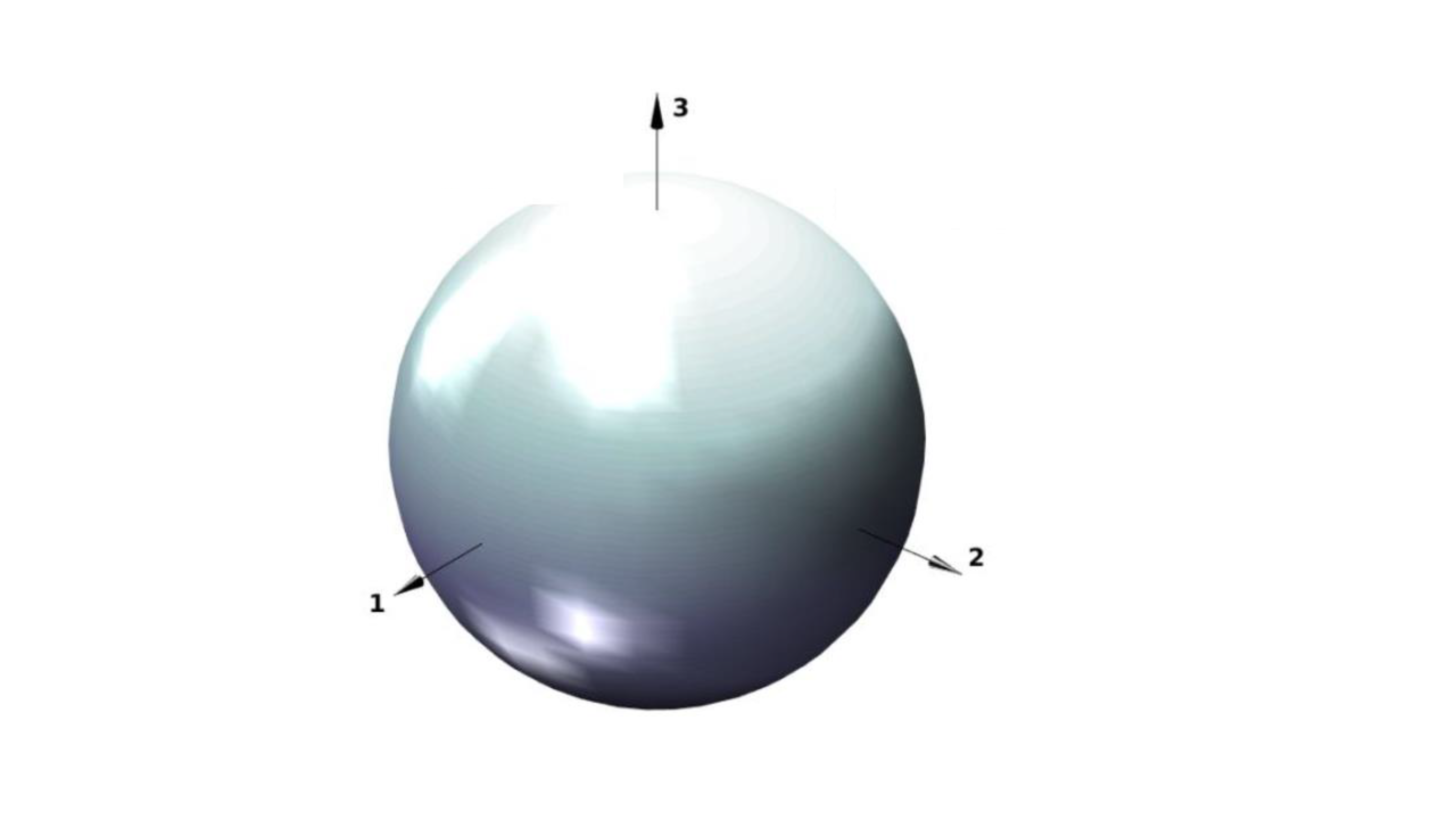} 
        \label{fig:3c}
    }
    \end{subfigure}
    
    \begin{subfigure}[$1C$ state with $v_{max}$ eigenvector alignment.]{
        \includegraphics[trim=200 40 300 50, clip, width=0.24\textwidth]{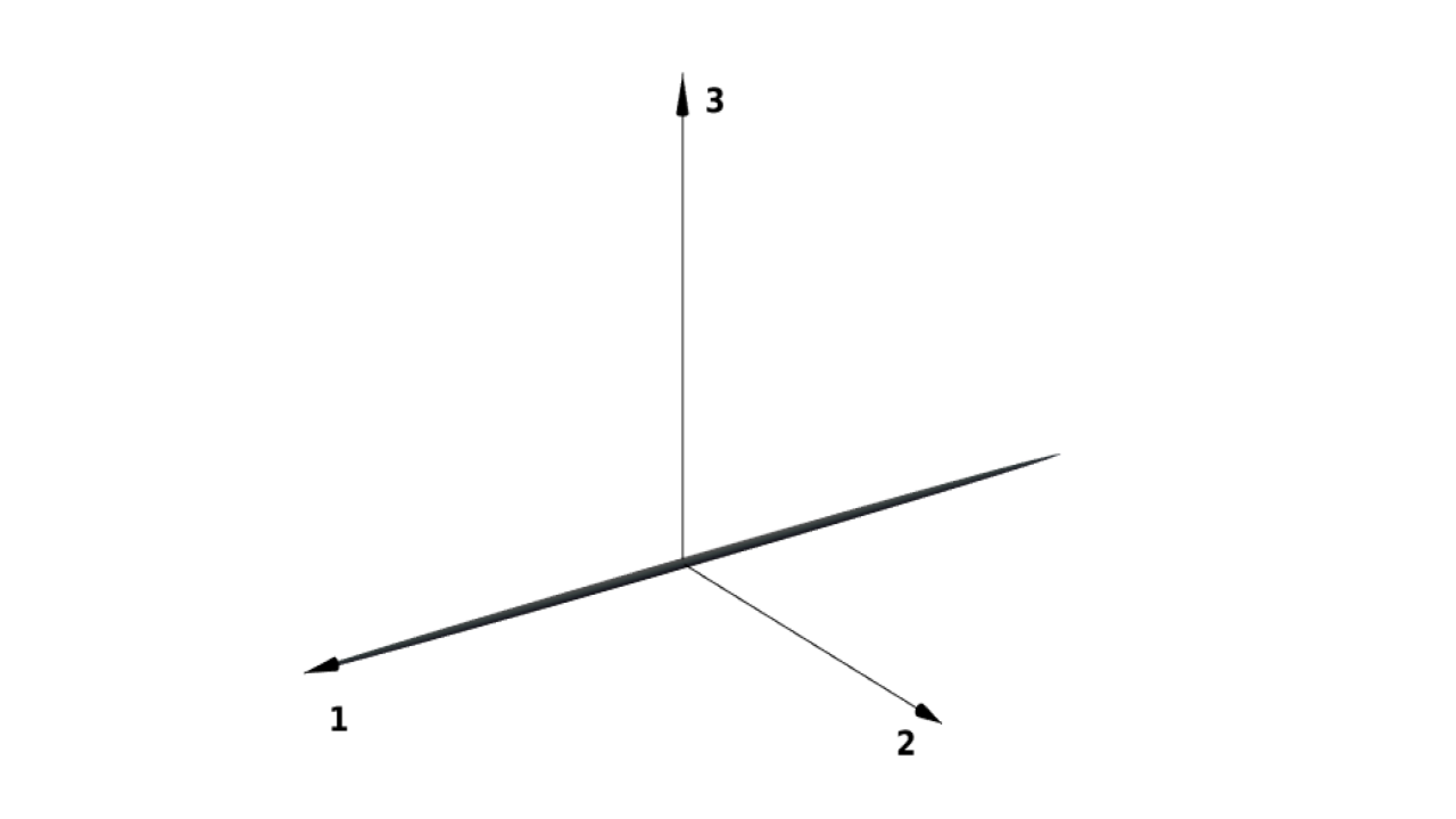}
        \label{fig:1c_max}
    }
    \end{subfigure}
    \hspace{10pt}
    \begin{subfigure}[$1C$ state with $v_{min}$ eigenvector alignment.]{
        \includegraphics[trim=220 40 300 50, clip, width=000.24\textwidth]{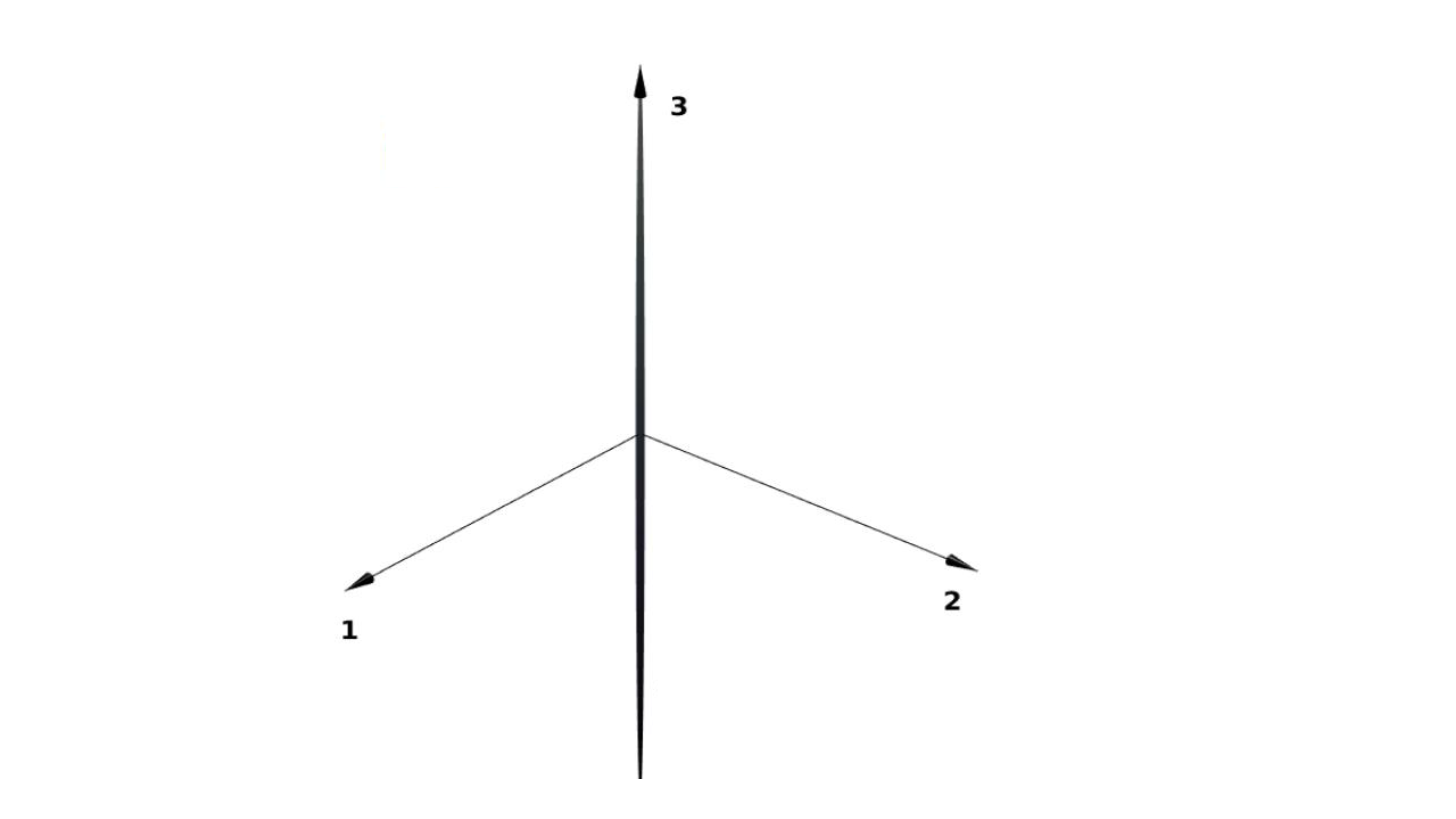}
        \label{fig:1c_min}
    }
    \end{subfigure}
    \caption{Graphical visualization of each eigenspace perturbation as Reynolds stress ellipsoids. \label{fig:all_perts}}
\end{figure}

The combinations of 3 eigenvalue perturbations (towards the $1C, 2C$, and $3C$ states) and 2 eigenvector perturbations ($v_{min}$ and $v_{max}$) result in 5 unique sets of perturbations.
Figure \ref{fig:all_perts} shows the resulting Reynolds stress ellipsoid shapes.
This number of perturbations is five and not six because the $3C$ eigenvalue perturbation results in a rotationally symmetric stress ellipsoid, which means the eigenvector perturbation does not change anything.
Table \ref{tab:perts} lists the eigenvalue and eigenvector combinations that make up each perturbation.
These five perturbed states correspondingly require five different RANS simulations, in addition to the baseline simulation with the unmodified version of the turbulence model, to provide information about the uncertainty introduced by the turbulence model.
For each simulation, the same perturbation is applied uniformly across the entire computational domain, at each pseudo-time step, until the simulation converges.
There are now five new realizations of the flow field and the flow field predicted by the unperturbed turbulence model.
The maximum and minimum values of any QoI predicted by these 6 (5 perturbed + 1 baseline) simulations create the interval bound for that QoI.
These interval bounds are used in the multi-fidelity framework as uncertainty estimates for CFD simulations.
Note that the additional five simulations for the perturbed turbulence model can be run simultaneously.
If sufficient computational resources are available, the entire process of RANS UQ can run in the same wall-clock time as a typical deterministic RANS simulation.

\begin{table}
\centering
    \def\arraystretch{1.2}
    \begin{tabular}{c|c}
        $\mathbf{\Lambda^*}$ & $\mathbf{Q^*}$ \\\hline
        $\Lambda_{1C}$ & $Qv_{max}$\\
        $\Lambda_{2C}$ & $Qv_{max}$\\
        $\Lambda_{3C}$ & $Qv_{max}$\\
        $\Lambda_{1C}$ & $Qv_{min}$\\
        $\Lambda_{2C}$ & $Qv_{min}$\\
    \end{tabular}
    \caption{Combinations of eigenvalue and eigenvector perturbations that are performed.}
    \label{tab:perts}
\end{table}

An inherent assumption is that the maximum and minimum value of any QoI occurs at the corners of the barycentric triangle and not somewhere in the interior.
In \cite{emory2014visualizing} the bounds were monotonically increasing as the perturbation magnitude increases.
This monotonicity means that the interval bound is at its largest at the corner when perturbing along that direction.
It is still possible that there could be a different direction of perturbation that would result in larger interval bounds.
For example, the largest interval could result from a perturbation towards one of the sides of the barycentric triangle.
For this work, the perturbations occur towards the triangle's vertices to exercise the limits of the componentiality of the turbulence field. 

The theoretical underpinnings of the eigenspace perturbation methodology have been discussed in detail by Mishra, and Iaccarino \cite{mishra_perturbations_2019}.
They prove that the eigenspace perturbations extend the isotropic eddy viscosity assumption to an anisotropic relation between the mean velocity gradients and the Reynolds stresses.
This anisotropy represents the most general relationship between the mean flow gradients, and the Reynolds stresses, $ R_{ij}=\nu_{ijkl}S_{kl}+\mu_{ijkl}W_{kl}$, where $S_{kl}$ and $W_{kl}$ represent the mean rates of strain and rotation respectively.
This extension enables the perturbed models to account for the effects of complex flow features such as flow separation, secondary flows, and highly anisotropic flows. 

The eigenspace perturbation methodology has exhibited substantial success in various engineering applications.
This methodology has been successfully applied to the analysis of uncertainty in flows through scramjets \cite{emory2011characterizing}, contoured aircraft nozzles \cite{mishra2019uncertainty,aiaajets, envelopingmodels, alonso2017scalable}, and turbomachinery designs \cite{emory2016uncertainty}.
The methodology has been used for the design optimization under uncertainty of turbine cascades \cite{razaaly2019optimization} and in aerospace designs \cite{cook2019optimization}.
In civil engineering applications, this methodology has been applied in the design of urban canopies \cite{garcia2014quantifying,ricci2015local}.
Owing to the structural similarity between the RANS and LES paradigms, this methodology has been extended and successfully applied for the uncertainty estimation of Large Eddy Simulations and scalar flux models \cite{gorle2013framework} as well.

It is important to note that this methodology provides no probability distribution information for the QoIs within these interval bounds and assuming any particular distribution would be inconsistent with the methodology.
The multi-fidelity GP requires data to be jointly normally distributed. Consequently, to use these interval bounds in the multi-fidelity GP framework, a Gaussian distribution of the QoIs within the bound is assumed.
The QoI is given by $y \sim \mathcal{N}(\mu,\sigma^2)$ where $\mu$ is the center of the bound, and $\sigma^2$ is calculated such that $95\%$ of the interval bound predicted by the RANS UQ methodology lies at $2\sigma$ from the mean, $\mu$.

\section{Implementation in SU2}

SU2 is an open-source software suite used to solve non-linear Partial Differential Equations (PDE) along with PDE-constrained optimization problems while utilizing unstructured meshes \cite{su21}.
The framework is general and meant to be extensible to any governing equations for multi-physics problems.
This work focuses on its Reynolds Averaged Navier-Stokes (RANS) solver capable of simulating compressible, turbulent flows found in aerospace engineering and design problems.
Extensive validation studies of the SU2 platform conducted across a diverse assortment of turbulent flows \cite{su22} ensures the solver's reliability.

SU2 is actively developed around the world.
It has been released under an open-source license and is freely available to the community so that developers may contribute to the source code and further improve the accuracy and capabilities of the suite.
To accomplish ease of use, the SU2 includes industry-standard solver technology for turbulent flows while also developing numerical solution algorithms that result in robust convergence rates.
Finally, to aid with gradient-based optimization, SU2 includes a discrete adjoint solver implementation for efficiently computing the gradient of any quantity of interest with respect to any design variable.

This section outlines details of implementing the eigenspace perturbation methodology into the CFD solver.
The first part of the discussion outlines the algorithms and the steps required to perform the eigenspace perturbation within the CFD solver.
The second part focuses on its implementation via computer code and its integration with the SU2 solver. 

\begin{figure}
\centering
\includegraphics[width=0.75\textwidth]{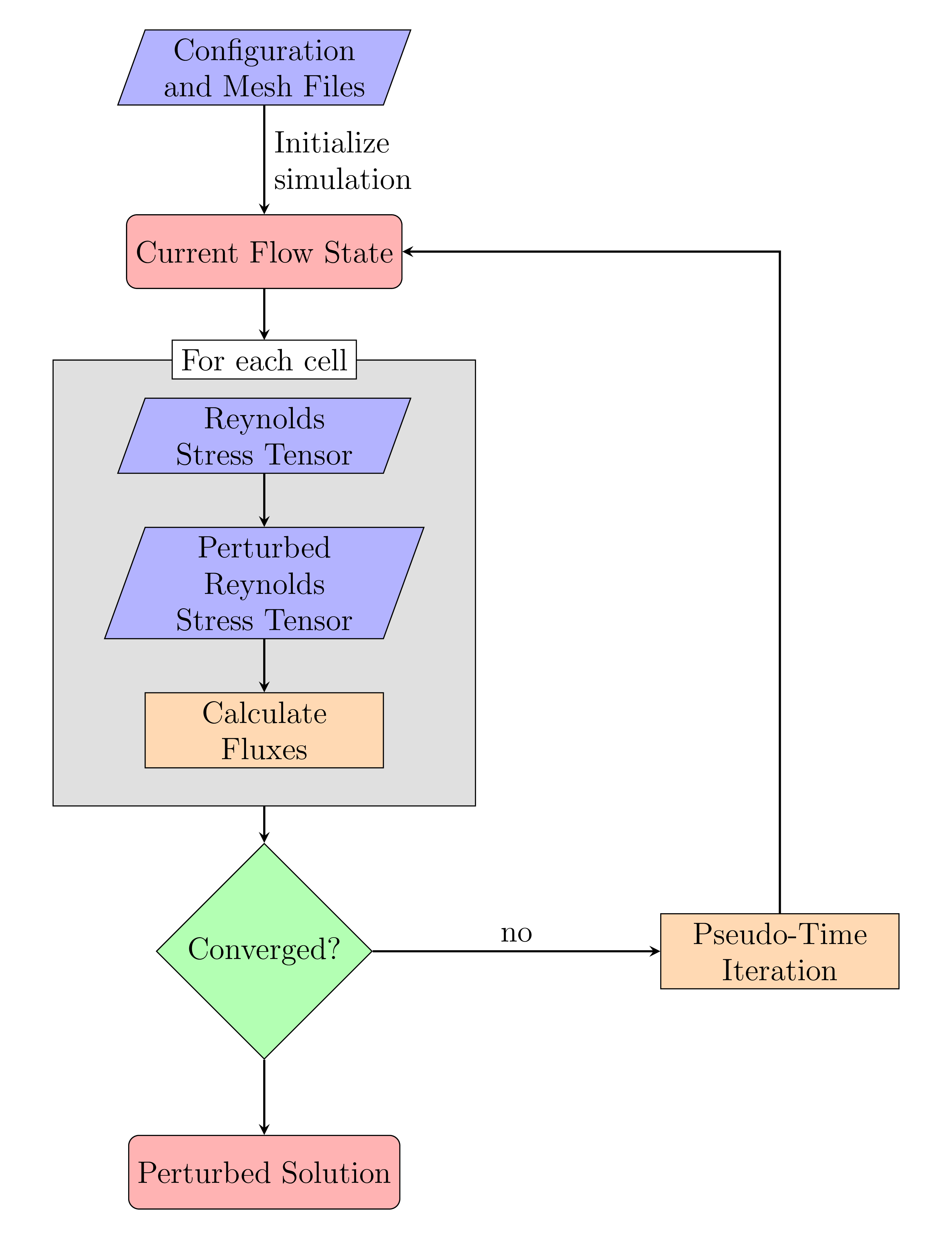}
\caption{Flow chart showing the implementation of EQUiPS within SU2 \label{fig:equips_overview}}
\end{figure}

Fig. \ref{fig:equips_overview} shows where the eigenspace perturbation fits within the context of the CFD solver.
SU2 carries out the perturbation for each cell of the grid at each pseudo-time step.
This process repeats until the flow simulation converges and outputs a perturbed solution. 

\begin{figure}
\centering
\includegraphics[width=0.75\textwidth]{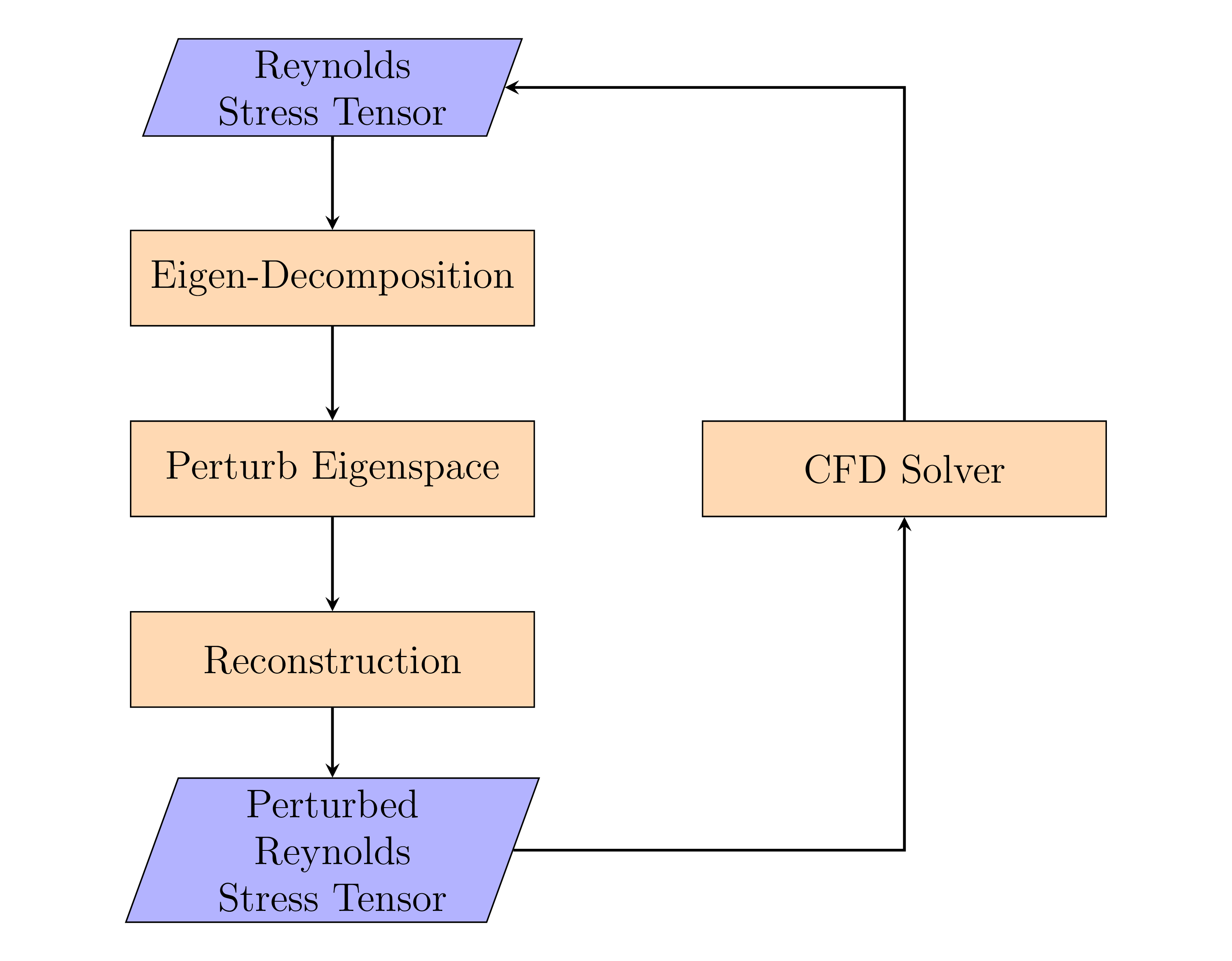}
\caption{Schematic outlining interaction between CFD solver and perturbation methodology, at each cell and iteration.\label{fig:perturbation_schematic}}
\end{figure}

Fig. \ref{fig:perturbation_schematic} focuses on the steps required to perform the eigenspace perturbation.
The first step is to use the flow field to calculate the Reynolds stress tensor, using Equation \ref{equ:rst}, at each cell in the computational domain.
The anisotropic part of the tensor, $b_{ij}$ in Equation \ref{equ:rst_decomp}, is decomposed into its eigenvalues and eigenvectors, as shown in Equation \ref{equ:eigendecomposition}.
The decomposition is done efficiently by leveraging the anisotropic tensor's symmetry.
First, the symmetric tensor is reduced to a symmetric tridiagonal form using the Householder's transform and accumulating orthogonal similarity transformations.
This step uses the TRED2 procedure outlined by \cite{tred2a}.
\cite{numres} presents the details of the routine.
Then, the eigenvalues and eigenvectors of this symmetric tridiagonal matrix are computed and sorted using the TQL2 procedure outlined in \cite{tred2b}.
These two steps provide a complete eigendecomposition of the Reynolds stress tensor at a cell in the domain.

The eigenvalues are used to project the stress state onto the barycentric anisotropy invariant map using Equation \ref{equ:barycentric_mapping}.
According to the type of perturbation, the stress state, denoted by the coordinates $\mathbf{x}$, is perturbed towards one of the vertices of the barycentric map.
An under-relaxation factor ensures solution stability.
For example, if the stress state is perturbed towards the 1-component state, the perturbed stress state is represented by the coordinates, 
\begin{equation}
    \mathbf{x^*} = \mathbf{x} + r\left ( \mathbf{x}_{1C} - \mathbf{x} \right )
\end{equation}
where $r$ is the under-relaxation factor.
$r$ is a user-tunable parameter. 
All the results shown in this thesis use the default value of $r=0.1$.

Using the barycentric coordinates of the perturbed stress tensor, $\mathbf{x^*}$, Equation \ref{equ:barycentric_mapping} is refactored to get the perturbed eigenvalues.
This results in 2 equations, one for each coordinate, with three unknown eigenvalues.
The fact that the anisotropy tensor has a zero trace i.e. 
\begin{equation}
    \lambda_1 + \lambda_2 + \lambda_3 = 0
\end{equation}
provides the third equation required to solve the system of equations for the perturbed eigenvalues.

Similarly, for the eigenvector perturbation, the eigenvectors of the Reynolds stress tensor at a cell are permuted to $v_{min}$ or $v_{max}$, depending on the specific perturbation.
The Reynolds stress tensor at each cell is reconstructed using these perturbed eigenvalues and eigenvectors.
Additionally, dependent quantities like the turbulence production tensor are re-constituted using this perturbed Reynolds stress tensor.
These are input back into the CFD solver for the next iteration.
This process continues till numerical convergence as determined by the convergence criteria. 

SU2's modular architecture enables integrating this framework without significant alterations to the main code.
The module is split into two parts: a high-level Python script that sequentially specifies the type of perturbation and C++ code that performs the perturbations during the execution of simulations.
For smooth operation, it is best to have performed a baseline simulation with SU2 and have achieved sufficient convergence.
It ensures that the mesh file and the input configuration file are well-posed and can provide converged perturbed solutions if run through the Python script.

The Python script takes an input configuration file and a mesh file identical to the ones used to run the baseline CFD simulation in SU2.
The script sets the necessary configuration options to run the EQUiPS framework.
These include the direction of the eigenvalue perturbation (aligned towards one of the $1C, 2C$ or $3C$ state on the barycentric triangle for a specific simulation), the magnitude of the eigenvalue perturbation ($\Delta_{B} \in [0,1]$) and the perturbed alignment of the Reynolds stress eigenvector ($v_{min}$ or $v_{max}$, as detailed earlier in this section).
It sequentially runs through the simulations for each perturbed state, creating a new directory for each new simulation, and outputting the results in the respective directories.
In conjunction with the converged baseline solution, these results can be post-processed to provide the necessary model-form uncertainty information. 

The implementation within the existing code base is limited to the two areas where the perturbations are injected into the simulation: the viscous and turbulent flux calculations.
The perturbed stress tensors replace the original ones in the flux calculations.
The new fluxes advance each node to the next pseudo-timestep.
In the SST turbulence model, \cite{sst}, the effects of the perturbations manifest themselves in the turbulence production term. 

As the solution converges, the Reynolds Stress also converges to its perturbed state.
Once a perturbed solution is converged, the Python script moves on to the next eigenspace perturbation.
It creates a new directory and configuration file to specify the new perturbation options.
This process continues until all the specified perturbed simulations are complete. 

In the spirit of \textit{versatility}, the implementation was designed to minimize the number of user-defined inputs.
The implementation allows the use of the framework without in-depth knowledge of its mechanics.
The details of the perturbations (componentiality, eigenvector permutation) are abstracted away to provide a clean user interface that does not deviate from the workflow of running regular CFD simulations. All the options used for the module can also be specified in the configuration file for the potential expert user without needing the Python script. 

\section{Validation} \label{sec:VandV_rans_uq}

Before the eigenspace perturbation methodology can be applied to flow simulations involving complex geometries, the methodology must be tested on a range of flow configurations often used to validate CFD codes. 
Validation refers to \say{The process of determining the degree to which a model is an accurate representation of the real world from the perspective of the intended uses of the model} \cite{computational_fluid_dynamics_committee_guide_1998}.
In the context of the eigenspace perturbation methodology, validation involves ensuring the predicted uncertainty estimates are a reasonable representation of the uncertainty injected by the turbulence model in use.
Flow configurations with either experimental or high-fidelity simulation data are required to validate the methodology.
It should meet the following expectations:

\begin{enumerate}
    \item If there is a discrepancy between the RANS simulation predictions and the experimental data, the uncertainty estimates should be large.
    \item If the RANS simulation predictions agree with the experimental data, the uncertainty estimates should be small. 
\end{enumerate}

In an ideal case, if the turbulence model was the only source of uncertainty and error, all the experimental data should lie within the uncertainty estimates from the methodology.
In practice, this is rarely the case.
Often errors stemming from insufficient numerical discretization or a discrepancy in experimental and simulation conditions affect the quality of the simulation prediction. 

This section presents the uncertainty estimation module results for a range of test cases detailed in Table \ref{tab:vandv_cases}.
The first two flows are benchmark cases widely used to test turbulence models.
They test the ability of the methodology to predict the behavior of highly separated flows.
The last four cases are flows of aerospace interest.
The jet flow cases test the methodology's performance for shear-layer flows, whereas the airfoil flows test its performance over a range of flow regimes.
All the simulations use the SST turbulence model\cite{sst}.

\begin{table}
\caption{\label{tab:vandv_cases} Details of testing and validation cases}
\begin{center}
\begin{tabular}{ccc}
Test Case& Rationale& Notes \\\hline
Turbulent flow over a backward facing step& Benchmark flow& 2D, incompressible simulation.\\
Turbulent flow through an asymmetric diffuser& Benchmark flow& 2D, compressible simulation.\\
Jet effluxes from the NASA Acoustic Response Nozzle& Engineering case & 3D, sub-sonic simulation.\\
NACA 0012 airfoil at a range of angles of attack& Engineering case& 2D, sub-sonic simulation.\\
Turbulent flow over a three-element High-lift airfoil& Engineering case& 2D, sub-sonic simulation.\\
\end{tabular}
\end{center}
\end{table}

\subsection{Turbulent flow over a backward facing step}

The inlet and outlet boundary conditions are chosen to yield the correct Mach number $\left ( M_{ref} = 0.128 \right )$ upstream from the step.
The figure is scaled by the step height $\left ( H \right ) $.
Note that the $x$ and $y$ axes have different scales.
The different scales help visualize the geometry since the geometry is $\approx 20\times$ longer in the $x$ direction when compared to the channel height.
The channel height after the step is $9H$. 

The flow is allowed to develop before flowing over the backward-facing step.
The sudden expansion in the channel height results in flow separation and subsequent reattachment downstream from the step.
The size of the resulting recirculation region or the reattachment length is the critical quantity that must be predicted accurately by a turbulence model.
Classical two-equation turbulence models underpredicted the re-attachment length by 10-25\% \cite{thangam1991}.
The underprediction occurs due to inaccurate predictions for the Reynolds stresses arising from the use of an isotropic eddy viscosity \cite{thangam1991}.
This configuration has been investigated experimentally by \cite{driver1985} and their data is used in our investigation.

The uncertainty estimates can account for the discrepancy between model predictions and experimental data at most locations.
While there is some discrepancy between the unperturbed RANS simulation and the experimental data, the uncertainty intervals can account for most of this discrepancy, with almost all experimental data points lying within the uncertainty estimates. 

\subsection{Turbulent flow through an asymmetric diffuser}

Turbulent flow in an asymmetric diffuser has some interesting features, such as separation over a smooth wall, subsequent reattachment, and redevelopment of the downstream boundary layer, which offer considerable challenges to eddy-viscosity-based models.
The flow is allowed to develop in the channel leading up to the diffuser.
The channel height, $H$, is used to normalize the geometry's length scale.
As before, the $x$ and $y$ axes have different scales to make it easier to represent the geometry.
The reference pressure, $P_{ref}$, was adjusted to ensure the channel centerline velocity just upstream of the diffuser is $22.6 m/s$. 
The simulation results are validated using experimental data of \cite{buice}. The data include mean velocities at various stations in the diffuser and skin friction data on the upper and lower walls. 


\subsection{Jet efflux of the NASA Acoustic Response Nozzle}

Reliable predictions of turbulent jets exhausting from aircraft nozzles are essential for various aerospace design applications.
However, these exhaust jets involve many complications, such as the shear layer between the jet efflux and the ambient air, complicated nozzle geometries, compressibility effects, and under or over-expanded flows.
These pose significant challenges to eddy-viscosity-based models.
Focusing on the mixing between the jet and the ambient fluid in the vicinity of the nozzle exit, RANS models predict a significantly lower rate of jet mixing compared with high-fidelity data.
Farther downstream of the jet's potential core, RANS models predict the far-field mixing rate to become substantially higher than is observed in experiments.
Similarly, the fidelity of RANS predictions is highly inconsistent, having higher fidelity for cold jets than heated, for axisymmetric than non-axisymmetric geometries, and even varying significantly over different quantities of interest.

In this test case, we investigate sub-sonic jets from the NASA Acoustic Response Nozzle.
These jet efflux cases have been studied experimentally by \cite{nasajet}, and extensive Particle Image Velocimetry (PIV) datasets are available.
The NASA Glenn Research Center generated the PIV dataset using the Small Hot Jet Acoustic Rig.
The tests were repeated on numerous instances during 2001-2007 with varying PIV configurations.
In addition to a robust mean, this corpus provides insight into the data uncertainty.
The raw data set consists of over 23,000 velocity fields available for statistical analysis at each location.
The researchers also shared a final consensus dataset, with the averaged mean over the raw data, and the standard deviations over the measurements at each point, termed as error bars in the data set by \cite{nasajet}. 
This data uncertainty attempts to account for any statistical bias and random errors in the measurements arising from PIV calibration, image analysis procedure, rig flow instrumentation, minor errors in the instantaneous measurements, and the size and number of ensembles used. 

Outlined here are the results for different subsonic jet conditions.
Table \ref{tab:table2} details the inlet ratios and the resulting Mach number at the jet throat.

\begin{table}
\caption{\label{tab:table2} Reference conditions for subsonic jet flow cases considered}
\begin{center}
\begin{tabular}{ccccc}
Test Case & Classification& $Ma$& ${T_T}/{T_{ref}}$& ${P_T}/{P_{ref}}$\\\hline
Case I  & heated, perfectly expanded & 0.376 & 1.764 & 1.102\\
Case II & cooled, perfectly expanded & 0.513 & 0.950 & 1.197\\
\end{tabular}
\end{center}
\end{table}


The solid black dots represent the mean experimental data, and the associated error bars represent the reported standard deviation of the PIV data.
The solid black line represents the baseline result using the unperturbed SST turbulence model, and the gray area is the uncertainty estimated by the eigenspace perturbation methodology.

The differences in the solid black line and black dots show that the baseline turbulence model has great difficulty in accurately capturing the flow physics in the jet wake. 
The uncertainty estimates, represented by the grey areas, account for a significant proportion of the discrepancy between the RANS simulations and the high-fidelity experimental data.
If the experimental data points themselves do not lie within the grey area, there is a non-trivial intersection between the error bars and the grey area. 




\subsection{NACA 0012 airfoil at a range of angles of attack} \label{sec:equips_naca0012}
In this test case, we consider the flow over a NACA 0012 airfoil at a range of angles of attack, varying from 0$^{\circ}$ to 20$^{\circ}$.
This design was chosen specifically due to its widespread adoption in the industry.
These include in conventional aircraft, for instance, the wingtips of the Cessna 140A, 207; helicopter designs such as the inboard and outboard blades of the Aerospatiale AS365, Boeing 600N, Lockheed 475; in addition to numerous horizontal and vertical axis wind turbines.
The simulation conditions are shown in Table \ref{tab:naca0012_cond}.
The experimental data \cite{ladson1988} report the coefficient of lift ($C_L$) and the surface pressure coefficient ($C_P$) for different angles of attack. 

\begin{table}
\centering
    \renewcommand{\arraystretch}{1.2}
    \captionsetup{justification=centering}
    \caption{Simulation conditions for the NACA0012 test case.} 
    \begin{tabular}{|c|c|}
        \hline
        Mach Number & $0.15$ \\ \hline
        Reynolds Number & $6\times10^6$ \\ \hline
        Reference chord length & $1$ m \\ \hline
        Freestream Temperature & $300~\text{K}$ \\ \hline
        $\alpha$ & $0^\circ \leq \alpha \leq 20^\circ$ \\ \hline 
    \end{tabular}
    \label{tab:naca0012_cond}
\end{table}

\begin{figure}
\center
\includegraphics[width=0.7\textwidth]{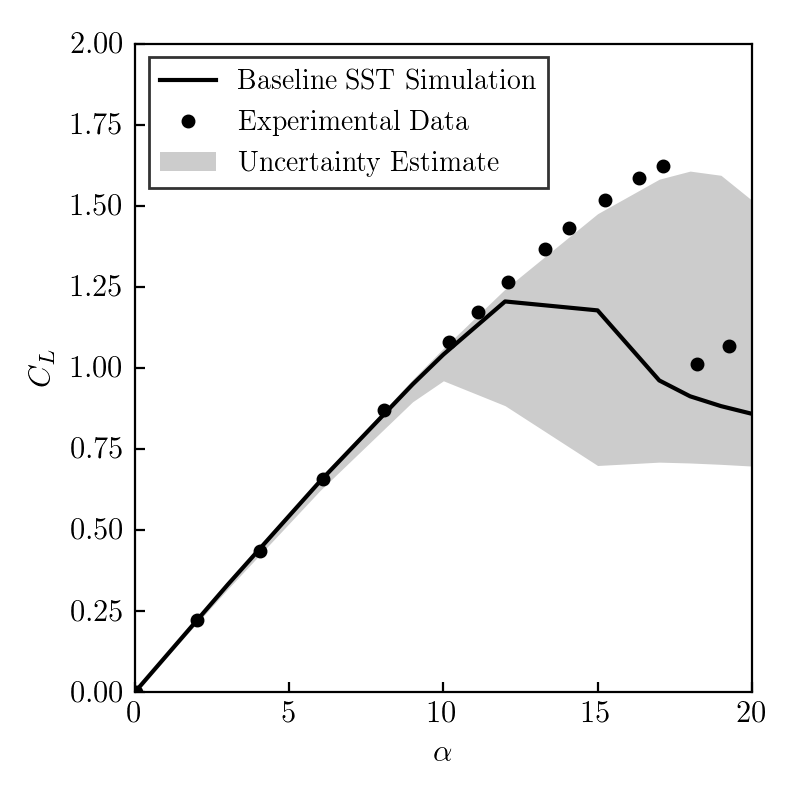}
\caption{Variation in the coefficient of lift $C_L$ with the angle of attack $\alpha$\label{fig:naca0012_cl_vs_alpha}}
\end{figure}

In Fig \ref{fig:naca0012_cl_vs_alpha}, we outline the variation in the coefficient of lift with the angle of attack $\alpha$.
There is almost no discernible difference between the RANS predictions and the experimental data at low angles of attack.
Accordingly, here the uncertainty bounds are negligible.
There is a substantial discrepancy between the RANS predictions and the high fidelity data at higher angles of attack.
For these values of $\alpha$, the uncertainty estimates are significant.
At all values of $\alpha$, the uncertainty estimates envelop the experimental data.

Away from the stall, at $\alpha=10^{\circ}$, the RANS predictions are in good agreement with the experimental data.
Accordingly, the uncertainty bounds are negligibly small.
At $\alpha=15^{\circ}$, there is a significant discrepancy between the RANS predictions and the experimental data.
Here, the uncertainty estimates are sizable, and they envelop the experimental data.
However, the uncertainty estimates account for this discrepancy, and all the experimental data points intersect the shaded interval estimates.

\subsection{Turbulent flow over a three-element High-lift airfoil}

In this test case, we investigate the turbulent flow over a McDonnell Douglas Aerospace (MDA) single-flap, three-element airfoil.
The flap rigging used corresponds to the 30P/30N designated by MDA.
The results correspond to the case with Mach number, $Ma=0.2$; Reynolds number, $Re=5 \times 10^6$ with an angle of attack of $\alpha=8^{\circ}$.
We use the experimental data of \cite{chin1993}.

\begin{figure}
\center
\includegraphics[width=0.8\textwidth]{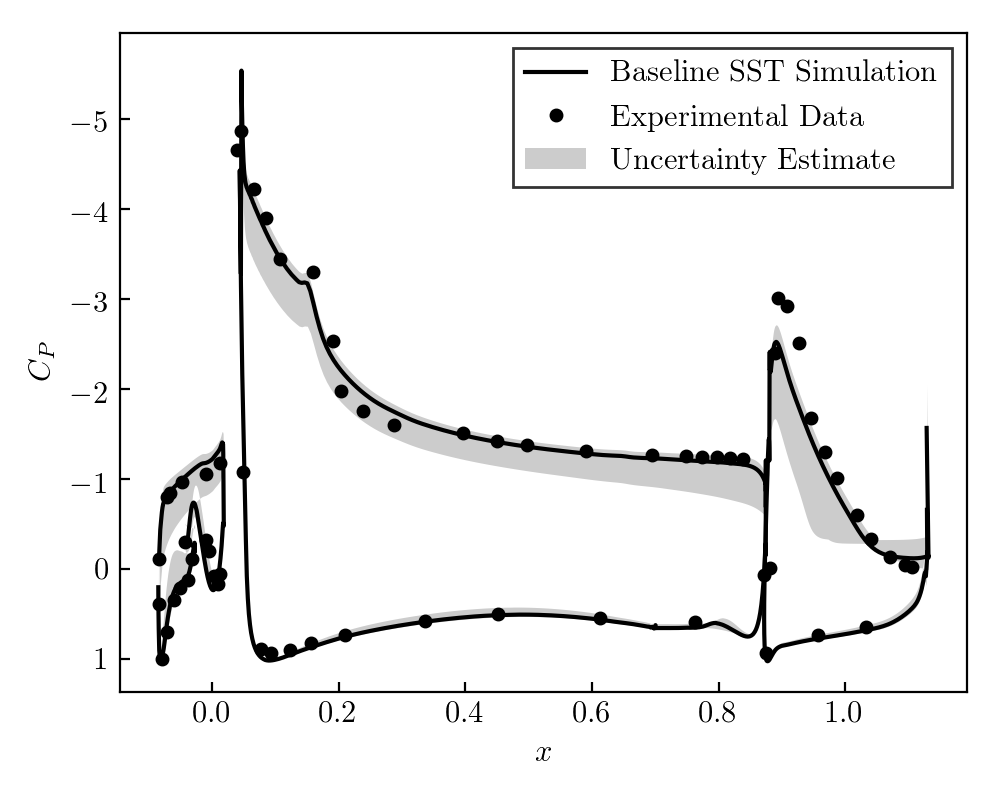}
\caption{Pressure coefficient for the 30P30N configuration\label{fig:30p30n}}
\end{figure}

We outline this case as a test against the false positive.
In cases where there is a significant discrepancy in the RANS predictions, the uncertainty bounds should exhibit the same.
However, in cases where the RANS predictions are accurate, having spurious uncertainty bounds that are significant in their extent would be misleading and would correspond to a false positive.
\cite{klausmeyer1997} have tested this flow case for a range of RANS models and have found the RANS predictions to be accurate.
In such a scenario, ideally, we would expect the uncertainty bounds to be negligible at most locations along the airfoil sections.
Figure \ref{fig:30p30n} outlines the distribution of the coefficient of pressure ($C_p$) on the surfaces of the different sections.
The discrepancy between the RANS predictions and the experimental data is minimal over the main element and the airfoil flap.
Accordingly, the uncertainty bounds are negligibly small over these zones.
The upper surface of the slat exhibits an appreciable amount of discrepancy between RANS predictions and the high-fidelity data.
The uncertainty estimates over this surface are substantial, and they envelop the experimental data.

\section{Numerical Discretization Error vs. Turbulence Modeling Uncertainty}\label{sec:num_vs_turb_error}

Solving continuous equations on a discrete domain creates numerical discretization error.
RANS CFD simulations involve solving the continuous RANS equations that define fluid flow on a discrete domain known as the mesh or grid.
Increasing the number of discrete points in the domain reduces the numerical discretization error.
As the discretization increases and the grid spacing tends towards $0$, the numerical error approaches $0$ as well. 
This property is the basis for the "Grid Convergence Study" method for quantification of numerical discretization error \cite{american_society_of_mechanical_engineers_standard_2009}.
Details of the methodology are reproduced here for clarity.

The methodology requires using a family of sequentially coarser grids using the same grid generation parameters.
The easiest method to create the grid family involves generating a very dense structured mesh which would result in minimal discretization error.
Then each successive coarser grid level removes every other grid line in each direction.
In 2D and 3D computations, this process results in the number of grid points reducing by a factor of 4 and 8, respectively, at each grid level.
It results in uniformly refined grids that isolate the effect of the discretization on the simulations. 

Three grids of successive refinement are required to calculate all the numerical error metrics. 
First, a representative grid size for the $i$-th grid is defined as
\begin{equation} \label{equ:grid_h}
    h_i = \left ( \frac{1}{N_i}\right )^A
\end{equation}
where $N_i$ refers to the number of degrees of freedom in the $i$th grid and $A=1/2$ for 2-D simulations and $A = 1/3$ for 3-D simulations.
SU2 uses node-centered numerics, so $N_i$ refers to the number of nodes in the grid.
The grids are ordered such that $i=1$ refers to the most refined grid (most number of nodes), and $i=3$ refers to the coarsest grid.
Correspondingly, a smaller value of $h$ indicates a finer mesh with more nodes. 
Next, grid ratios are defined as 
\begin{equation}
    r_{21} = \frac{h_2}{h_1},~r_{32} = \frac{h_3}{h_2},
\end{equation}
and solution differences are defined as
\begin{equation}
    \epsilon_{21} = \phi_2 - \phi_1,~\epsilon_{32} = \phi_3 - \phi_2,
\end{equation}
where $\phi_i$ is the solution on the $i$-th grid.
Often for aerodynamics-related problems $\phi$ represents a force or moment coefficient such as $C_L$ or $C_m$.

These quantities are used to compute the observed order of convergence, $p$, by solving the following equations using a fixed point iteration:
\begin{align}
    p & = \frac{1}{\ln{\left ( r_{21} \right )}} \left ( \ln{\left \vert \frac{\epsilon_{32}}{\epsilon_{21}} \right \vert } + q(p) \right )
    \\
    q(p) & = \ln{ \left ( \frac{r_{21}^{p} - s }{r_{32}^{p} - s}\right )}
    \\
    s & = 1\times sign \left ( \frac{\epsilon_{32}}{\epsilon_{21}}\right ).
\end{align}
The order of convergence should be close to the order of the numerical method used to solve the simulations. 
All of the RANS calculations made using SU2 use second order numerical methods, so we expect the observed order to be close to $2$.

Using the grid ratios and the apparent order of convergence, the infinite grid solution is extrapolated as,
\begin{equation}
    \phi_{ext}^{21} = \frac{r_{21}^p\phi_1 - \phi_2}{r_{21}^p - 1}.
\end{equation}

The approximate relative fine-grid error: 
\begin{equation}
    e_a^{21} = \left \vert \frac{\phi_1 - \phi_2}{\phi_1} \right \vert,
\end{equation}
can be used to calculate the fine-grid convergence index:
\begin{equation}
    GCI_{fine}^{21} = \frac{1.25e_a^{21}}{r^p_{21}-1}.
\end{equation}
Here $1.25$ is an empirically recommended Factor of Safety (FoS) based on hundreds of CFD case studies \cite{roache1998verification}.
Furthermore, the GCI can be used to express the $95\%$ confidence interval on the fine grid solution. 
The solution can be expressed as
\begin{equation} \label{equ:num_error_bars}
    \phi \approx \phi_1 \pm \left( GCI_{fine}^{21} \right)\left \vert \phi_1 \right \vert 
\end{equation}

Armed with these metrics to quantify the numerical discretization error in the CFD simulations, comparisons to the uncertainty introduced by turbulence models can be made. 
The eigenspace perturbation methodology discussed in Section \ref{sec:equips_rans_uq} only estimates uncertainties introduced by turbulence models.
The RANS CFD simulations required to quantify the uncertainties are run on grids that introduce some degree of discretization error.
The following sections explore the relationship and the relative magnitudes of the two quantities.

\subsection{NACA0012 Airfoil} \label{sec:num_err_naca0012}

The same NACA0012 case presented in Section \ref{sec:equips_naca0012} is used here. 
The 5th and 6th Drag Prediction Workshops use this case as a verification study \cite{levy2013summary,roy2017summary}.
The grids used for those verification studies are used here.
Table \ref{tab:naca0012_meshes} details the mesh metrics.

\begin{table}
    \renewcommand{\arraystretch}{1.2}
    \centering
    \begin{tabular}{ c|c|c|c|c } 
         Mesh Level & Nodes & Surface Nodes & Wall spacing & Approx. $y^+$  \\ 
         \hline
         L1 & $14,687,744$ & $4,097$ & $1.0\times10^{-7}~m$ & 0.025\\
         L2 & $3,673,856$ & $2049$ & $2.0\times10^-7~m$ & 0.05\\
         L3 & $919,424$ & $1,025$ & $4.0\times10^{-7}~m$ & 0.1\\
         L4 & $230,336$ & $513$ & $8.0\times10^{-7}~m$ & 0.2\\
         L5 & $57,824$ & $257$ & $1.6\times10^{-6}~m$ & 0.4\\
         L6 & $14,576$ & $129$ & $3.2\times10^{-6}~m$ & 0.8\\
         L7 & $3,584$ & $65$ & $6.4\times10^{-6}~m$ & 1.6\\
        
    \end{tabular}
    \caption{Details of the grid family used to perform numerical discretization error quantification for the NACA0012 case.}
    \label{tab:naca0012_meshes}
\end{table}

To be consistent with the results form the eigenspace perturbation methodology, the SST turbulence model is used for these simulations. 
The $x$-axis for the figures is the representative mesh size $h$ (Equation \ref{equ:grid_h}). 
Going from right to left on the $x$-axis corresponds to increasing grid refinement, from the L7 mesh to the L1 mesh. 
As the grid refinement increases the coefficients converge towards the infinite grid solution.

Ideally, to compare the numerical discretization error and the turbulence modeling uncertainty, the finest three meshes should be in the asymptotic region of convergence for the quantities of interest.
For this case the $GCI_{fine}^{43} = 5.0 \times 10^{-5}$ which represents a $95\%$ confidence interval of $1$ drag count ($10^-4$). 
This value is a suitably low numerical discretization error that would not adversely affect the RANS CFD simulations. 
The finer meshes, L2 and L1, are computationally expensive to perform simulations.
So to compare the numerical discretization error and the uncertainties predicted by the eigenspace perturbation methodology, we use the L3, L4, and L5 meshes. 
The L6 mesh, which has significant numerical error as evidenced by its under-prediction of $C_L$ and over-prediction of $C_D$, is also used to see how the RANS UQ methodology handles insufficient discretization that might not capture relevant flow features. 

The RANS UQ methodology is applied to each of the meshes in question, L3-L6.
The same $\alpha$ sweep that was used in Section \ref{sec:equips_naca0012}, $0^\circ \leq \alpha \leq 20^\circ$, is used for these simulations. 
As the mesh refinement increases, going from L6 to L3, the simulation results tend to agree more with the experimental data. 
As expected, increasing the number of grid points in the mesh reduces the numerical discretization error in the simulations. 
The color of the uncertainty area corresponds to the mesh level represented.
It is hard to distinguish between the uncertainty areas from the L3-L5 meshes because all of these uncertainty areas are almost coincident.
The notable exception is the uncertainty estimate for the L6 mesh in blue. 
These two observations indicate that while a minimum grid refinement that captures the relevant flow features is required, further grid refinement does not significantly impact the uncertainty estimate predicted by the eigenspace perturbation methodology.
This conclusion is significant as it means that uncertainty quantification can be performed using coarser meshes that reduce the computational cost of performing the six simulations required to engender the uncertainty estimate.

The L3-L5 meshes are used to compute the numerical error metrics. 
The error bars on the simulation data points represent the $95\%$ confidence interval on the numerical discretization error as defined by Equation \ref{equ:num_error_bars}.
The blue shaded area is the turbulence uncertainty estimate. 
For most of the points, the discretization error bars are too small to be visible at this scale.
The turbulence uncertainty is greater than the discretization error bars for most simulations, even those at lower angles of attack. 
However, talking about the simulations at higher angles of attack requires nuance. 
The mesh family is unlikely to be in the asymptotic range of convergence in the stall region of the $\alpha$ sweep, and so the numerical discretization error metrics are not necessarily valid for those simulations. 
The solutions for $\alpha = 12^\circ$ suffered from oscillatory convergence, and the ones for $\alpha = 20^\circ$ yielded a negative apparent order.
The error bars for these two cases are omitted as they are not valid.

From this exploration of a 2D case, it seems that, given sufficient grid refinement to capture the relevant flow physics, the uncertainty arising from turbulence modeling remains independent of further grid refinement. 
The following section explores a 3D case to confirm these observations.

\subsection{ONERA M6 Wing}

A classic CFD validation case, the ONERA M6 wing, is chosen as the 3D case to investigate the relationship between numerical discretization errors and turbulence modeling uncertainties. 
It is a swept, untwisted, semi-span wing that uses a symmetric airfoil, the ONERA D, as its cross-sectional shape. 
The simulations conditions are presented in Table \ref{tab:ONERAM6_test_cond}.
A family of structured meshes is available from the NASA Turbulence Modeling Resource website.
Table \ref{tab:oneram6_meshes} details the subset of the structured meshes used.
The finest mesh level from this family, the L1 mesh, has $\ge 64$ million points. 
It would be very computationally expensive to perform the eigenspace perturbation simulations, and so it is omitted from this study.

\begin{table}
\centering
    \renewcommand{\arraystretch}{1.2}
    \captionsetup{justification=centering}
    \caption{Simulation conditions for the ONERAM6 wing.} 
    \begin{tabular}{|c|c|}
        \hline
        Mach Number & $0.84$ \\ \hline
        Reynolds Number & $14.6\times10^6$ \\ \hline
        Reference chord length & $0.805$ m \\ \hline
        Freestream Temperature & $300~\text{K}$ \\ \hline
        $\alpha$ & $3.06^\circ, 6.06^\circ$ \\ \hline 
    \end{tabular}
    \label{tab:ONERAM6_test_cond}
\end{table}

\begin{table}
    \renewcommand{\arraystretch}{1.2}
    \centering
    \begin{tabular}{ c|c|c } 
         Mesh Level & Nodes & Approx. $y^+$  \\ 
         \hline
         L2 & $8,677,681$ &  $1.0$\\
         L3 & $1,087,665$ &  $2.0$\\
         L4 & $136,793$ & $4.0$\\
        
    \end{tabular}
    \caption{Details of the grid family used to perform numerical discretization error quantification for the ONERA M6 case.}
    \label{tab:oneram6_meshes}
\end{table}

The transonic flow over the wing creates a strong $\lambda$-shock on the wing's upper surface.
Accurate predictions for the relative locations of the two shocks and their meeting point requires sufficient mesh refinement. 
It is also a problematic feature for RANS CFD solvers to predict correctly as the turbulence model can introduce significant uncertainty in the results.
The black squares represent the baseline RANS CFD predictions, and the gray shaded region represents the estimated uncertainty that the turbulence model injects. 
The finest mesh data point, shown on the left of these plots, has the numerical discretization error bars associated with it. 
Using just three meshes results in error metrics for the finest mesh only.
The uncertainty estimated by the EQUiPS module is significantly larger than the errors introduced due to insufficient numerical discretization.
Additionally, the mesh refinement does not significantly impact the size of the uncertainty estimates.

When the angle of attack is increased to $\alpha = 6.06^\circ$, a different story unfolds. 
The error metrics for the discretization error are no longer valid as the grids are not sufficiently large to capture all the relevant flow physics. 
Consequently, increasing the grid refinement changes the uncertainty estimates drastically. 


In summary, the 2D and 3D investigations into the relationship between numerical discretization errors and the turbulence modeling uncertainty estimates yield similar results and provide guidelines for the grids that should be used with the EQUiPS module.
A minimum level of grid refinement that can capture all the relevant flow physics is required.
More specifically, the grids should be in the asymptotic region of grid convergence for the particular flow configuration in question.
This requirement ensures that the numerical discretization error is small and that the turbulence modeling uncertainty estimate is the dominant source of error in the simulations. 
Once a grid of sufficient refinement is used, additional refinement does not significantly impact the turbulence modeling uncertainty estimate.
These conclusions are kept in mind when creating grids for use with the EQUiPS module. 

\section{Application to NASA CRM} \label{sec:crm_rans_uq}
Section \ref{sec:VandV_rans_uq} demonstrated the eigenspace perturbation methodology has been demonstrated on various test cases.
Here, the methodology is applied to a full-configuration aircraft.
The results in this section were first published in \cite{mukhopadhaya2020multi}  and are reproduced here.
The aircraft configuration is identical (without accounting for aeroelastic deflections of the model) to the one tested in the wind tunnel.
It closely resembles a Boeing 777 aircraft with a fully redesigned wing.
Table \ref{NASA_CRM_test_cond} describes the transonic simulation conditions.
Note that this range of angles of attack at a free-stream Mach number of 0.85 lead to non-linear physical phenomena and flow separation that can significantly increase the uncertainties in the RANS predictions.
Separated flow exists for $\alpha > 4^\circ$ at this Mach number.
All of the necessary RANS CFD simulations were conducted using the SU2~\cite{su2_aiaajournal} solver.
The SST turbulence model~\cite{sst,menter2003ten} was used and the previously defined perturbations, Equations \eqref{equ:eigenvalue_pert} - \eqref{equ:vmin_vmax}, were applied to it. 
 
\begin{table}
\centering
    \renewcommand{\arraystretch}{1.2}
    \captionsetup{justification=centering}
    \caption{Simulation conditions for the NASA CRM.} 
    \begin{tabular}{|c|c|}
        \hline
        Mach Number & $0.85$ \\ \hline
        Reynolds Number & $5\times10^6$ \\ \hline
        Reference chord length & $7.00532$ m \\ \hline
        Freestream Temperature & $310.928~\text{K}$ \\ \hline
        $\alpha$ & $-2^\circ \leq \alpha \leq 12^\circ$ \\ \hline 
        $N$ (mesh elements) &  $11.8\times10^6$ \\ \hline
    \end{tabular}
    \label{NASA_CRM_test_cond}
\end{table}

\begin{figure}
    \centering
    \begin{subfigure}[Surface mesh of the NASA CRM.] {
        \includegraphics[trim=80 130 100 160, clip, width=.47\textwidth]{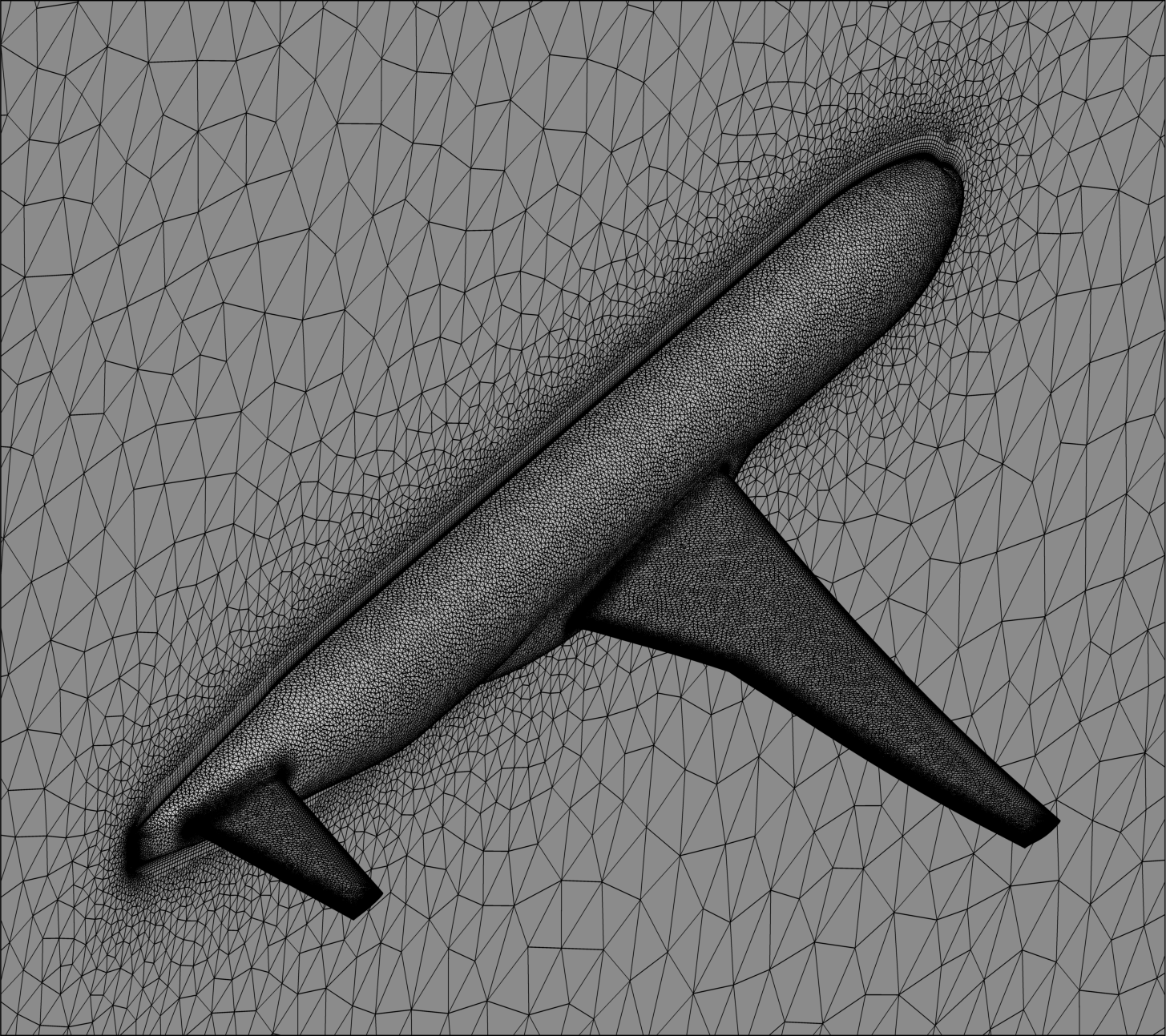} }
    \end{subfigure}
    \hfill
    \begin{subfigure}[Close up of the nose cone showing boundary layer cells on the symmetry plane.]{
        \includegraphics[trim=80 130 100 160, clip, width=.47\textwidth]{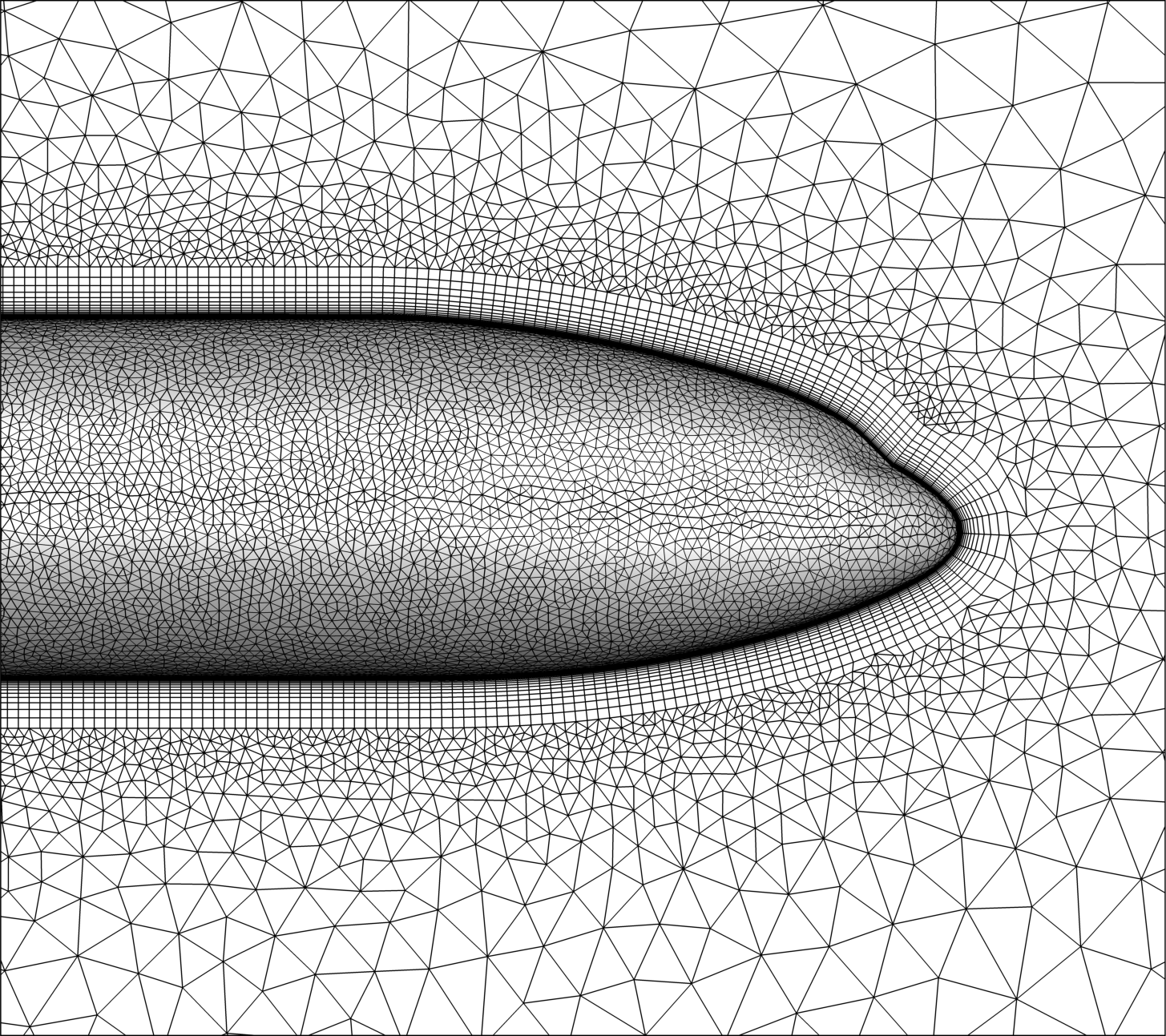} 
    }
    \end{subfigure}
    \hfill
    \begin{subfigure}[Details of the wing surface mesh.]{
        \includegraphics[trim=80 80 80 100, clip, width=.5\textwidth]{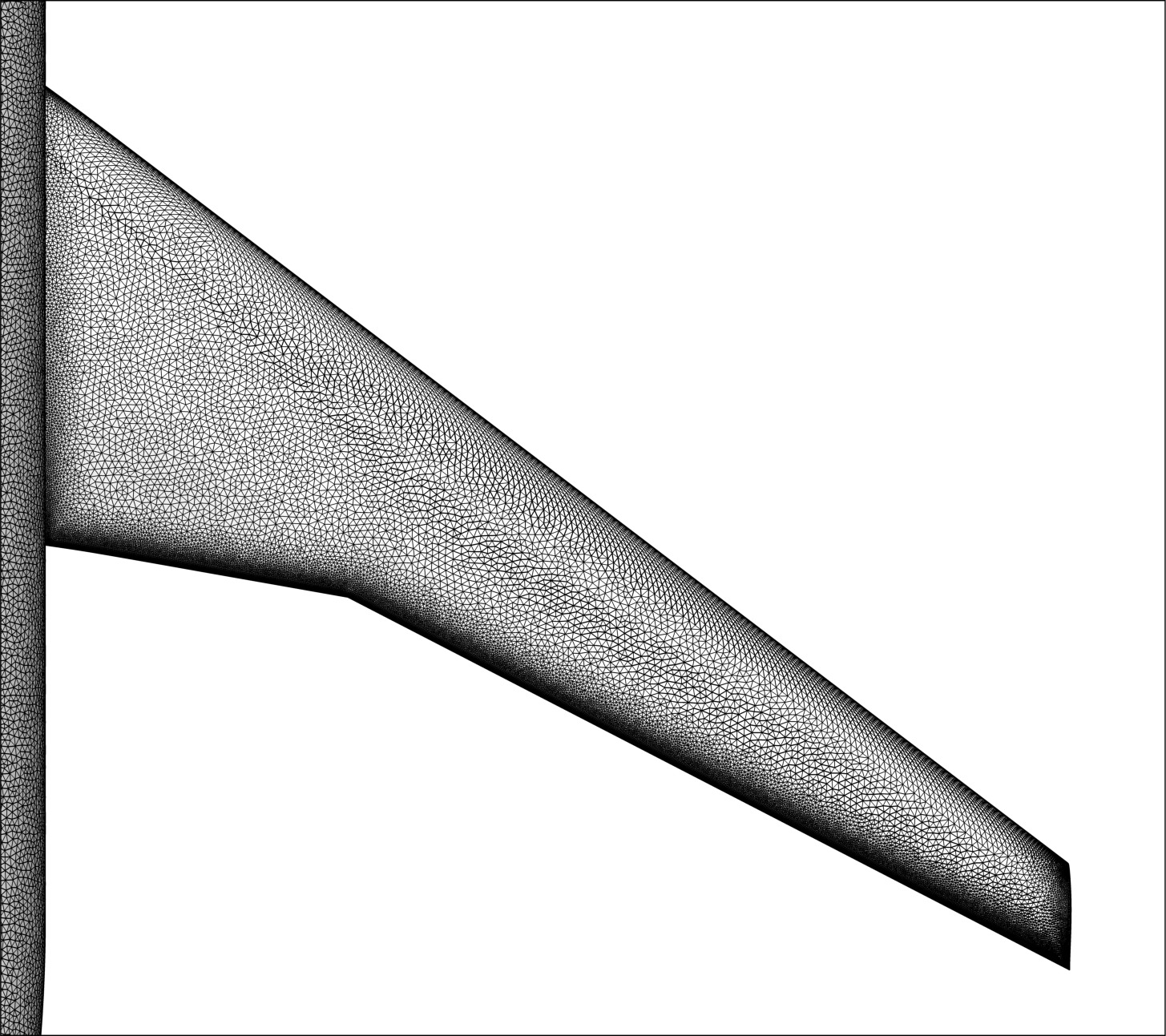} 
    }
    \end{subfigure}
    \caption{Images of the NASA CRM mesh that was used for the CFD simulations.\label{fig:crm_mesh}}
\end{figure}

Figure \ref{fig:crm_mesh} shows details of the unstructured mesh used for the CFD simulations.
The computational domain consists of $11.8\times10^6$ mixed elements ($4.6\times10^6$ nodes), which corresponds to a coarse mesh based on the grid convergence studies performed for multiple solvers and grid topologies \cite{vassberg_summary_2010}.
The difference between the grid-converged drag at a fixed $C_L$ of 0.5 and the present study is $\approx4\%$; this difference can be used as a rough metric for the discretization error.
This error could be reduced through additional grid resolution studies, but it is not a primary factor for the present demonstration.

\begin{figure}{
\centering
    \includegraphics[trim=0 0 0 0, clip, width=.7\textwidth]{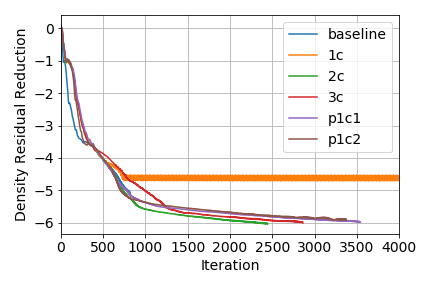} 
    \caption{Convergence history of the density residual for RANS UQ CFD simulations at $\alpha = 2.35^\circ$, Baseline SST turbulence model + $5$ perturbed simulations. \label{fig:convergence_history}}
    \hfill
}
\end{figure}

Figure \ref{fig:convergence_history} shows the convergence history for the simulations required to characterize the RANS uncertainty at a particular operating condition ($\alpha = 2.35^\circ$).
Most cases achieve six orders of magnitude reduction in the density residual.
For the $1C, v_{max}$ case, the residuals stall, but the solution is said to be converged once the force coefficients stabilize to six orders of magnitude (less than $10^{-6}$ change in force coefficients over $100$ non-linear iterations).

In this figure, the solid black line represents the predictions made by the baseline SST turbulence model, the grey area represents the uncertainty estimates predicted by the eigenspace methodology, and the black crosses represent the wind tunnel data.
These wind tunnel data points have error bars associated with them, but these are barely discernible on the scale of the plot. 

The performance of the UQ module is illustrated by comparing the variation of the coefficients of lift ($C_L$), drag ($C_D$), and longitudinal pitching moment ($C_m$) due to the angle of attack ($\alpha$), as predicted by the CFD simulations to those experimentally determined.
At low angles of attack, the flow remains well attached to the aircraft body; therefore, the turbulence model does not introduce significant uncertainty in its predictions.
Accordingly, the interval bounds predicted by the UQ module are relatively small.
At higher angles of attack when there is flow separation over portions of the aircraft, turbulence models struggle to make accurate flow predictions due to the unsteady nature of the flow features and the structural limitations of the isotropic eddy viscosity enforced by the Boussinesq assumption.
The growing uncertainty estimates predicted by the module reflect this shortcoming.


This plot illustrates the paradigm change that RANS UQ methodologies such as the one described in this paper can bring about.
A drag polar is often used in aerospace engineering to understand an aircraft's behavior across its operating range at a given Mach number.
Design engineers use the deterministic values represented by the solid black line in the drag polar plot to determine the optimal operating condition for an aircraft and the aircraft performance characteristics.
Traditionally, only the solid black line is available to the aircraft designer, and they use conservative factors of safety to build operating margins into the design.
The grey areas representing the possible variability of the drag polar explicitly quantify the uncertainties injected by the turbulence model.
Instead of relying on a single deterministic value to design the aircraft around, the uncertainty in the performance prediction can inform the most robust optimal operating condition, and the reliability of the design choices can be quantified.

Since the model-form uncertainty introduced by the turbulence model is negligible at low angles of attack, the CFD predictions of the force and moment coefficients should adhere more closely to the experimental data.
Error in RANS CFD simulations can originate from multiple sources, including, but not limited to: discretization error due to insufficient mesh resolution, modeling error due to the turbulence models, or any errors due to geometric discrepancies in simulated and real-world objects.
If there were only turbulence-model-related uncertainties in the simulations, we would see all the experimental data points lie within the grey uncertainty bounds predicted by the RANS UQ methodology.
However,  we observe a significant deviation of the simulation data from the wind tunnel experiments.
This deviation manifests as a bias and a difference in the slope of the lift coefficient curve.
As was discovered at the 4th Drag Prediction Workshop (DPW) \cite{levy2013summary}, the wind tunnel model of the NASA CRM underwent aeroelastic deformation that affected the values recorded for the force and moment coefficients.
The deformations were more extensive than expected, and as a result, the shape of the model analyzed using numerical simulations was different from that of the wind tunnel model.
New computational geometries that accurately reflected the deformed wind tunnel shape were used for the subsequent workshop \cite{morrison20166th}.
For this work, we use the older computational geometry (without the wind-tunnel-induced deformations) to exercise the ability of the multi-fidelity method to learn biases existing in the data. 

On a swept wing like the one on the NASA CRM model, the added aeroelastic twist results in a decrease in the angle of attack of the wing-tip region relative to the numerically analyzed rigid shape.
This discrepancy leads to a lower $C_L$ than the one calculated using CFD.
Moreover, the increase in the overall twist of the wing unloads the wing-tip sections, effectively displacing the center of lift of the wing upstream and leading to the observed increased values of $C_M$ when compared to those calculated using CFD. 
These differences introduced new uncertainties (of an aeroelastic nature) in the numerical predictions that escape the UQ analysis.  

This discussion serves as an important reminder that regardless of the level of model (in)adequacy of a simulation method, there may be unforeseen uncertainties and errors introduced in the predictions that can cause results to deviate from real-world experiments.
In this particular case, the unexpected aeroelastic deformation of the CRM model in the wind tunnel resulted in slightly different geometries being numerically and experimentally analyzed.
The biases and errors introduced due to such unknowns are not explicitly derived here, but they are present in the data.
The auto-regressive formulation of the multi-fidelity GP will learn these biases and errors from the data.
If the high-fidelity data is truly the most accurate representation of the modeled physical system, it is good that the multi-fidelity GP can learn the bias between the lower-fidelity simulations and the high-fidelity data and compensate for it.
However, if there is an error in the high-fidelity data, the multi-fidelity GP will learn based on the erroneous data and still assume it to be the most accurate source. 
This observation highlights the importance of the hierarchy of data sources in this formulation.
This work employs a clearly defined hierarchy based on the physics captured by each information source.
Wind tunnel experiments form the highest fidelity level, followed by RANS CFD simulations, while vortex-lattice methods form the lowest fidelity level.
In general, the hierarchy could be defined based on the uncertainty of the predictions or expert-informed trust in the information source.

Another critical point is that the interval predictions from the RANS UQ methodology only estimate the uncertainty in simulations due to the turbulence model used.
The methodology cannot estimate other sources of error as it does not rely on any high-fidelity data.
For example, discretization error due to insufficient mesh quality can bias the performance predictions from CFD simulations. 
The RANS UQ methodology does not predict this bias.
Instead, we rely on the multi-fidelity GP framework to learn this and any other biases from the differences between the low- and high-fidelity data. 

As mentioned at the end of Section \ref{sec:equips_rans_uq}, these bounds contain no probability distribution information.
Nonetheless, for multi-fidelity modeling, it is assumed that the distribution of the QoIs within the bounds is Gaussian and symmetric about the middle of the interval.
Additionally, the standard deviation ($\sigma$) of the Gaussian distribution is defined such that the extents of the interval estimates are $2\sigma$ away from the middle of the interval.
This definition means that for any CFD data point with RANS UQ uncertainty estimate, the middle of the predicted interval is regarded as the mean of the Gaussian distribution of the prediction, and the extent of the interval bounds are $\pm 2\sigma$ away from the mean.

\begin{figure}
    \centering
    \begin{subfigure}[$\alpha = 1^\circ$.] {
        \label{fig:01aoa}
        \includegraphics[trim=40 300 150 280, clip, width=.45\textwidth]{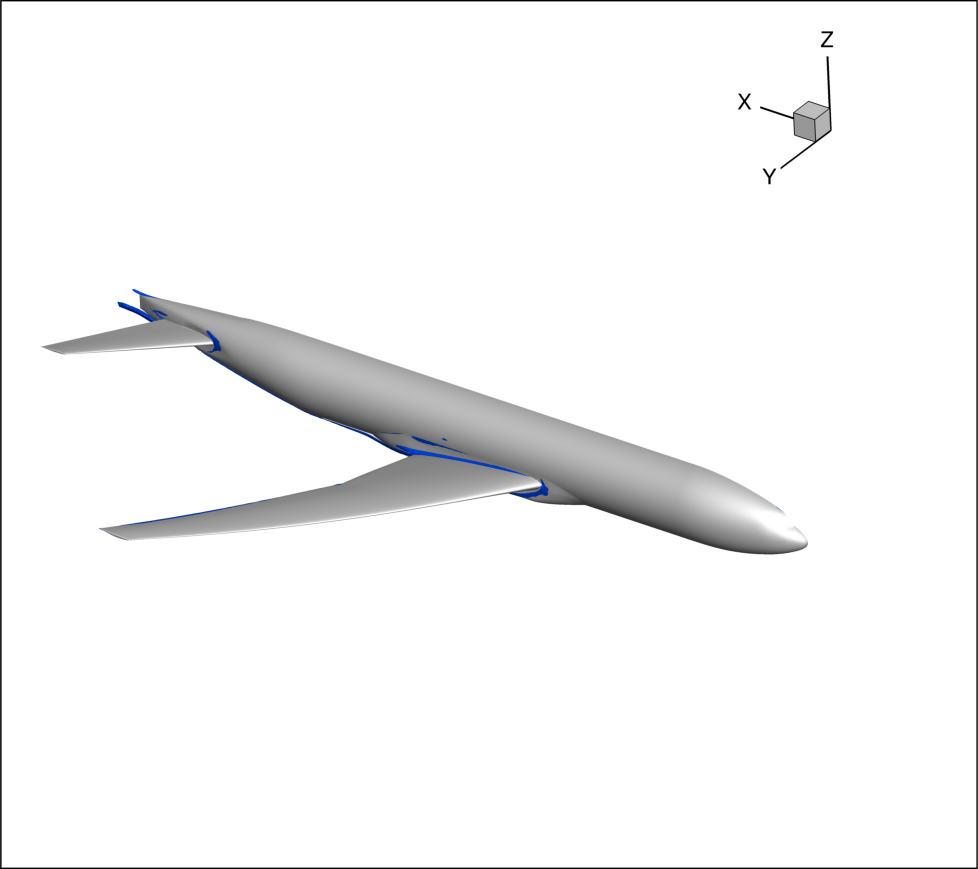} }
    \end{subfigure} 
    \hfill
    \begin{subfigure}[$\alpha = 2.35^\circ$.]{
        \label{fig:02aoa}
        \includegraphics[trim=40 300 150 280, clip, width=.45\textwidth]{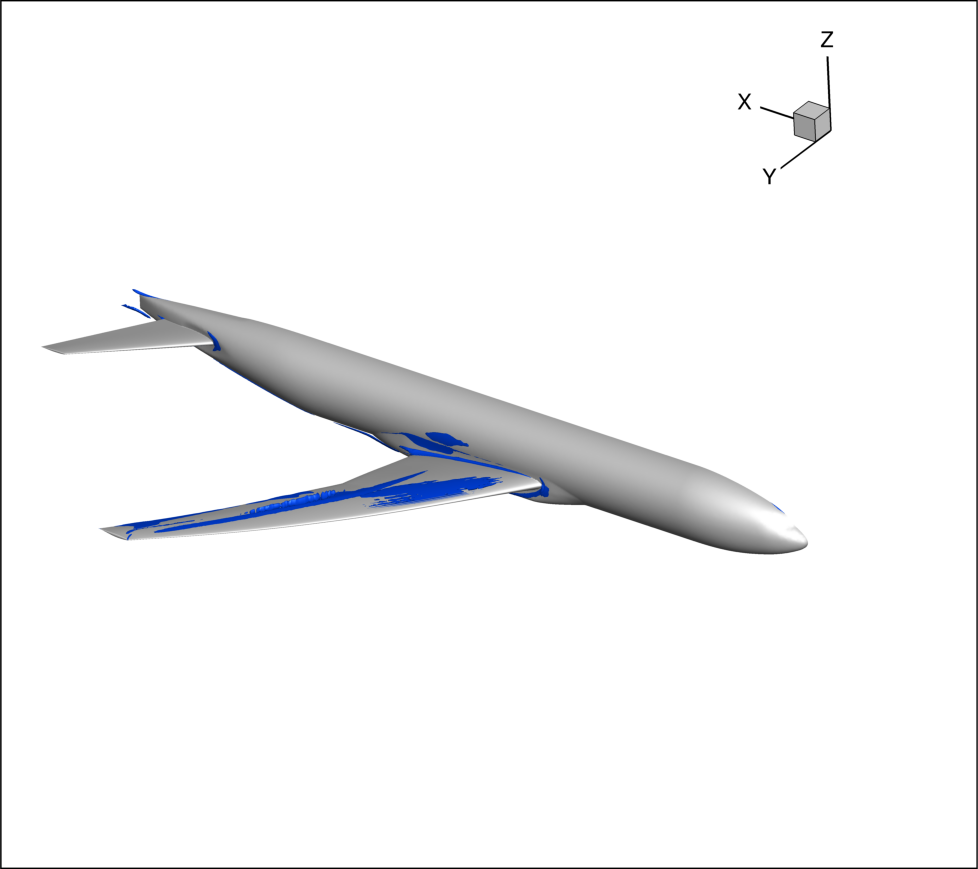} 
    }
    \end{subfigure}
    \hfill
    \begin{subfigure}[$\alpha = 3^\circ$.]{
        \label{fig:03aoa}
        \includegraphics[trim=40 300 150 280, clip, width=.45\textwidth]{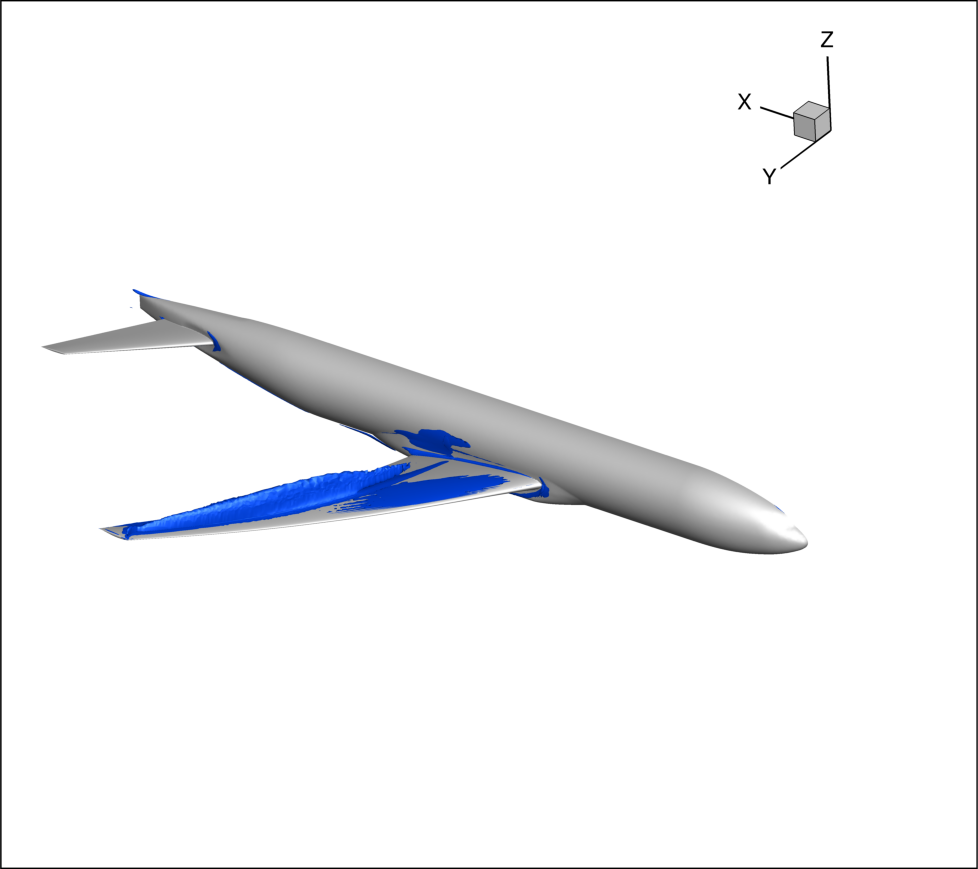} 
    }
    \end{subfigure}
    \hfill
    \begin{subfigure}[$\alpha = 4^\circ$.]{
        \label{fig:04aoa}
        \includegraphics[trim=40 300 150 280, clip, width=.45\textwidth]{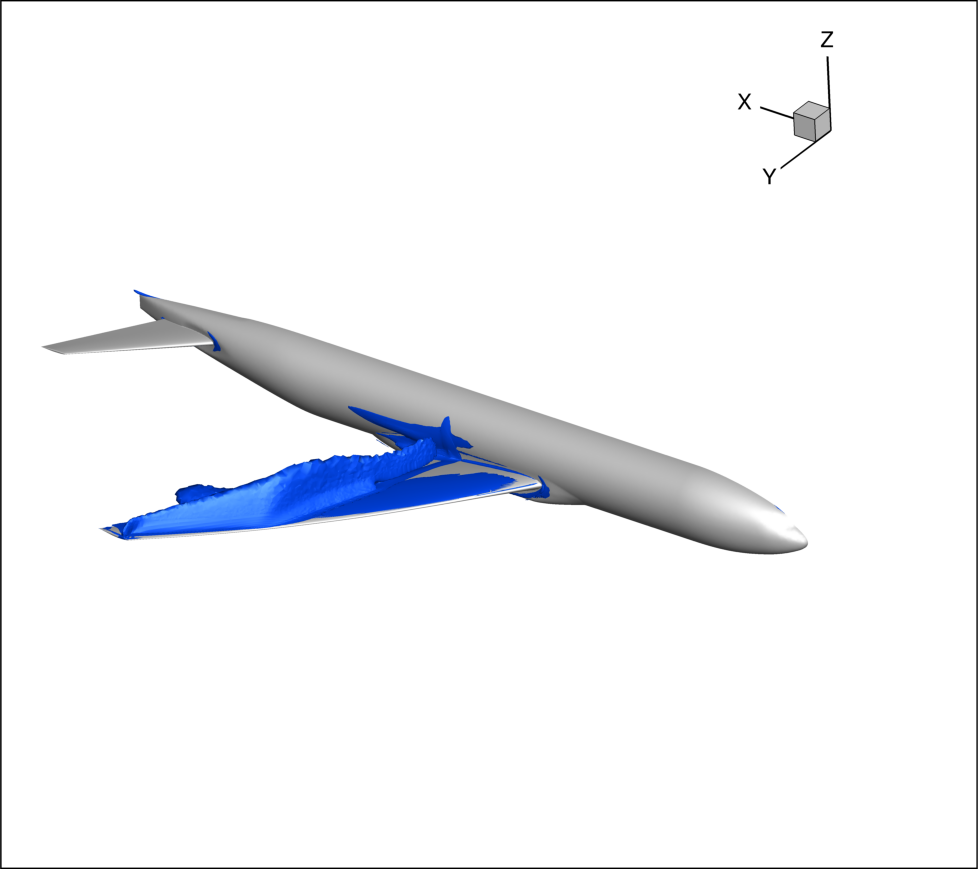} 
    }
    \end{subfigure}
    \hfill
    \caption{Isosurfaces representing areas where local Mach variability $M_v = 0.2$ at various angles of attack. \label{fig:mach_isosurface}}
\end{figure}

The RANS UQ methodology does not only provide interval estimates on integrated quantities like $C_L$, $C_D$, and $C_m$.
Since the eigenspace perturbations result in different realizations of the entire flow field, the data can be post-processed to provide valuable insight into the mechanics of the turbulence model and the regions of the flow field that contribute to the resulting uncertainties.
Figure \ref{fig:mach_isosurface} depicts iso-surfaces of areas where the local Mach number varies by greater than $0.2$ across all the perturbed simulations.
This Mach variability $(M_v)$ is defined at every point in the computational domain as $M_v = max(M_i) - min(M_i)$ where $i$ refers to each realization of the flow field ($5$ perturbed + $1$ baseline flow fields) and $M_i$ represents the Mach number at each point in that flow field. 

At low angles of attack, Figure \ref{fig:01aoa}, the Mach variability is low and limited to small regions in the flow field.
The small regions indicate that the eigenspace perturbations do not cause significant changes in the flow, resulting in smaller uncertainty bounds.
As the angle of attack increases, as shown in Figures \ref{fig:02aoa} and \ref{fig:03aoa}, larger areas of variability appear where the shock is expected: at the upper surface of the wing and away from the leading edge.
These regions denote an uncertainty in the shock location.
This area grows rapidly until it reaches the leading edge in Figure \ref{fig:04aoa}, signaling large uncertainty estimates and reduced confidence in the CFD predictions.
Such visualizations allow us to analyze the relationship between the dominant flow features and their role in introducing uncertainty in the turbulence models. 

Similarly, the variability in any other flow quantity can be analyzed.
Knowing which areas contribute to the uncertainty in performance predictions can aid design decisions.
These results can inform sensor placement when moving to experimental campaigns.
For example, a higher density of pressure sensors can be used in areas with large variability in the pressure coefficient ($C_P$).
Flow visualization techniques can be focused on areas with large velocity variability.
From the perspective of turbulence modeling, these variability visualizations can shed light on the types of flow features that are hard to predict using RANS simulations.
This additional data processing provides more qualitative applications for the RANS UQ methodology. 

\chapter{Multi-Fidelity Gaussian Process Regression} \label{chap:mf_gp}
Most computational or experimental analysis techniques provide realizations of a quantity of interest (QoI) at discrete points in the domain of interest. 
Each analysis, referred to as a function evaluation, has a monetary and computational cost. 
In practice, higher-fidelity function evaluations are more costly than lower-fidelity ones.
If the converse were true, there would be no reason to use a lower-fidelity analysis in place of a higher-fidelity one.
Fidelity here refers to how closely an analysis technique mirrors real-life physics. 

In theory, the discretization of the domain can be fine enough that it results in a nearly continuous representation of the QoI.
Additionally, using only the highest-fidelity analysis techniques would minimize the uncertainty or error in the QoI predictions. 
In reality, this is not tractable. 
Often cost minimization is a priority. 
This constraint equates to representing QoIs as sparsely as possible and lower-fidelity analysis techniques replacing higher-fidelity ones wherever valid.  

Numerous statistical methods that use these discrete realizations to create continuous representations of the QoI, have been developed.
These representations, called surrogate models, assume that the underlying function of interest varies smoothly between the discrete data points.
It allows us to squeeze the most out of the available, limited function evaluations.
Popular surrogate modeling techniques include radial basis functions \cite{park1991universal}, Gaussian processes (GP) \cite{krige1951statistical,matheron1963principles,rasmussen_gaussian_2006}, stochastic collocation \cite{loeven2007probabilistic}, and polynomial chaos expansions (PCE) \cite{oladyshkin2012data,blatman2011adaptive}.

Of these, the use of multi-fidelity data has been developed for GP \cite{kennedy_predicting_2000,gratiet_multi-fidelity_nodate} and, more recently, for PCE \cite{ng2012multifidelity, palar2018global}.
Both methods handle the multi-fidelity data by using correction terms trained on the difference between the low-fidelity and high-fidelity data.
GP have a slight advantage in this regard as they use an additive and a multiplicative term in the correction, whereas PCE only uses an additive term. 
PCE have a distinct advantage in sensitivity analyses as Sobol indices can be directly post-processed \cite{sudret2008global,crestaux2009polynomial}.
On the other hand, GP can handle uncertain inputs and directly estimate the error in its modeling.
This feature can be used for adaptive sampling techniques that suggest additional function evaluations to reduce the uncertainty in the model \cite{xu2011adaptive}.
For this work, GP are the surrogate model of choice due to its advanced handling of multi-fidelity data that have associated uncertainties and the direct estimation of uncertainty in its predictions.

This chapter introduces the fundamental equations of GP regression used to handle single-fidelity data that can have uncertainties associated with it.
It builds upon this by introducing multi-fidelity GP regression that combines data from different sources to build a single, superior surrogate model for the QoI.
These multi-fidelity GP equations are then used to create probabilistic aerodynamic databases representing the performance characteristics of a model aircraft, the NASA CRM. 
The benefits of using multi-fidelity data vs. single-fidelity data are emphasized as well.

\section{Gaussian Process Regression} \label{sec:gpr}
The basic building block of the multi-fidelity framework is Gaussian Process (GP) regression \cite{rasmussen_gaussian_2006}.
It is a supervised learning technique used to build a surrogate model for an unknown function $y = f(\mathbf{x})$ given $n$ observed input-output pairs $\mathcal{D} = (\mathbf{x}_i, y_i)$ for $i \in\{1,...,n\}$.
The function can be non-deterministic and have Gaussian noise, $\sigma$, associated with its observations.
The unknown function can have multi-dimensional inputs but is assumed to have a scalar output.
These input-output pairs can be arranged in matrices $X$, $\mathbf{y}$ and $\sigma$.
If the function has an $m$ dimensional input then $X$ is an $\left (n \times m \right)$ matrix of inputs, and $\mathbf{y}$ and $\sigma$ are $\left (n \times 1 \right)$ vectors of outputs and associated uncertainties, respectively.

In the context of this work, while computer simulations are deterministic, they can have modeling errors and uncertainties. 
These are treated as Gaussian noise in the function of interest with a $\sigma$ proportional to the errors and uncertainties.
Experimental data is not deterministic as factors such as natural variation in environmental conditions, or sensor limitations, can introduce uncertainties in the data. 
These are also estimated as Gaussian noise.

Since these observations can be imperfect, each observation is assumed to carry some Gaussian noise associated with it such that $y_i \sim \mathcal{N}(E(f(\mathbf{x}_i)),\sigma^2(\mathbf{x}_i))$.
Assuming that all the observations in $\mathcal{D}$ have a joint Gaussian distribution, a GP can be used as a surrogate model for the data.
A GP is completely defined by its mean function, $ \mu(\mathbf{x}) $, and a kernel function $k(\mathbf{x,x';\theta})$ that is parameterized by some hyperparameters $\theta$.
For this study, the squared exponential function is used as the kernel function:
\begin{equation}
    k\left (\mathbf{x,x'} \right ) = \sigma_f^2 \exp \left ( -\sum_{d=1}^{m}\frac{\left ( x_d - x'_d \right )^2}{2l_d} \right ),
\end{equation}
where $m$ is the dimension of the input.
The hyperparameters for this kernel function are the signal variance $\sigma_f^2$, and the length scales $l_d$.
The kernel function is used to create a kernel matrix $K \in \mathbb{R} ^{ n \times n}$ where $K_{ij} = k\mathbf{\left( x_i, x_j \right )}$.

To enable the GP to estimate functions with a non-zero mean, the mean of $f(\mathbf{x})$ is represented using $p$ fixed basis functions, $\mathbf{h(x)}$, and learned regression coefficients $\beta$.
At a minimum, these basis functions include a constant term but can have multiple polynomial terms.
With these in mind, the surrogate model evaluated at a location of interest $\mathbf{x}_*$ can be represented as a mean value, $\mathbf{h(\mathbf{x}_*)}^T\beta$, plus a zero-mean GP.

The $n_*$ sample locations and the basis functions can also be arranged in matrices $X_* \in \mathbb{R} ^{ n_* \times m}$ and $H \in \mathbb{R} ^{ p \times n_*}$ such that each row of $X_*$ is a $m$-dimensional sample location and each column of $H_*$ is a $p$-dimensional result of the basis functions at the locations in $X_*$.

Combining the GP regression equations for noisy observations with those incorporating explicit basis functions and writing in the matrix notation, the surrogate model is defined its mean and error estimate:
\begin{equation} \label{equ:mu_gpr}
    \mu(X_*) = H_*^T\hat{\beta} + K(X_*,X)[K(X,X)+\text{diag}(\sigma_i)]^{-1} (\mathbf{y}-H^T\hat{\beta}), 
\end{equation}
\begin{equation} \label{equ:sig_gpr}
    \sigma^2(X_*,X_*) = K(X_*,X_*) - K(X_*,X)[K(X,X)+\text{diag}(\sigma_i)]^{-1} K(X,X_*), 
\end{equation}
where $\hat{\beta} = (H^TV^{-1}H)^{-1}H^TV^{-1}y$ is the best linear estimator for the regression coefficients and $V = K(X,X) + \text{diag}(\sigma_i)$ represents the kernel matrix at the observed points $\left ( K(X,X) \right )$ and includes the Gaussian noise that is associated with each observation $\left ( \sigma_i \right )$.
The prediction from the surrogate model is defined by the mean $\mu(X_*)$ and the uncertainty associated with these predictions is represented by the diagonal of the $\sigma^2(X_*,X_*)$ function.

To fully define the GP, the hyperparameters of the kernel function need to be learned from the data.
The marginal likelihood of the model as a function of the hyperparameters is expressed as:
\begin{equation}
    p(\mathbf{y}|\theta,X) = \int p(\mathbf{y}|f,X) p(f|\theta,X) df.
\end{equation}
Often the log of the marginal likelihood is used. 
In the case of a Gaussian process with a Gaussian prior the log marginal likelihood function is
\begin{equation} \label{equ:hyp_param_sf}
    \log~p(\mathbf{y}|\theta,X) = -\frac{1}{2} \log|V| - \frac{1}{2}\alpha^T V^{-1}\alpha - \frac{n}{2}\log 2\pi,
\end{equation}
where $\alpha = \left ( \mathbf{y}-H^T\hat{\beta} \right )$ \cite{rasmussen_gaussian_2006}.
The hyperparameters are chosen to maximize this log marginal likelihood. 

For consistency across sections, the following low- and high-fidelity analytic functions will be used to show the functioning of the GP regression process: 
\begin{align} \label{equ:lf_function}
    f_{LF}(x) &= 0.5 \left ( 6x - 2\right )^2 \sin{ \left (12x -4 \right )} + 10 \left ( x - 0.5 \right ) -5.
\\ \label{equ:hf_function}
    f_{HF}(x) &= 2 f_{LF}(x) - 20x + 20 + \sin {\left ( 10 \cos{ \left ( 5x \right )}\right )}.
\end{align}
The high-fidelity function differs from the low-fidelity function by a few polynomial terms and has a high-frequency variation absent in the low fidelity approximation. 

Using only the high-fidelity function, we show a basic example of how single-fidelity GP regression uses discrete function evaluations to create a continuous representation of the QoI.
The solid red line represents this function. 
The data points, shown as black circles, are uniformly distributed between $0$ and $1$.
The mean prediction of the GP is the solid black line, and the gray area represents the $2\sigma$ error estimate.
The error estimate from the GP regression goes to zero near these data points, where the uncertainty in the value of the underlying function is zero. 
The error estimate increases between the data points, where the uncertainty in the modeling parameters introduces uncertainty in the surrogate model prediction. 
The larger, near-constant size error estimate shows that GP regression respects these uncertain data inputs, and the GP incorporates it into the prediction of the underlying function.

An essential feature of GP regression used extensively in this work is the ability to create samples of the GP that are potential representations of the underlying function and respect the error estimates from the GP. 
A sample mean, $\mu_S(X_*)$ at $X_*$ locations is generated as
\begin{equation} \label{equ:gp_sampling}
    \mu_S(X_*) = \mu(X_*) + \sigma^2(X_*,X_*) U,
\end{equation}
where $U$ is a $n_* \times 1$ vector of random variables drawn from a standard normal distribution $\mathcal{N}(0,1)$.
Each colored line is a separate sample and represents a potential candidate for the underlying function being estimated, based on the limited provided data. 

\section{Multi-Fidelity Gaussian Process Regression} \label{sec:mf_modeling}

It is often the case that simulations or experiments of sufficiently high fidelity are too expensive to perform over the entire domain of interest for a modeled problem.
Often, there are lower-fidelity approximations available that can be evaluated quickly to perform parameter studies.
The multi-fidelity GP aims to use data from different fidelity levels to create a surrogate model that can best approximate the highest-fidelity function and its uncertainty while reducing the required number of high-fidelity function evaluations. 

Assume there are $s$ information sources $f_t(\mathbf{x})$, where $t\in\{1,2,...,s\}$, and the function at the highest fidelity level, $f_s(\mathbf{x})$, is being approximated using a Gaussian Processes, $Z_s(\mathbf{x}) \sim \mathcal{N}(\mu_{s}(\mathbf{x}), \sigma_s^2(\mathbf{x}))$.
An auto-regressive formulation of the multi-fidelity framework is used. This was first put forward in \cite{kennedy_predicting_2000} and was improved upon by \cite{gratiet_multi-fidelity_nodate} to reduce computational cost and improve predictions.
The GP approximation at the $t$ fidelity level is modeled as

\begin{equation}
    Z_t(\mathbf{x}) = \rho_{t-1}(\mathbf{x})Z_{t-1}(\mathbf{x}) + \delta_t(\mathbf{x}),
\end{equation}
\begin{equation}
    \rho_{t-1}(\mathbf{x}) = \mathbf{g}_{t-1}^T(\mathbf{x})\beta_{\rho_{t-1}},
\end{equation}
where $\mathbf{g}_{t-1}(\mathbf{x})$ is a set of $q$ basis functions, similar to $\mathbf{h}(\mathbf{x})$ in the previous section, $\beta_{\rho_{t-1}}$ is the learned regression coefficients, and $\delta_t(\mathbf{x})$ is modeled using a GP.
A way to interpret these terms is to consider $\delta_t(\mathbf{x})$ the additive bias and $\rho_{t-1}(\mathbf{x})$ the multiplicative bias between fidelity levels $t$ and $t-1$.
To account for the different fidelity levels and their corresponding data, the subscript $t$ is added to the notation introduced in Section \ref{sec:gpr}.
For example, $X_t$ refers to all the input data at level $t$. Additionally, the term $\Sigma_t = \text{diag}(\sigma^2_{i,t})$ is introduced, which refers to the noise in the outputs $\mathbf{y_t}$. 

In Appendix B of \cite{gratiet_multi-fidelity_nodate}, Gratiet presents the predictive equations for the case when the design sets are not nested ($\mathcal{D}_t \notin \mathcal{D}_{t-1}$) and the data has no process noise, such that $\Sigma_t$ is a null matrix.
This work extends those equations to include process noise $\Sigma_t \neq \emptyset$, which produces the following representations for the mean and covariance equations for fidelity level $t \neq 1$ as

\begin{equation}\label{equ:mu_Zt}
\begin{split}
    \mu_{t}(X_*) = & ~ \rho_{t-1} \left ( X_* \right ) \mu_{t-1} \left (X_* \right ) + H_*^T\beta_t \\
    & + \left [ \left ( \rho_{t-1} \left (X_* \right ) \rho_{t-1} \left (X_t \right )^T \right ) \odot \sigma^2_{t-1} \left(X_*,X_t \right) + K_{t} \left(X_*,X_t \right)\right] \\ 
    & \times \left [ \left ( \rho_{t-1} \left (X_t \right ) \rho_{t-1} \left (X_t \right )^T \right ) \odot \sigma^2_{t-1} \left(X_t,X_t \right) + V_t \right ]^{-1} \\ 
    & \times \left (\mathbf{y}_t - \rho_{t-1} \left (X_t \right ) \odot \mu_{t-1} \left (X_t \right) - F_t^T \beta_t \right),
\end{split}
\end{equation}
and
\begin{equation}\label{equ:sig_Zt}
\begin{split}
    \sigma^2_{t}(X, \Tilde{X}) = & ~ \left (\rho_{t-1} \left ( X \right ) \rho_{t-1} ( \Tilde{X})^T \right ) \odot \sigma^2_{t-1} (X, \Tilde{X}) + K_t(X, \Tilde{X}) - \\
    & \left [ \left ( \rho_{t-1} \left (X \right ) \rho_{t-1} \left (X_t \right )^T \right ) \odot \sigma^2_{t-1} \left(X,X_t \right) + K_{t} \left(X,X_t \right)\right] \\ 
    & \left [ \left ( \rho_{t-1} \left (X_t \right ) \rho_{t-1} \left (X_t \right )^T \right ) \odot \sigma^2_{t-1} \left(X_t,X_t \right) + V_t \right ]^{-1} \\ 
    & \left [ \left ( \rho_{t-1} \left (X_t \right ) \rho_{t-1} ( \Tilde{X} )^T \right ) \odot \sigma^2_{t-1} (X_t, \Tilde{X} ) + K_{t} ( X_t, \Tilde{X} ) \right], 
\end{split}
\end{equation}
where $(X, \Tilde{X})$ are generic input arguments, $V_t = K_{t} \left(X_t,X_t \right) + \Sigma_t $, and $\rho_{t-1} (X) = G_{t-1}(X)^T \beta_{\rho_{t-1}}$.
$G_{t-1}(X)$ is a $q \times n$ matrix where each column is a $q$-dimensional result of the basis functions for the corresponding $m$-dimensional row of input samples in $X$, and $\beta_{\rho_{t-1}}$ are learned regression coefficients.
For the lowest fidelity level, $t=1$, the regular GP regression equations, Equations \ref{equ:mu_gpr} and \ref{equ:sig_gpr}, are used.
For a set of sample locations $X_*$, the mean predictions at fidelity level $t$ is given by $\mu_t(X_*)$ and the variance in the predictions is given by the diagonal of $\sigma^2_t(X_*,X_*)$. 

To fully define the GP of each fidelity level, the regression coefficients ($\beta_{\rho_{t-1}}$ and $\beta_t$) and the hyperparameters of the kernel functions of each fidelity level need to be learned from the data.
The parameter estimation equations from  \cite{gratiet_multi-fidelity_nodate} are extended for noisy observations:

\begin{equation} \label{equ:param_est_mf}
    \begin{bmatrix}
    \beta_t & \beta_{\rho_{t-1}}
    \end{bmatrix} = \left [ J_t^T \left ( K_t(X_t,X_t) + \Sigma_t \right )^{-1} J_t \right ]^{-1} \left [ J_t^T \left ( K_t(X_t,X_t) + \Sigma_t \right ) ^{-1} \mathbf{y_t} \right ], 
\end{equation}
with $J_1 = H_1$ and for $t > 1$, $J_t =  \begin{bmatrix} G_{t-1} \odot \left ( \mu_{t-1} \left ( X_t \right )  \mathbf{1}_{q_{t-1}} \right ) & F_t \end{bmatrix}$.
$\mathbf{1}_{q_{t-1}} \in \mathbb{R} ^{q_{t-1} \times n_t} $ is a matrix of ones.
The hyperparameters of the kernel functions are learned by minimizing the log marginal likelihood of each fidelity level:
\begin{equation} \label{equ:hyp_param_mf}
    \log~p(\mathbf{y}_t|\theta_t,X_t) = -\left (\frac{1}{2} \log|V_t| + \frac{1}{2} \alpha_t^T V_t^{-1} \alpha_t + \frac{n_t}{2}\log 2\pi \right),
\end{equation}
where $\alpha_t = \left (\mathbf{y_t} - \rho_{t-1} \beta_{\rho_{t-1}}-F_t\beta_t \right )$.

The potential advantage of using multi-fidelity data to estimate a function of interest is shown by using the analytic functions defined by Equations \ref{equ:lf_function} and \ref{equ:hf_function} as the low-fidelity approximation and the high-fidelity function of interest, respectively. 
For this case, the number of high-fidelity data points is restricted to $4$. 
There are not enough data points for the GP regression to learn all the nuanced trends in the high-fidelity function of interest. 
The low-fidelity data does not have some of the higher-frequency information present in the high-fidelity data, but it approximates the general trends reasonably well. 
In this case, the low-fidelity data bolsters the scant $4$ high-fidelity data points, and the multi-fidelity GP that combines both sets of data provides a more accurate representation of the underlying function of interest. 
It is also essential to notice the large error estimates in the GP prediction for both cases. 
More high-fidelity data would be required to reduce the uncertainty in the GP modeling.

As mentioned earlier in this section, the recursive formulation put forth by Gratiet \cite{le_gratiet_recursive_2014} improves on the work initially done by Kennedy and O'Hagan \cite{kennedy_predicting_2000} by reducing the computational complexity of the training and sampling steps of the multi-fidelity GP.
The Gratiet implementation achieves this improvement by splitting the dataset into individual fidelity levels instead of agglomerating the data from all levels into one set of equations.
This step results in inverting smaller matrices, which dramatically improves the computational cost of the process.
The same analytic functions used thus far, Equations \ref{equ:lf_function} and \ref{equ:hf_function}, are used to train the GP regressions.
In this case, the number of high-fidelity data points ($n_{HF}$) and low-fidelity data points ($n_{LF}$) had a constant ratio: $\frac{n_{HF}}{n_{LF}} = 0.2$.
These savings in computational time increase with more fidelity levels and higher-dimensional functions. 

\section{Application to NASA CRM} \label{sec:mf_gp_nasa_crm}

To provide more quantitative applications of the RANS UQ methodology introduced in Chapter \ref{chap:rans_uq}, we demonstrate its inclusion in the multi-fidelity GP framework.
In the multi-fidelity setting, the CFD data is augmented with low-fidelity simulations using the Athena Vortex Lattice (AVL) code \cite{drela2008athena}, and high-fidelity experimental data from the wind tunnel campaigns \cite{rivers_further_2012,rivers_experimental_2010} used in the preceding section.
For the vortex-lattice simulations, the uncertainty information is provided by subject matter experts (industry users and academics), while for the experimental data, the uncertainty intervals used are those described in the wind-tunnel campaign reports. 

For the ease of illustration, $C_L$, $C_D$, and $C_m$ are initially considered one-dimensional functions of $\alpha$ ($m = 1$).
The methodology described earlier in Section \ref{sec:mf_modeling} is generally applicable to functions of many variables.
Aerodynamic databases are often multi-dimensional functions, usually with a maximum of $5$ input variables (angle of attack, side-slip angle, Mach number, altitude, and dynamic pressure).
The end of this section explores multi-dimensional, multi-fidelity databases. 

For each figure, the solid black line represents the mean predicted by the GP, and the grey area represents the $\pm 2\sigma$ error estimate as predicted by the GP.
The left column of the figures shows the AVL data and a single-fidelity GP fit on that data.
In the middle, we introduce the SU2 RANS CFD data with uncertainty bounds informed by the RANS UQ methodology and show the two-fidelity GP fit.
On the right, we introduce a limited set of wind tunnel data points to inform the highest fidelity and show the resulting three-fidelity fit.
For each QoI, the build-up of the database is shown.
Table \ref{table:data_points} shows the distribution of data points across the fidelity levels.

\begin{table}
\centering
    \renewcommand{\arraystretch}{1.2}
    \captionsetup{justification=centering}
    \caption{Number of data points of each fidelity that are used} 
    \begin{tabular}{c|c}
        Data Source & Data Points \\ \hline
        Low Fidelity (AVL) & 23 \\ 
        Medium Fidelity (CFD) & 11 \\  
        High Fidelity (Wind Tunnel) & 5 \\  
    \end{tabular}
    \label{table:data_points}
\end{table}

These data points also have uncertainties associated with them, but these are very small and are indistinguishable on this scale in the figures.
The second and third columns of figures show the mean and standard deviation predictions made by the multi-fidelity GP methodology from Section \ref{sec:mf_modeling}.

The multi-fidelity GP can learn the biases between the different fidelity levels and provide predictions that fit very well with the highest fidelity.
The error is calculated using the $N$ unused highest-fidelity data points:

\begin{equation}\label{equ:rmse}
    RMSE = \sqrt{\frac{\sum_{i=1}^{N}\left ( \mu_{s,i} - y_{s,i} \right )^2}{N}},
\end{equation}
where $y_{s,i}$ is the $i$-th data point of the highest ($s$) fidelity, and $\mu_{s,i}$ is the highest-fidelity prediction from the GP at the same input conditions.  

Since the QoIs are simple functions of $\alpha$, not many high-fidelity data points are required to capture functional dependence accurately.
The differences between the prediction accuracy for a single- vs. multi-fidelity fit are not significant.
Nonetheless, the trends are as expected.
When high-fidelity data is scarce, the multi-fidelity predictions perform better for all QoIs since the low-fidelity data helps provide the general trends learned by the multi-fidelity GP.
As the number of high-fidelity data points increases, the RMSE converges since there is enough data for both fits to reproduce the functional dependence accurately.
When high-fidelity data cover the domain very well, the single-fidelity fits do marginally better since the low-fidelity data (in the multi-fidelity fits) does not provide any helpful information and can introduce noise in the predictions.

Another strength of this multi-fidelity GP methodology is apparent when the high-fidelity data is localized to a specific part of the domain.
Such a situation might arise if resources are limited and it is not feasible to perform high-fidelity evaluations over the entire domain of interest.
It might also be the case that the lower-fidelity simulations are reasonably accurate in a specific part of the domain and, consequently, introduce smaller uncertainties in these domain regions.
For the NASA CRM, it is mentioned in Section \ref{sec:crm_rans_uq} that at low angles of attack, where the flow remains attached to the aircraft, RANS CFD simulations are pretty successful at predicting performance metrics.
In this case, an engineer might conclude that the highest-fidelity evaluations are not necessary at $\alpha < 5^\circ$ and that the lower-fidelity sources achieve sufficient accuracy.

Such a situation is simulated by creating a multi-fidelity GP that uses AVL and SU2 data that spans the entire domain of interest but uses wind tunnel evaluations only at high angles of attack $(\alpha > 5^\circ)$.
This situation is a manufactured one, where we choose to ignore some of the wind tunnel data to illustrate the ability of the multi-fidelity GP framework to perform reliably without high-fidelity information that spans the domain of interest.
The wind tunnel data used to train the models is restricted to high angles of attack, but the unused wind tunnel data is also included in the plots to discern the quality of the predictions.
The right column presents the RMSE of the single- and multi-fidelity GP predictions when using localized high-fidelity data.
The case shown in the left and middle columns is highlighted in the error comparison in the right column.
From the RMSE comparison, it is clear that having accurate low-fidelity data at low angles of attack informs the GP prediction in that region and allows it to follow the trend of the physical phenomena more accurately than when only using the localized high-fidelity data.

The performance of the multi-fidelity predictive capability in multiple dimensions is explored by considering the same aerodynamic coefficients from before ($C_L$, $C_D$, and $C_m$) as functions of both, $\alpha$ and Mach number.
Two sources of information, AVL simulations and wind tunnel data create two-fidelity, two-dimensional GP.

These samples of the GP represent candidates for possible functions that the provided data would explain. 
The error estimate from the GP dictates the slight variations between the samples.
This estimate, in turn, depends on the uncertainty in the input data and the learned GP parameters. 
This GP sampling procedure is used extensively in later chapters to create multiple representations of the same aircraft. 
These samples can be put through the same flight simulations to analyze how the uncertainties in data affect predicted flight performance.

Predictions from the two-fidelity GP are compared to those made from single-fidelity GP that use only the wind tunnel data.
When trying to represent two-dimensional functions, the multi-fidelity fit retains its advantage for longer, with the single-fidelity fit taking $\approx 50$ high-fidelity data points to achieve similar accuracy.
If the number of high-fidelity points is increased beyond that, the two fits behave identically.
For these results, the high-fidelity data was spread evenly across the domain of interest: $-2^\circ \leq \alpha \leq 12^\circ$ and $0.7 \leq \text{Mach} \leq 0.87$.
As the number of input dimensions increases, more data points would be required to capture the functional trends.
Leveraging the multi-fidelity improvement in these high-dimensional spaces would be beneficial in reducing time spent collecting high-fidelity data where a lower-fidelity might suffice.

\chapter{Probabilistic Aerodynamics and Controls Databases} \label{chap:aero_db}
At any point during an aircraft's flight, the airflow over the aircraft exerts aerodynamic forces and moments on the airframe.
These are a function of the aircraft's geometry, orientation (angle of attack, angle of sideslip), and operating conditions (dynamic pressure, Mach number, altitude).
While designing the aircraft, calculating these forces and moments at various points in its operating envelope helps predict the aircraft's behavior and performance characteristics.
Most aerodynamic analyses, be it computational or experimental, are geared towards creating a database that catalogs these values as a function of the aircraft's orientation and operating conditions.

The industry standard is to have a lookup table populated by data from aerodynamic analyses performed during the design process.
They get updated as the design progresses, and the results from the higher-fidelity analysis techniques replace the lower-fidelity data.
The forces and moments are described as multi-dimensional functions depending on up to $5$ input variables: angle of attack, sideslip angle, Mach number, dynamic pressure, and altitude.
Often only a subset of the $5$ input variables is used.
Discrete analyses in this multi-dimensional domain provide data points that populate the table. 
Using values between analysis locations requires interpolation.
These databases are deterministic and have no characterization of the uncertainties present in the analysis techniques. 

Previous work by Wendorff et al. \cite{wendorff_combining_2016} introduces the concept of probabilistic aerodynamic databases that uses multi-fidelity data and its associated uncertainties in a Gaussian Process regression framework to create a non-deterministic representation of the database.
He used a combination of sensitivity and uncertainty analysis to develop an adaptive sampling technique to find the best location to perform the subsequent analysis to minimize the uncertainty in the objective function at minimum analysis cost.
Its application was demonstrated using the NASA CRM performing a longitudinal FAA certification maneuver. 

The current work further matures the probabilistic aerodynamic database concept.
Comprehensive, multi-fidelity, multi-dimensional aerodynamics and controls databases are created for a full-configuration generic T-tail transport aircraft.
The databases define the aircraft's lateral and longitudinal dynamics.
Uncertainties in analysis techniques are considered, with the CFD uncertainties provided by the eigenspace perturbation methodology introduced in Chapter \ref{chap:rans_uq}.
The single- and multi-fidelity GP regression equations from Chapter \ref{chap:mf_gp} are used to create probabilistic surrogate models for these databases. 
This chapter presents the details of these databases, including data sources, data generation, and visualizations. 

\section{Aerodynamics and Controls Databases}

An aircraft flies through a multitude of operating conditions during a mission. During takeoff and landing, the aircraft flies through dense air in its high-lift configuration at low speeds and high angles of attack. 
While cruising, the aircraft flies steady and wings-level at high altitude with little controller input. 
Understanding an aircraft's predicted behavior in all these operating conditions is required for successful certification of the aircraft. 

This work builds extensive aerodynamics and controls databases by combining multiple sources of information and their associated uncertainties.
The generic T-tail transport aircraft is used to demonstrate this capability, and consequently, all the work shown in this chapter will concern itself with this configuration.

Table \ref{tab:aero_db} lists the coefficients used to create the aerodynamic database.
Force and moment coefficients are two-dimensional functions of the angle of attack ($\alpha$) and the sideslip angle ($\beta$).
Stability derivatives are one-dimensional functions of $\alpha$. 
\begin{table}
    \renewcommand{\arraystretch}{1.2}
    \centering
    \begin{tabular}{ c|c|c|c } 
         Coefficient & Description & Input Variables & Group \\ 
         \hline
         $C_L$ & Coefficient of lift & $\alpha, \beta$  & \multirow{3}{5em}{Force coefficients}\\ 
         $C_D$ & Coefficient of drag & $\alpha, \beta$  \\
         $C_{SF}$ & Coefficient of side force & $\alpha, \beta$  \\ \hline
         $C_l$ & Coefficient of rolling moment & $\alpha, \beta$  & \multirow{3}{5em}{Moment coefficients} \\
         $C_m$ & Coefficient of pitching moment & $\alpha, \beta$  \\
         $C_n$ & Coefficient of yawing moment & $\alpha, \beta$  \\ \hline
         $C_{m_q}$ & Coefficient of pitching moment due to pitch rate & $\alpha$  & \multirow{5}{5em}{Stability derivatives}\\
         $C_{l_p}$ & Coefficient of rolling moment due to roll rate & $\alpha$ \\
         $C_{l_r}$ & Coefficient of rolling moment due to yaw rate & $\alpha$ \\
         $C_{n_p}$ & Coefficient of yawing moment due to roll rate & $\alpha$ \\
         $C_{n_r}$ & Coefficient of yawing moment due to yaw rate & $\alpha$
         \\
    \end{tabular}
    \caption{List of aerodynamic coefficients that are used to make up the aerodynamic database}
    \label{tab:aero_db}
\end{table}

Similarly, Table \ref{tab:control_db} lists the coefficients used to create the controls database.
Except for the case with elevator deflections, all coefficients are three-dimensional functions of $\alpha$, $\beta$, and control surface deflection ($\delta_*$).
Since the flaps change the baseline aerodynamics of the aircraft, its effect on force coefficients is included in the controls database.
For other control surfaces, only their effect on moment coefficients is used. 

\begin{table}
    \renewcommand{\arraystretch}{1.2}
    \centering
    \begin{tabular}{ c|c|c } 
         Control Surface & Coefficients Used & Input Variables \\ 
         \hline
         Ailerons & $C_l, C_m, C_n$ with deflected ailerons & $\alpha, \beta, \delta_a$  \\
         Elevator & $C_l, C_m, C_n$ with deflected elevator & $\alpha, \delta_e$  \\
         Rudder & $C_l, C_m, C_n$ with deflected rudder & $\alpha, \beta, \delta_r$  \\
         Flaps & $C_L, C_D, C_{SF}, C_l, C_m, C_n$ with deflected flaps & $\alpha, \beta, \delta_f$  \\
         Spoilers & $C_l, C_m, C_n$ with deflected spoilers & $\alpha, \beta, \delta_s$  \\
    \end{tabular}
    \caption{List of controls surface coefficients that are used to make up the controls database}
    \label{tab:control_db}
\end{table}

It is important to note that actual control derivatives, such as $C_{l_{\delta_a}}$, are not calculated.
Calculating the derivative using finite differencing is susceptible to numerical issues.
Rather the change in the moment coefficient due to the deflected control surface is used. For example: 

\begin{equation}\label{equ:control_derivative}
    C_{l_{\delta_a}}(\alpha,\beta,\delta_a) = C_l(\alpha,\beta,\delta_a) - C_l(\alpha,\beta,\delta_a=0)
\end{equation}

The discrete data for each coefficient, and its associated uncertainty, is used to train a Gaussian process.
Individual multi-fidelity Gaussian processes model each coefficient.

\section{Multi-fidelity Data Generation} \label{sec:data_gen}

As with the NASA CRM example in Section \ref{sec:mf_gp_nasa_crm}, the Generic T-tail Transport (GTT) test case uses three data sources as well.
The lowest fidelity uses analyses from the Athena Vortex Lattice (AVL) code, the middle fidelity uses RANS CFD analyses using SU2, and the highest fidelity uses experimental data from wind tunnel campaigns \cite{cunningham_generic_2018,cunningham_preliminary_2018}.
The wind tunnel data was made available through our partnership with The Boeing Company and NASA. 
The lower fidelity data from AVL and SU2 had to be generated.
The following subsections detail the data generation process. 

Aerodynamic and controls data needs to be supplemented with geometry and engine data for the simulation process. 
The published mass and aerodynamic reference dimensions for the GTT aircraft are used \cite{cunningham_generic_2018}.
The engine data had to be taken from the engine specifications for the CRJ700 aircraft as there were no published values for the GTT aircraft.

\subsection{Lowest Fidelity: Vortex Lattice Simulations in AVL} \label{sec:data_gen_avl}
\begin{figure}
    \center
    \includegraphics[width=0.65\textwidth]{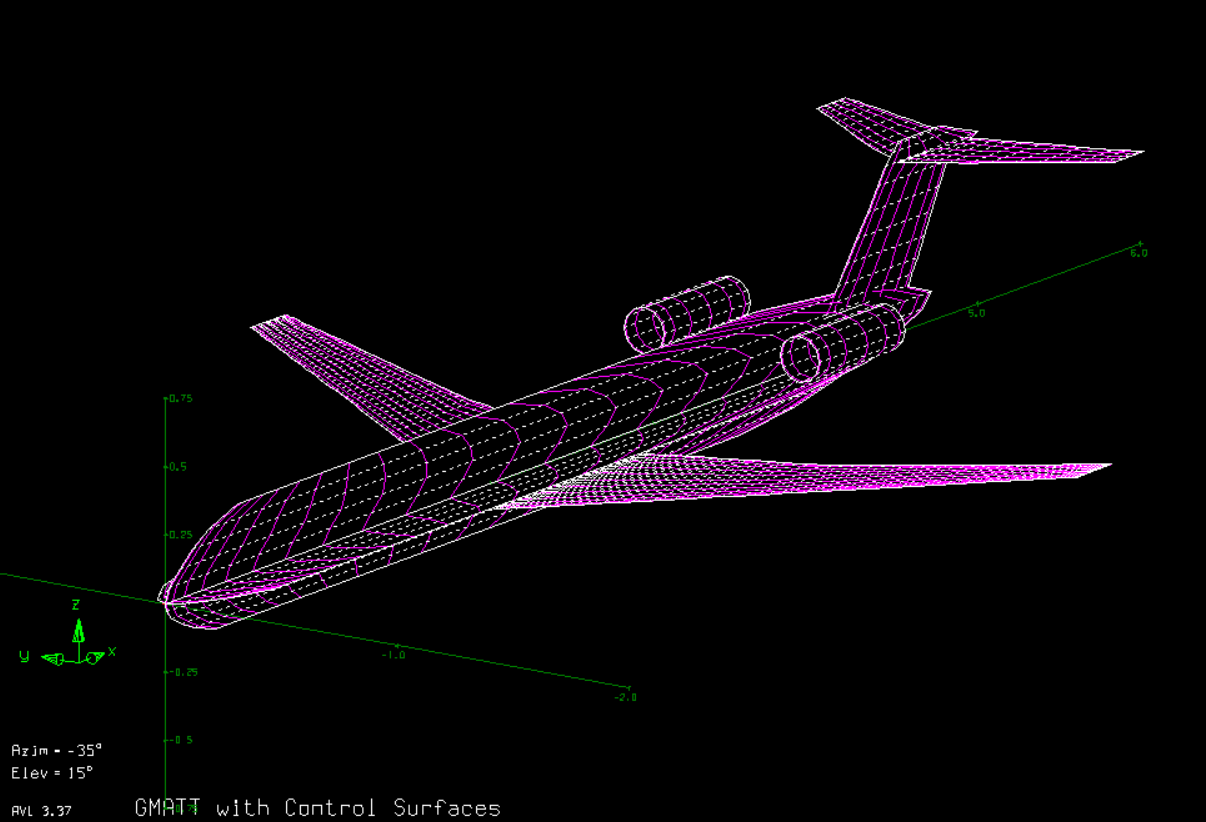}
    \caption{Representation of the GTT design in AVL \label{fig:gmatt_avl}}
\end{figure}

Creating the AVL model required extracting geometry information from Computer-Aided Design (CAD) files for the wind tunnel models. 
Details such as wing, nacelle, and tail locations, airfoil and fuselage cross-sections, and control surface locations were extracted from the provided geometry files. 
These are input into the AVL model to generate the model seen in Figure \ref{fig:gmatt_avl}. 

Performing simulations in AVL are relatively inexpensive, taking on the order of a few seconds per analysis. 
Quick analysis means that large databases can be generated quickly. 
For the baseline aerodynamics, an angle of attack sweep of $-4 ^\circ \leq \alpha \leq 25 ^\circ$ with an evaluation at each angle and an angle of sideslip sweep of $-20 ^\circ \leq \beta \leq 20 ^\circ$ with an evaluation every $4^\circ$, is used. 
This discretization results in 330 required evaluations to define the aerodynamics database.
The controls databases use an identical discretization in $\alpha$ and $\beta$. 
Table \ref{tab:avl_data_points} shows the control surface deflection sweeps and the total number of AVL evaluations required to generate the controls database.

\begin{table}
    \renewcommand{\arraystretch}{1.2}
    \centering
    \begin{tabular}{ c|c|c|c } 
         Control Surface & Deflection Sweep & Number of Deflection Angles & Number of AVL Evaluations \\ 
         \hline
         Ailerons &  $-25^\circ \leq \delta_a \leq 25^\circ$ & 11 & $330 \times 11 = 3630$\\
         Elevator &  $-30^\circ \leq \delta_e \leq 30^\circ$ & 13 & $330 \times 13 = 4290$\\
         Rudder & $0^\circ \leq \delta_r \leq 35^\circ$ & 11 & $330 \times 11 = 3630$\\
         Flaps & $0^\circ \leq \delta_f \leq 60^\circ$ & 10 & $330 \times 10 = 3300$\\
         Spoilers & $0^\circ \leq \delta_s \leq 60^\circ$ & 7 & $330 \times 7 = 2310$\\
         \hline
         & & \textbf{Total Evaluations} & 17160
         
    \end{tabular}
    \caption{List of control surface deflection sweeps and number of AVL evaluations require create the controls database at the lowest fidelity.}
    \label{tab:avl_data_points}
\end{table}

The uncertainty in the AVL analyses is specified using a standard deviation value for each data point. 
AVL uses an extended vortex lattice method to calculate aerodynamic characteristics of lifting surfaces and uses a slender-body model for nacelles and fuselages. 
These simulations solve the incompressible potential flow equations, which means they ignore the effects of compressibility and viscosity. 
They cannot model flow separation.
As a result, they are more accurate at lower angles of attack $(\alpha)$, angles of sideslip $(\beta)$, and control surface deflection angles $(\delta_*)$.
To represent these shortcomings in the uncertainty estimates, the standard deviation values are calculated as
\begin{equation}\label{equ:avl_sd}
    \sigma_{AVL}^{C_*(X)} = \max \left \{ 0.1C_*(X=0),~0.002 \left \Vert X \right \Vert \mathrm{range}(C_*) \right \},
\end{equation}
where $C_*$ refers to any coefficient and $X$ refers to the input variables.
Lets unpack Equation \ref{equ:avl_sd}.
The standard deviation is the maximum between two terms.
The first term establishes the minimum standard deviation for any coefficient in AVL and is $10\%$ of the coefficient value at $X =0$. 
Here $X=0$ means $\alpha=0, \beta=0$ and $\delta_* = 0$.
The second term takes into account the inaccuracy of AVL at higher $\alpha$, $\beta$, and $\delta_*$ by using the $L^2$ norm of the input variables and multiplying it with the range of the coefficient.
Using the range allows the equation to generalize to all coefficients, which can differ in absolute values.
The $0.002$ factor is a normalizing factor that yields reasonable uncertainty estimates for coefficients.

Since the maneuver of interest focuses on the roll-capability of the aircraft, let us look at the data that AVL generates for the rolling moment coefficient ($C_l$).

\subsection{Medium Fidelity: RANS CFD Simulations in SU2} \label{subsec:gtt_cfd_data_gen}

The medium-fidelity level uses RANS CFD simulations performed in SU2.
The eigenspace perturbation methodology provides the uncertainty information.
These simulations are computationally expensive, so the number of evaluations was limited to a coarse grid-sampling across the $\alpha$ and $\beta$ domain.
Simulations were run for $\alpha \in \left \{ -4^\circ,2^\circ,8^\circ,14^\circ,20^\circ \right \}$ and $\beta \in \left \{ 0^\circ,5^\circ,10^\circ,15^\circ,20^\circ \right \}$.
The results were suitably reflected to create the data for negative sideslip angles. 
Simulation conditions are outlined in \textcolor{red}{Table \ref{tab:gtt_test_cond}}. 
All simulations use the SST turbulence model.

\begin{table}
\centering
    \renewcommand{\arraystretch}{1.2}
    \captionsetup{justification=centering}
    \caption{Simulation conditions for the GTT aircraft.} 
    \begin{tabular}{|c|c|}
        \hline
        Mach Number & $0.1655$ \\ \hline
        Reynolds Number & $1.98\times10^6$ \\ \hline
        Reference chord length & $0.53986$ m \\ \hline
        Freestream Temperature & $295~\text{K}$ \\ \hline
    \end{tabular}
    \label{tab:gtt_test_cond}
\end{table}

Modeling the stability and control derivatives using CFD is challenging.
For stability derivatives, running unsteady simulations with forced oscillations is required \cite{mcmillin_computational_2019}. 
For control derivatives, modeling the control surface deflections increases the grid points required to resolve surface gaps and the resulting shear layers in the flow. 
Both requirements balloon the computational cost of running simulations to predict these derivatives. 
The high cost precludes the use of CFD data in the controls database and stability derivatives.
As a result, the CFD data and associated uncertainties are only used for the baseline force and moment coefficients $\left ( C_L, C_D, C_{SF}, C_l, C_m, \text{and }C_n\right )$

Pointwise is used to generate meshes from the GTT geometry. 
First, a family of three unstructured, half-span meshes is generated.
A grid convergence study using these meshes helps estimate the numerical discretization error \cite{american_society_of_mechanical_engineers_standard_2009}. 
Table \ref{tab:gtt_meshes} details of the mesh metrics.
The half-span mesh is mirrored along the $x-z$ plane to generate the full-span mesh used for all simulations involving non-zero $\beta$.
The surface mesh utilizes anisotropic stretching of grid elements along the chord of lifting surfaces.

\begin{table}
    \renewcommand{\arraystretch}{1.2}
    \centering
    \begin{tabular}{ c|c|c|c|c } 
         Mesh Level & Nodes & Surface Nodes & Wall spacing & Approx y+  \\ 
         \hline
         L3 & $2,281,668$ & $46,621$ & $5\times10^-6 m$ & 1.0 \\
         L2 & $6,722,768$ & $99,644$ & $3.3\times10^-6 m$ & 0.67 \\
         L1 & $19,946,697$ & $209,488$ & $2.3\times10^-6 m$ & 0.23 \\
         
    \end{tabular}
    \caption{Details of the mesh family used to perform numerical discretization error quantification for the GTT configuration.}
    \label{tab:gtt_meshes}
\end{table}

Section \ref{sec:num_vs_turb_error} covered the relationship between the numerical discretization error and the turbulence modeling uncertainties. 
The modeling uncertainty was shown to be independent of the numerical discretization error, given a minimum level of mesh refinement that can accurately capture the physics. 
This grid convergence study aims to determine the coarsest usable mesh that does not introduce significant numerical discretization error.
The black squares represent the baseline RANS CFD predictions, and the gray shaded region represents the estimated uncertainty that the turbulence model injects. 
On the left side of the plots, the finest mesh data point has the numerical discretization error bars associated with it. 
Since three meshes are used, only the most refined mesh has error metrics associated with it. 
The figure shows that the uncertainty estimated by the EQUiPS module is less than the errors introduced due to insufficient numerical discretization.
While the mesh refinement does not significantly impact the size of the turbulence uncertainty estimates, the L3 mesh is too coarse to capture $C_L$ and $C_m$ accurately. 
The L2 mesh strikes a good balance between computational cost and capturing the correct coefficients and turbulence uncertainty estimates.  
For this reason, the L2 mesh is used to generate the RANS CFD data. 


The eigenspace perturbation methodology provides the uncertainty information.
The methodology requires $6$ simulations ($1$ baseline $+~5$ perturbed) at each flight condition. 
With $\alpha \in \left \{ -4^\circ,2^\circ,8^\circ,14^\circ,20^\circ \right \}$ and $\beta \in \left \{ 0^\circ,5^\circ,10^\circ,15^\circ,20^\circ \right \}$, a total of $5\times5\times6 = 150$ RANS CFD simulations are run to create the medium-fidelity data and associated uncertainties. 


\subsection{Highest Fidelity: Wind/Water Tunnel Experiments}

\begin{figure}
    \centering
    \begin{subfigure}[NASA Langley Research Center 12-Foot Low-Speed Tunnel] {
        \includegraphics[trim=0 19 0 0, clip, width=.47\textwidth]{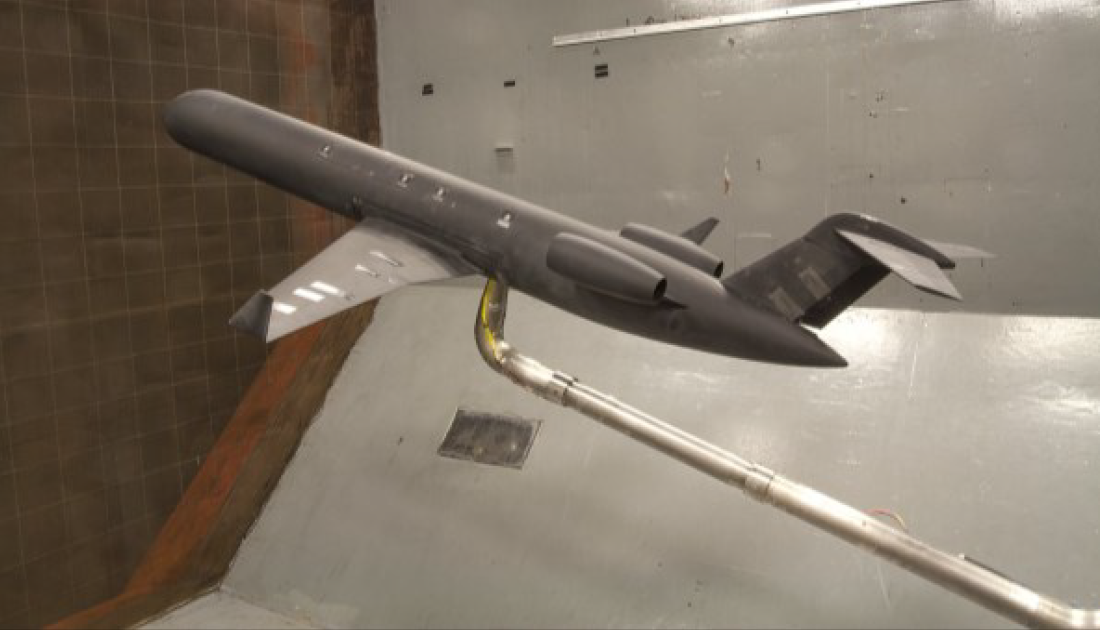} }
    \end{subfigure}
    \hfill
    \begin{subfigure}[Boeing North American Aviation Research Tunnel]{
        \includegraphics[trim=0 104 0 104, clip, width=.47\textwidth]{images/gtt_naart.png} 
    }
    \end{subfigure}
    \hfill
    \begin{subfigure}[Boeing Flow Visualization Water Tunnel]{
        \includegraphics[trim=0 0 0 0, clip, width=.5\textwidth]{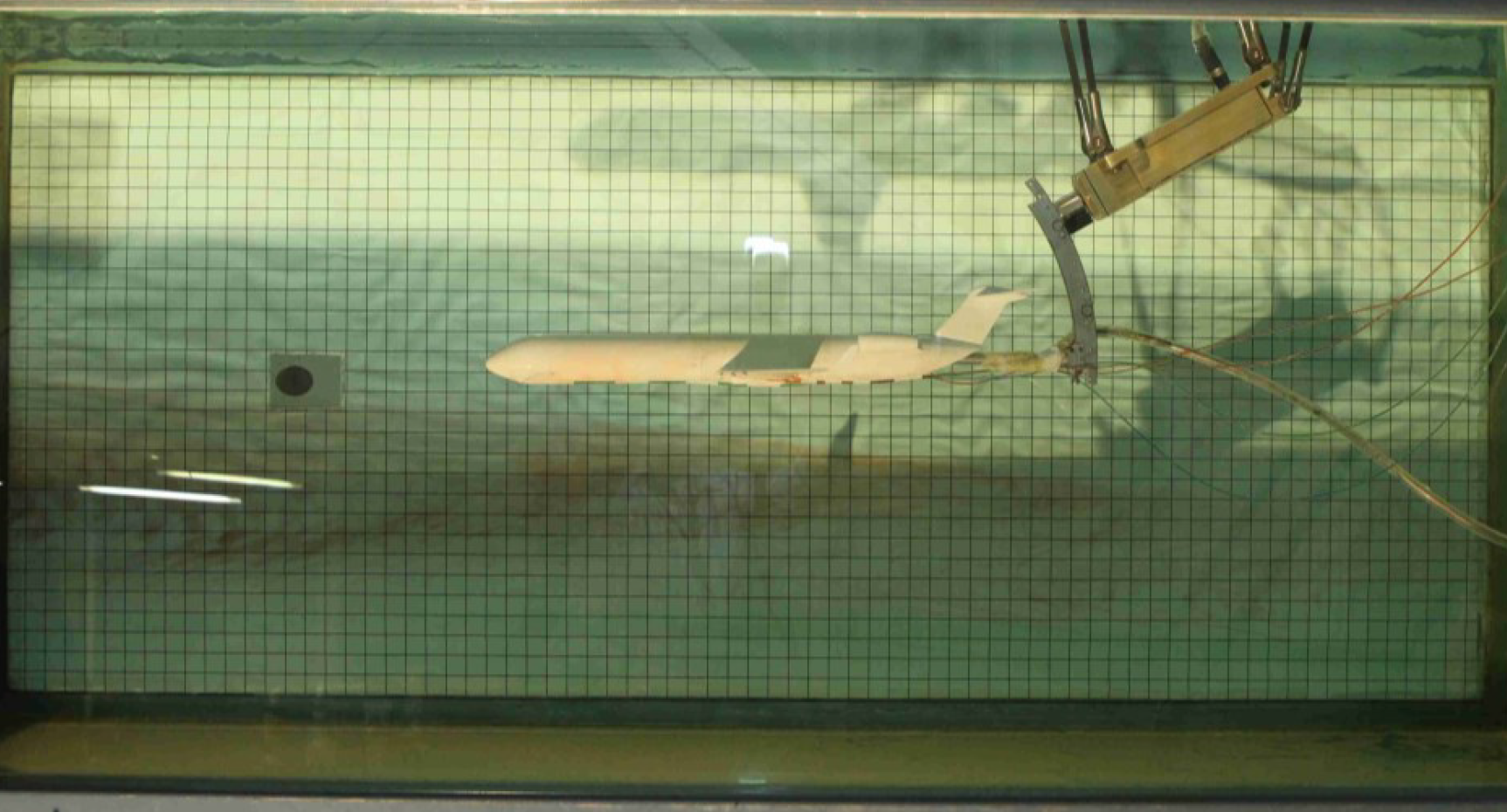} 
    }
    \end{subfigure}
    \caption{The GTT model mounted in the various experimental setups that were used.\label{fig:gtt_exp_images}}
\end{figure}

The primary motivation for using the GTT configuration for this work is the wealth of experimental wind tunnel data that is available \cite{cunningham_generic_2018,cunningham_preliminary_2018}. 
Three experimental campaigns in different facilities explored various aspects of the aircraft performance characteristics:
\begin{enumerate}
    \item NASA Langley Research Center 12-Foot Low-Speed Tunnel (12-Foot LST): Focus on stability derivatives and $\alpha$ sweeps,
    \item Boeing North American Aviation Research Tunnel (NAART): Focus on control derivatives and $\alpha$ and $\beta$ sweeps,
    \item Boeing Flow Visualization Water Tunnel (FVWT): Focus on stability derivatives at various values of $\alpha$.
\end{enumerate}
Figure \ref{fig:gtt_exp_images} shows the GTT model in each of the three experimental setups.
This work only uses data from the NAART and the FVWT experimental campaigns.
The data from the 12-Foot LST is a subset of the data from the other two sources.  

Ideally, raw sensor reading from the wind tunnel experiments could be post-processed to determine the uncertainty in the data.
There are well-established practices to calculate the systematic and precision uncertainty from experimental data \cite{coleman1995engineering}.
In the absence of the raw data, the uncertainty in the aerodynamic coefficients and stability derivatives is estimated as 
\begin{equation}\label{equ:wt_aero_sd}
    \sigma_{WT}^{C_*(X)} = \max \left \{ 1\times 10^{-4},~0.05~\mathrm{range}(C_*) \right \},
\end{equation}
whereas for control derivatives the uncertainty estimate calculation is broken up for each deflection angle
\begin{equation} \label{equ:wt_control_sd}
    \sigma_{WT}^{C_*(\alpha,\beta,\delta=\Delta)} = \max \left \{ 1\times 10^{-4},~0.05~\mathrm{range}(C_*(\alpha,\beta,\delta=\Delta)) \right \}.
\end{equation}

A few guiding principals are employed to reach Equations \ref{equ:wt_aero_sd} and \ref{equ:wt_control_sd}.
The first term in the $\max$ operation forces a minimum uncertainty of $1\times10^{-4}$.
This minimum uncertainty ensures that the covariance matrices in the Gaussian Process are positive-definite.
In contrast to uncertainties in AVL simulations(Equation \ref{equ:avl_sd}), no great variation in uncertainty is expected due to increased $\alpha, \beta$ or $\delta$ in experimental data. 
Consequently, the uncertainty is not dependent on the magnitude of the input variables, as is the case for uncertainties in AVL simulations. 
Using the range operator allows for a relatively consistent uncertainty estimation across all coefficients, which vary in absolute magnitude. 
For example, pitching moment coefficients are on the order of $10^-1$ while rolling and yawing moment coefficients are on the order of $10^-2$.
The $0.05$ factor means that the expected uncertainty can be estimated as $5\%$ of the range of values for the particular coefficient. 
Control surface deflections cause step changes in the control derivatives as calculated using Equation \ref{equ:control_derivative}.
This behavior necessitates the breakup of the uncertainty estimation for each control surface deflection angle. 
It prevents the higher control derivatives at higher deflection angles from influencing the uncertainty estimates at lower deflection angles.


\section{Databases for the Generic T-tail Transport Aircraft} \label{sec:gtt_dbs}

Thus far, the examples of probabilistic databases have focused on the baseline aerodynamics of the aircraft in question. 
This section will focus on controls databases that define the aircraft's response to a control surface deflection by providing the imparted rotational moments. 
As mentioned in Section \ref{subsec:gtt_cfd_data_gen}, the difficulty in modeling control surface deflections precludes the use of CFD simulation data in building these controls databases. 
Consequently, these databases can be, at most, two-fidelity ones with AVL simulations and wind tunnel experiments as the data sources. 
This section presents several visualizations of the GP regressions.
Single-fidelity GP results using AVL data and WT data in isolation are contrasted with the two-fidelity GP results that use both sets of data to showcase the advantages of using multi-fidelity data.

In the interest of brevity and since the maneuver of interest focuses on the roll-authority of the aircraft, these visualizations will present the roll moment imparted due to aileron deflection ($C_{l_{\delta_a}}$).
This variable is just one of the 18 coefficients (Table \ref{tab:control_db}) that define the aircraft's control behavior, but it is the most pivotal for the maneuver of interest. 

\subsection{Single-fidelity Databases}

With controls databases, a new input variable defining the control surface deflection angle ($\delta_*$) is used in addition to $\alpha$ and $\beta$ .
This additional variable makes visualizing the results of the 3-dimensional GP in 3D space challenging. 
Both sets of figures show four surfaces or lines in the first three subfigures. 
Each line or surface corresponds to the GP prediction at a particular aileron deflection angle.
For each of the figures, the lines/surfaces are stacked in order of increasing deflection angle, where the angles are $\delta_a \in \{-25^\circ, -10^\circ, 10^\circ, 25^\circ\}$. 
These deflection angles are chosen because the wind tunnel experiments were carried out with these aileron deflection angles.

Explicitly, the lowest surface in both figures corresponds to $\delta_a = -25^\circ$.
At the lowest level, the AVL data does capture the general trend of increasing $\delta_a$ leading to increased rolling moment, but there is a clear contrast in the linear trends learned from the AVL data and the non-linear trends represented in the wind tunnel data. 



There is a distinctly linear relationship between the rolling moment and the aileron deflection. 
While the AVL-based GP respects the uncertainty in the data across the domain, the WT-based GP predicts increased uncertainty between data points. 
In these areas, the uncertainty in the model parameters is surpassing the uncertainty in the data.
This trend indicates the need for additional data between the current set of aileron deflection angles for which wind tunnel data is available.
Using the multi-fidelity framework addresses this need.

\subsection{Multi-fidelity Databases}
The lower-fidelity AVL data can augment the limited wind tunnel data points to create a better surrogate model.
When using the abundant AVL data to supplement the sparse WT data in the aileron deflection dimension, the ballooning of the uncertainty estimates between data points disappears. 
The uncertainty estimates across the domain are reduced without using any more high-fidelity data.

This feature is a significant advantage of using multi-fidelity data, but there is a caveat.
The lower fidelity data needs to capture some relevant trends in the high-fidelity data. 
If the lower-fidelity trends contradict those seen in the high-fidelity data, then the low-fidelity data only corrupts the GP predictions. 
The multi-fidelity GP is trained on the difference between the low-fidelity data and the high-fidelity data. 
If there is no correlation between the low-fidelity and the high-fidelity data, then the difference between the two sets of data provides no useful information and only adds noise to the system. 
This issue, in turn, would prevent the multi-fidelity fit from performing as well or better than the single-fidelity fits. 

With the numerous multi-fidelity GP defining the aerodynamics and controls databases for the GTT aircraft created, flight simulations can be run using deterministic samples taken from these models.
These samples introduce minor variations in the aircraft's performance metrics that respect the GP error estimates resulting from the uncertainty in the underlying analysis data.

\chapter{Certification by Analysis} \label{chap:cba}
The increased reliance on simulation techniques for design analysis begs the question, can simulated analyses ever replace experimental ones?
Is it possible to complete parts of the flight certification process without first building a prototype of the aircraft and then putting it through flight testing?
This goal is of particular interest to the aerospace industry and is often referred to as Certification by Analysis (CbA).
To achieve this, improving simulation accuracy is paramount. 
Simulation capabilities would have to be as accurate, if not more accurate, than what is possible with flight testing.  
There are many required improvements to simulation capabilities \cite{slotnick_cfd_nodate} that will take years to develop.

In the interim, quantifying the uncertainties in the analysis techniques allows us to handle current shortcomings rigorously, visualize their effect on flight performance predictions, and understand their effect on predicted performance during flight testing.
While this will not be replacing real-world airworthiness testing any time soon, it provides aircraft designers a method to estimate the likelihood that a design will pass or fail a certification test. 

An aircraft in the design process has some performance characteristics associated with it.
These can be represented as probabilistic aerodynamics and controls databases using multi-fidelity GP regression introduced in Chapter \ref{chap:mf_gp}.
Stochastically sampling these databases creates multiple representations of the same aircraft that vary slightly in their performance.
The slight variations arise out of, and respect, the uncertainties in the analysis techniques used. 
Flight simulation software runs each of these aircraft samples through a certification maneuver of interest.
The results of the numerous simulations can be post-processed to calculate what percentage of the aircraft samples passed or failed the certification test.
This procedure provides a quantifiable metric for the probability that the aircraft design will succeed in the airworthiness test.

In addition to a probability of success/failure, by looking at the variation in the flight characteristics of the different samples, this framework provides data on how current uncertainties in performance predictions affect test results.
Based on these results, the design might need to be changed to ensure a better success rate, or higher-fidelity analysis methods with less uncertainty might need to be used to reduce the spread in the flight characteristics. 
These are direct benefits of the early handling of uncertainties in the design process to preempt possible issues in the future. 

This chapter presents the first steps in creating a rigorous methodology for CbA.
Contributions made in uncertainty quantification (Chapter \ref{chap:rans_uq}), multi-fidelity modeling (Chapter \ref{chap:mf_gp}), and probabilistic aerodynamics and controls databases (Chapter \ref{chap:aero_db}) are combined to create this framework.
Section \ref{sec:maneuver} presents a flight certification maneuver of interest that is taken directly from the official flight testing guide used by the Federal Aviation Administration (FAA) \cite{romanowski_flight_2018}.
Section \ref{sec:sim_procedure} detail the use of probabilistic aerodynamics and controls databases to simulate the maneuver of interest.
Finally, Section \ref{sec:cba_results} presents the results of performing the virtual flight testing and compares the use of different fidelity levels and amounts of data.

\section{Maneuver of Interest} \label{sec:maneuver}

A certification maneuver is needed to demonstrate CbA.
The Generic T-tail Transport, as the name suggests, is based on a commercial transport class aircraft.
It is most closely related to the Bombardier CRJ 700.
A few guiding principles were used in selecting the maneuver of interest:
\begin{itemize}
    \item Real-world certification requirement that the aircraft would be required to pass.
    \item Availability of data required to simulate the maneuver adequately.
    \item Preferably a lateral or directional maneuver that can leverage the multi-dimensional experimental data.
    \item Emphasis on control input to utilize control derivative data.
    \item A quantifiable metric to determine success or failure of the certification maneuver.
\end{itemize}

Based on these guidelines and with the help of industry experts at The Boeing Company, a maneuver was picked from the FAA's \textit{Flight Test Guide for Certification of Transport Category Airplanes} \cite{romanowski_flight_2018}.
Within Chapter \textit{5.3 Directional and Lateral Control} of the guide, the \textit{Lateral Control: Roll Capability \S 25.147(d)} maneuver was chosen.   
The testing procedure is paraphrased as: 
\begin{enumerate}
    \item Airplane starts in a trimmed state for steady straight flight at maximum takeoff speed.
    \item Establish a steady $30^\circ$ banked turn.
    \item Roll the airplane to a $30^\circ$ bank angle in the other direction.
    \item Aircraft must have sufficient roll authority to perform the $60^\circ$ change in bank angle in no more than $11$ seconds. 
    \item The aircraft must be able to do this with one engine inoperative, specifically the one that makes this maneuver more difficult.

\end{enumerate}

Two additional parameters mentioned in the guide relate to the maneuver being unchecked: the roll maneuver does not need to stop immediately after achieving the $30 ^\circ$ bank angle, and the aircraft should avoid excessive sideslip and bank angle during recovery.
Since these are not easily quantifiable parameters, they are not explicitly enforced in the flight simulation.
Regarding the airplane flap configuration, the guide specifies \textit{"Wing flaps in the most critical takeoff position."}
This statement is vague, and while some simulations deployed the flaps at $15^\circ$ per the CRJ700 Pilot's Handbook, most simulation results presented do not use flap deflections. 
It was infeasible to run RANS CFD simulations with flaps deployed due to the absence of an accurate geometry with deployed flaps and the significant increase in computational cost that it would merit.  
While engine data is not readily available and, consequently, some modeling was required, the maneuver satisfies all the guiding principles and provides an excellent platform to showcase the culmination of all the different parts of the work.

\begin{figure}
    \center
    \includegraphics[width=0.95\textwidth]{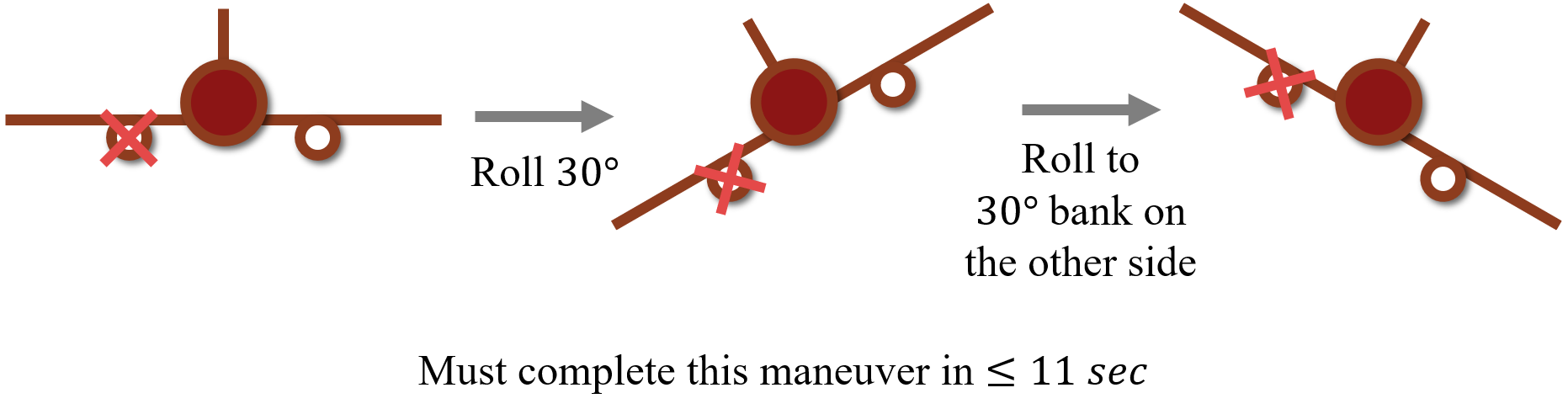}
    \caption{Graphical representation of Roll Capability certification maneuver. \label{fig:roll_maneuver}}
\end{figure}

Figure \ref{fig:roll_maneuver} shows a visual representation of the maneuver. 
For this image, the positive x-axis and the airplane's nose are pointing out of the page.
For an aircraft that rolls from $\phi = +30^\circ$ to $-30^\circ$, having an inoperative right engine is the more difficult maneuver.
This difficulty arises from the thrust of the left engine inducing a yaw moment and, consequently, a roll moment that resists the roll maneuver.
Accordingly, the right engine is labeled as inoperative with the "x" over it.

\section{Simulation Procedure} \label{sec:sim_procedure}

With the flight certification maneuver identified, a method to run a current aircraft design through the maneuver-of-interest needs developing.
Probabilistic, multi-fidelity aerodynamics and controls databases represent an aircraft's predicted performance characteristics.
The databases are used to create multiple deterministic samples of the aircraft.
The uncertainty in the data that informs the databases manifests itself in slight variations in the samples of the databases.
Each sample aerodynamic database has information about the forces and moments on an aircraft at various points in the flight envelope.
Each sample controls database contains information about the moments induced on the aircraft due to various control surface deflections. 
These samples are run through flight simulation software that can integrate the force and moment information, combine it with the effects of control inputs, and perform a time-accurate maneuver defined in the simulation software. 

This part of the work leans heavily on The Boeing Company's expertise in flight simulation and control law mixing.
Due to proprietary and patent restrictions, the exact implementation of the flight simulation code is unavailable, but enough information is provided to outline the simulations' overarching methodology and workflow.

Instead of a complete $6$ Degree of Freedom (DoF) flight simulation, a slightly simplified $5$ DoF flight simulation tool, ignoring displacements in the y-axis (body-fitted coordinate system), is used.
While this is slightly less accurate than a $6$ DoF simulation, it is well suited for Monte Carlo analysis, requiring the rapid analysis of hundreds of aircraft databases.
Another simplification is that the maneuver is not performed in a closed-loop, trajectory-following manner. 
Instead, the maneuver is first processed into a set of rotational accelerations required to complete the maneuver, and then the control inputs needed to provide those accelerations are calculated. 
If the aircraft can perform the maneuver without over-saturating any of the control surface deflections, the maneuver is considered a success.

For this section, the results use the single-fidelity aerodynamics and controls databases created using experimental data from the NAART and FVWT experimental campaigns (Section \ref{sec:gtt_dbs}). 
Multi-fidelity databases are used and compared for the results in Section \ref{sec:cba_results}.

\subsection{Maneuver Simulation}

The first step is to convert the maneuver of interest, in this case, the \textit{Lateral Control: Roll Capability \S 25.147(d)} maneuver, into an appropriate trajectory for the aircraft to follow. 
The trajectory is defined as a time history of the aircraft's orientation required to meet the parameters of the maneuver. 
The roll angle of the aircraft primarily defines this particular maneuver.
There are other parameters included in the trajectory definition, such as the pitch angle required to maintain a constant altitude, but the roll angle defining the trajectory of the maneuver is the focus here.
The aircraft starts with steady level flight, rolls to an angle $\psi = +30^\circ$, and then completes the roll maneuver from $\psi = +30^\circ$ to $=-30^\circ$ in 11 seconds, as required by the certification maneuver. 

The second step of the simulation process is to use the aircraft's geometric properties and the aerodynamics database to calculate the accelerations required to follow the trajectory.
This step involves time-stepping through the trajectory definition and calculating the directional and rotational accelerations that would put the aircraft in the appropriate orientation at the next time step. 
There are a few distinct sections of the roll acceleration plot. 
For the first second, to establish steady trimmed flight, the aircraft stays level, and the roll acceleration stays at zero.
A sizeable positive acceleration is required to start the roll to a bank angle of $+30^\circ$. 
It tapers off to nearly zero at $13$ seconds, after which a moderate negative roll acceleration stabilizes the aircraft at that bank angle. 
The heart of the certification maneuver starts at $16$ seconds, as indicated by the negative peak for the roll acceleration.
The required acceleration reduces as the bank angle approaches zero and then peaks again once it is past $\psi=0^\circ$.
Finally, around the $27$ second mark, a sharp positive roll acceleration is required to stabilize the aircraft at the $-30^\circ$ bank angle. 
The acceleration definitions from this step feed into the next one.

The final step in the simulation procedure involves using the controls database of the aircraft, and The Boeing Company's patented control law mixer \cite{control_law_patent} to compute the control surface deflections needed to provide the accelerations that the maneuver demands.
This process requires time-stepping through the acceleration definitions from the previous simulation step and calculating the control inputs required to meet the acceleration demands at that time step. 
The primarily linear relationship between aileron deflection and the resulting roll acceleration leads to the control surface deflections mimicking the trends seen in the roll acceleration plot.

Other parameters in the trajectory definition, such as minimizing sideslip or maintaining altitude, necessitate yaw and pitch accelerations.
Again, the linear relationship between the rotational accelerations and the corresponding control surface deflections results in the acceleration trends being closely matched by the surface deflection trends. 
As long as the surface deflections commanded by the flight simulator do not exceed the physical limit of the available control surface deflection, the maneuver can be completed and considered a success.

\subsection{Engine Failure Simulations}
As stated in Section \ref{sec:maneuver}, the certification maneuver must be completed with one engine inoperative.
Specifically, the inoperative engine should be the one that makes the maneuver more difficult.
In the coordinate system defined by the $x$-axis pointing out of the nose and the $z$-axis pointing down towards the ground, the roll from $\psi = +30^\circ$ to $\psi = -30^\circ$ is made more difficult with the right engine inoperative. 
The asymmetric thrust causes the aircraft to yaw towards the inoperative engine, creating a rolling moment in the direction opposite to the roll maneuver.

The figure plots three configurations for the time history of the rotational accelerations (left column) and the control surface deflections (right column).
The blue line represents the nominal configuration with both engines operating. 
The red line represents the case with an inoperative right engine (Right Engine Out).
The yellow line represents the case with an inoperative left engine (Left Engine Out).

The configuration that makes the maneuver more difficult would be the one that requires higher control surface deflection angles. 
These figures show that the right engine out case (red line) requires higher aileron deflections, similar rudder deflections, and slightly higher elevator deflections. 
This observation confirms that having an inoperative engine on the right side of the aircraft makes the maneuver more difficult.

\subsection{Evaluating Success or Failure} \label{subsec:success_failure}

The metric for the successful completion of the certification maneuver is defined in the FAA's guide \cite{romanowski_flight_2018} as being able to perform the $60^\circ$ change in bank angle under 11 seconds. 
Directly enforcing this criterion would require a closed-loop, trajectory-following flight simulation. 
In such a simulation, the aircraft continuously calculates and adjusts the control input required to complete the maneuver without exceeding the physical limits on the control systems. 
In the flight simulation procedure used, the time taken to complete the maneuver is predefined.
There is minimal variation in actual time taken for the aircraft to complete the maneuver because the control inputs are calculated to complete the maneuver in the predefined time.

Reframing the metric for success works around this challenge. 
Since the time to complete the maneuver is predefined, the metric of success becomes the level of saturation of the control surfaces required to complete the maneuver. 
Specifically, three success metrics are defined, one for each rotational acceleration: pitch, roll, yaw. 
These are calculated for a simulated maneuver as
\begin{align}
    \rho_{pitch} = 1- \max\left \{ \frac{\left \vert \delta_e^{req}(t) \right \vert }{\delta_e^{lim}} \right \}
    \\
    \rho_{roll} = 1- \max\left \{ \frac{\left \vert \delta_a^{req}(t) \right \vert }{\delta_a^{lim}} \right \} \label{equ:roll_metric}
    \\
    \rho_{yaw} = 1- \max\left \{ \frac{\left \vert \delta_r^{req}(t) \right \vert }{\delta_r^{lim}} \right \}
\end{align}
where $\delta_e$, $\delta_a$, and $\delta_r$ correspond to elevator, aileron, and rudder deflection angles, respectively.
The superscript $req$ indicates the value that is calculated by the simulator, and the superscript $lim$ indicates the maximum allowable control surface deflection angle. 
The maximum saturation over the entire duration of the flight simulation is taken and subtracted from $1$.
If at any point during the flight simulation a control surface is completely saturated, say $\delta_e^{req} = \delta_e^{lim}$, the corresponding metric is $0$, in this case $\rho_{pitch} = 0$.
Consequently, a flight maneuver is considered a success if all three success metrics are $>0$.
Table \ref{tab:gtt_defl_limits} lists the deflection limits for each of the control surfaces.

\begin{table}
\centering
    \renewcommand{\arraystretch}{1.2}
    \captionsetup{justification=centering}
    \caption{Control surface deflection limits on the GTT aircraft.} 
    \begin{tabular}{|c|c|}
    \hline
        Control Surface & Deflection limits \\ \hline
        Ailerons & $\pm 25^\circ$ \\ \hline
        Elevator & $\pm 20^\circ$ \\ \hline
        Rudder & $\pm 30^\circ$ \\ \hline
        Spoilers & $\pm 60^\circ$ \\ \hline
        Flaps & $\pm 45^\circ$ \\ \hline
    \end{tabular}
    \label{tab:gtt_defl_limits}
\end{table}

As a concrete example, the maneuver definition is tweaked to force the over-saturation of the ailerons.
The maneuver is made more aggressive by shortening the time taken to perform the $60^\circ$ change in bank angle from $11 sec$ to $2.5 sec$.
Over-saturation of the aileron deflection during the aggressive maneuver is visible, with the blue line extending past the saturation limit. 
For this case, the absolute maximum deflection angle for the right aileron is $34^\circ$ which exceeds the saturation limit by $9^\circ$
The resulting roll metric $\rho_{roll} = -0.36$


It is important to note that aircraft analysis data is only available within the saturation limits of the control surfaces. 
When the simulator commands a control surface deflection that is greater than the limit, it estimates the effect of that deflection by extrapolating from the provided data. 
While this extrapolation might be prone to error, it does provide an approximation for how much additional control authority would be required to complete the maneuver as planned. 

This feature highlights an advantage of using this simulation method and the control-surface-based metrics in the design process. 
When running a potential aircraft design through this virtual flight testing framework, the metrics provide a direct estimation of the amount of additional control authority that would be required to complete the maneuver successfully. 
If a simulation yields $\rho_{roll} = -0.25$, that means that the aircraft would require approximately $25\%$ more aileron control authority. 
This increment could be achieved by increasing the control surface deflection limits or increasing the ailerons' size. 
Additionally, by breaking up the calculations of the accelerations and the control surface deflections, different control laws can be tested without repeating the first step of the simulation. 
Perhaps instead of increasing the size of the aileron, more significant spoiler deflections can be mixed into the control law, which would increase the control authority in the roll direction. 

\subsection{Monte Carlo Analysis} \label{subsec:mc_analysis}

The GP models provide a probabilistic representation of the aerodynamics and controls databases. 
Random sampling from these models creates instances of the databases that have slight variations depending on the uncertainties present in the GP model.
The sampling methodology is explained in Section \ref{sec:gpr} using Equation \ref{equ:gp_sampling}.
Each sample can be run through the flight simulator independently.


Using the Monte Carlo method allows us to rigorously understand the effect of the uncertainties in the databases on the certification of the aircraft.
It is a brute-force method of characterizing the effect of input uncertainties on an output quantity of interest (QoI). 
Independently and randomly sampled inputs are run through the function in question to study the distribution of the output QoI. 
In the context of this work, the input uncertainty is the uncertainty in the analysis data used to create the GP models. 
The function in question is the simulation of the flight certification maneuver.
The output QoI is the success/failure of the aircraft in performing a flight certification maneuver, quantified by the metrics introduced in Section \ref{subsec:success_failure}.
Running flight simulations on hundreds of randomly sampled databases results in hundreds of success metrics that can be treated as random variables.
Applying statistical methods to the results characterizes the distribution of the success metrics. 

For the statistical analysis of the success metrics, cumulative distribution functions (CDF) are calculated.
The CDF for a random variable $X$ evaluated at $x$ represents the probability that $X$ will take a value less than or equal to $x$.
For example, $1000$ samples of the GTT aerodynamics and controls databases perform the flight certification maneuver with engine failures. 

None of the metrics for any of the simulations go below $0$. 
Even when accounting for uncertainties in the analyses, the current representation of the aircraft would pass the certification maneuver $100\%$ of the time.
This result is expected since the GTT aircraft is modeled on the CRJ 700, a currently certified aircraft.
These plots reiterate that the REO case is the most restrictive for the certification. 
All remaining results will only use the REO case for the simulations. 

With Monte Carlo simulations, enough samples must be used to ensure that the posterior distributions are converged.
The convergence of the distribution is tracked using the variance of the metrics.
While the REO case takes longer to converge, the variance is unchanging when using $500-1000$ samples. 
For this reason, all remaining results will use $1000$ Monte Carlo samples to represent the CDF for the simulation results. 


\subsection{Modifications to the Simulation} \label{subsec:sim_mods}

This result is accurate even when accounting for the uncertainties in the analyses that define the GP model. 
It makes the discussion of the success/failure of a certification maneuver difficult if the aircraft succeeds every time.
We modify the simulation parameters to force some failures in the maneuvers.
The maximum deflection limits on the control surfaces are changed to make the maneuver more challenging.
Table \ref{tab:gtt_defl_limits_new} lists these changes.

The aileron and rudder limits are made more restrictive. 
These changes directly increase the control surface saturation, thereby moving the roll and yaw metrics closer towards $0$.
Spoiler deflections aid in the roll authority and are ignored moving forward.
Since flaps affect the baseline aerodynamics and lower fidelity methods cannot model them well, they are ignored moving forward. 
This change impacts the pitch metric indirectly. 
Ignoring flap deployment reduces the $C_L$ of the aircraft at all angles of attack, consequently forcing the aircraft to trim at a higher angle of attack to maintain steady level flight. 
It increases dependence on the elevator during the maneuver and shifts the pitch metric closer towards $0$.

\begin{table}
\centering
    \renewcommand{\arraystretch}{1.2}
    \captionsetup{justification=centering}
    \caption{New control surface deflection limits on the GTT aircraft.} 
    \begin{tabular}{|c|c|c|}
    \hline
        Control surface & Old deflection limits & New deflection limits \\ \hline
        Ailerons & $\pm 25^\circ$ & $\pm 15^\circ$ \\ \hline
        Elevator & $\pm 20^\circ$ & $\pm 20^\circ$ \\ \hline
        Rudder & $\pm 30^\circ$ & $\pm 20^\circ$   \\ \hline
        Spoilers & $\pm 60^\circ$ & Not Used       \\ \hline
        Flaps & $\pm 45^\circ$ & Not used          \\ \hline
    \end{tabular}
    \label{tab:gtt_defl_limits_new}
\end{table}

\section{Results} \label{sec:cba_results}
This section discusses running the GTT aircraft through the modified flight certification maneuver.
There are two overarching goals of this work: first, to provide a framework that brings flight testing earlier in the aircraft design process, and second, to create the most accurate flight simulation results while minimizing the cost of the underlying analyses. 
To this end, this section first focuses on comparing simulations run using purely high-fidelity experimental data to those run using low-fidelity and multi-fidelity data that would be available early in the design process.
Then the cost-reduction aspect is explored by harnessing the benefits of multi-fidelity databases that use small subsets of the high-fidelity data. 
For all the results in this section, simulations are conducted with the right engine inoperative, and all modifications discussed in Section \ref{subsec:sim_mods} are applied.

\subsection{Simulations using Single- vs. Multi-Fidelity Databases} \label{subsec:sf_vs_mf_cba}

In the nascent stages of aircraft design, engineers use low-fidelity methods, such as AVL simulations, to perform rapid analyses that can span vast regions of the design domain quickly. 
These have higher uncertainties in their performance predictions due to simplifications made in modeling the flow physics in favor of speed of analysis.
As parts of the aircraft design get fixed, engineers employ higher-fidelity methods such as RANS CFD simulations to perform design optimizations.
These are computationally more expensive than AVL simulations and provide a more accurate representation of the aircraft's potential performance. 
Once designs get finalized, manufactured prototypes are experimentally tested in wind tunnels. 
These provide high-fidelity data that can recreate the expected flight conditions and have correspondingly low uncertainties associated with their performance predictions.
After finalizing all design decisions and manufacturing a full-scale prototype, real-world flight testing certifies the aircraft's airworthiness.

Mimicking this sequence of design analyses, single- and multi-fidelity databases that use, in order, low-fidelity (AVL), low- + medium-fidelity (AVL + SU2), and low- + medium- + high-fidelity (AVL + SU2 + WT) data are created. 
Section \ref{sec:data_gen} presented an example of this build-up.
Flight simulations using these lower- and multi-fidelity databases bring aspects of the real-world flight testing procedure into the virtual environment where engineers perform the early design analyses.
To assess the quality of the results, they are compared to flight simulations carried out using high-fidelity databases created using only experimentally obtained wind tunnel data. 
It is important to note that while wind tunnel data has its own set of models that introduce errors compared to flight tests, it is the highest-fidelity data available and is considered the truth function being emulated. 

It is important to remember that limited aerodynamic data is available from SU2. 
This limitation means that the two-fidelity results only use SU2 data for the force and moment coefficients in the aerodynamics database.
The controls databases for the two-fidelity results are only single-fidelity and are identical to the controls databases used for the AVL-only results. 
Consequently, the three-fidelity database is only three-fidelity for the force and moment coefficients; the rest are two-fidelity.
The overall results are still referred to as two-fidelity when using AVL+SU2 data and three-fidelity when using AVL+SU2+WT data to avoid verbose explanations.


With the details of the simulation procedure explained in Section \ref{sec:sim_procedure}, the results here will be presented primarily through the CDF plots of the success metrics resulting from the Monte Carlo analysis defined in \ref{subsec:mc_analysis}.

Early in the conceptual design stages, the aircraft performance would be determined using low-fidelity analyses, and the flight simulations would yield the results represented by the blue line. 
The significant uncertainties attributed to the AVL analyses yield a large variance in the metrics, indicated by the wide range of values that the CDF spans for every metric.
AVL over-predicts the effectiveness of all control surfaces with the CDF for each metric lying to the right of the other results, indicating lower control surface saturation during the simulations. 

Progressing with the aircraft design, we add RANS CFD data to the GP models.
The red line represents the results of flight simulations using AVL+SU2 data. 
Since no additional data improves the controls databases, the simulations preserve the significant variance in success metrics seen in the AVL results. 
Nevertheless, the additional accuracy in the baseline aerodynamics shifts all of the two-fidelity results closer towards the three-fidelity and high-fidelity results.

With the addition of the wind tunnel data to the multi-fidelity fit (yellow line), the flight simulation results should be very close, if not identical, to those using the single-, high-fidelity data (black line). 
The trim angles of attack, pitch metrics, and yaw metrics line up identically. 
However, there is a notable difference in the roll metric results. 
Not only is the mean value for the three-fidelity database (dashed yellow line) lower than that for the single-, high-fidelity database, but the variance in the roll metric is also lower. 
The reason for this difference was hinted at in Section \ref{sec:gtt_dbs} when comparing the single- and multi-fidelity databases for $C_{l{\delta_a}}$. 


The database samples, $10$ of which are shown by the individual colored lines,  respect the error estimate from the GP and have correspondingly higher variability between those data points.
When supplemented with lower-fidelity AVL data (blue circles), the multi-fidelity GP can use the trends learned from the lower fidelity to reduce the uncertainty between the same set of wind tunnel data points. 
As a result, the error estimate from the GP is lower and is barely discernible at the scale of the figure. 
The samples show significantly less variation from the mean database and are almost all coincident in this case. 

This observation brings to question the assumption of using the single-, high-fidelity database results as the truth function being emulated. 
The increased error estimates between data points do not indicate a failure of the GP model in learning the correct trends; instead, it suggests that data sampling is sparse and reducing the uncertainty in the model requires more data.
The advantage of using multi-fidelity data is clear. 
The uncertainty in the model can be reduced without any new, expensive, high-fidelity function evaluations.

\begin{table}
\centering
    \renewcommand{\arraystretch}{1.2}
    \captionsetup{justification=centering}
    \caption{Mean, variance, and failure rate for the flight certification maneuver when using different aircraft databases.} 
    \begin{tabular}{|c|c|c|c|c|c|c|c|c|c|}
    \hline
        Case & \multicolumn{3}{c|}{Pitch Metric} & \multicolumn{3}{c|}{Roll Metric} & \multicolumn{3}{c|}{Yaw Metric} \\ \hline
         & $\mu$ & $\sigma^2$ & Failure & $\mu$ & $\sigma^2$ & Failure & $\mu$ & $\sigma^2$ & Failure \\ \hline
        AVL & 0.56 & 8.21e-04 & 0.0\% & 0.23 & 1.83e-02 & 6.0\% & 0.36 & 1.70e-02 & 1.6\% \\ \hline
        AVL+SU2 & 0.23 & 4.59e-03 & 0.0\% & 0.16 & 2.45e-02 & 15.1\% & 0.04 & 2.87e-02 & 37.1\% \\ \hline
        AVL+SU2+WT & 0.35 & 1.76e-03 & 0.0\% & 0.01 & 3.04e-03 & 39.2\% & 0.08 & 4.73e-04 & 0.0\% \\ \hline
        WT & 0.36 & 2.87e-04 & 0.0\% & 0.05 & 1.13e-02 & 30.8\% & 0.08 & 4.74e-04 & 0.0\% \\ \hline
    \end{tabular}
    \label{tab:sf_vs_mf_perf_stats}
\end{table}

$\mu$ and $\sigma^2$ represent the sample mean and variance of the metrics.
The trends noticed visually in the CDF plots are confirmed by the numbers in the table. 
Looking at the simulation results from the perspective of flight certification, the main QoI is the failure rate for the air-worthiness maneuver. 
The failure rate is defined as the percentage of cases for which the metric $\rho_* < 0$.
The modifications in the maneuver simulations listed in Section \ref{subsec:sim_mods} result in higher failure rates compared to those shown in Section \ref{sec:sim_procedure}.
None of the databases have any failures in the pitch metric. 
AVL overestimates the control effectiveness of all the control surfaces as mentioned earlier.
This factor underestimates the roll metric failure rate when using only AVL data for the controls databases. 
The overestimated control effectiveness is also seen in the yaw metric, but the high uncertainty in the databases leads to more failures. 

These trends urge caution when relying on purely low-fidelity analyses to make control-sizing decisions early in the design process.
The overestimation of control-effectiveness by these low-fidelity methods needs addressing to correlate with high-fidelity simulations better. 
It does not mean that performing early flight simulations are not helpful, just that necessary corrections would need to be derived and applied to make meaningful inferences from the data. 
Perhaps constraints on acceptable performance at early design stages would need to be different from those used for later stages. 
For example, acceptable performance in the conceptual design stage could require successful flight simulations with half the full range of motion for the control surfaces planned for the actual design.
This suggestion is already employed here as the simulation modifications made in Section \ref{subsec:sim_mods} reduced the allowable range of motion for the ailerons and the rudder. 

Aircraft manufacturers often use such rules-of-thumb and factors of safety.
Historical data using previously certified aircraft and all the analyses performed during their design process can be used to create these corrections. 
Since most of the required data is proprietary, deriving these corrections is outside of the scope of this work. 

\subsection{Simulations using Reduced Data}

A significant advantage of using multi-fidelity data to create surrogate models is that the underlying function in question can be approximated better with fewer high-fidelity function evaluations.
This has been published on extensively, \cite{kennedy_predicting_2000,gratiet_multi-fidelity_nodate,perdikaris_multi-fidelity_2015, ghoreishi_gaussian_2018}, to name a few. 
Earlier in this work, Section \ref{sec:mf_modeling} shows this property for analytic functions and Section \ref{sec:mf_gp_nasa_crm} shows it for single- and multi-dimensional aerodynamic databases.
With the NASA CRM databases, the advantage is evident in both cases, when high-fidelity evaluations are uniformly spaced and when localized to areas where the low-fidelity data is inadequate. 
This section explores the effect of the flight simulation procedure on this advantage.

To explore this, we create three new sets of databases for the GTT:
\begin{enumerate}
    \item WT-Coarse: Single-fidelity GP only using a sparse subset of the available wind tunnel data.
    \item 3F-Coarse: Multi-fidelity GP using all the AVL and SU2 data, but uses a sparse subset of the wind tunnel data (same subset at WT-Coarse).
    \item 3F-Local: Multi-fidelity GP  using all the AVL and SU2 data, and supplementing it with a subset of the wind tunnel data localized to high values of $\alpha, \beta,$ and $\delta_*$.
\end{enumerate}

The sparse subset of the wind tunnel data uses data for every $4^\circ$ change in angle of attack.
This coarsening equates to using $\approx20\%$ of all the experimental data available for the aerodynamics database, and $\approx40\%$ of the experimental data available for the controls database. 
This level of data reduction would equate to significant cost savings in experimental analyses while allowing an even spread of data over the domain of interest. 
The WT-Coarse and the 3F-Coarse databases use the same sparse subset of wind tunnel data.

With the 3F-Local database, the idea is to localize the high-fidelity evaluations to certain domain regions.
It simulates a situation where if high-quality low-fidelity data is available in certain parts of the domain, high-fidelity evaluations in those areas would be an unnecessary expense. 
Instead, engineers can focus the limited resources on areas where the low-fidelity data is inadequate or has very high uncertainty associated with it. 
Both AVL and RANS CFD simulations are inaccurate when there is significant flow separation over the aircraft. 
This occurs at high values of $\alpha, \beta,$ and $\delta_*$.
To this end, for the 3F-Local database, high-fidelity data is limited to areas of the domain where $\alpha > 10^\circ, \beta > 10^\circ,$ and $\delta_* > 15^\circ$.

Statistics of the metric distrubutions are shown in Table \ref{tab:red_data_perf_stats}.
The different databases are compared to the same single-, high-fidelity database used in the previous section.

An overarching trend across the four plots is that the variance in the metrics increases when using a subset of the high-fidelity data. 
Having fewer data points increases the uncertainty in the GP predictions.
It leads to more significant variability in the database samples and, consequently, greater variance in the output metrics from the flight simulations. 


\begin{table}
\centering
    \renewcommand{\arraystretch}{1.2}
    \captionsetup{justification=centering}
    \caption{Mean, variance, and failure rate for the flight certification maneuver when using different subsets of high-fidelity data.} 
    \begin{tabular}{|c|c|c|c|c|c|c|c|c|c|}
    \hline
        Case & \multicolumn{3}{c|}{Pitch Metric} & \multicolumn{3}{c|}{Roll Metric} & \multicolumn{3}{c|}{Yaw Metric} \\ \hline
         & $\mu$ & $\sigma^2$ & Failure & $\mu$ & $\sigma^2$ & Failure & $\mu$ & $\sigma^2$ & Failure \\ \hline
        WT & 0.36 & 2.87e-04 & 0.0\% & 0.05 & 1.13e-02 & 30.8\% & 0.08 & 4.74e-04 & 0.0\% \\ \hline
        WT-Coarse & 0.34 & 4.60e-04 & 0.0\% & -0.03 & 7.87e-03 & 55.2\% & 0.06 & 1.54e-03 & 8.0\% \\ \hline
        3F-Coarse & 0.36 & 4.97e-04 & 0.0\% & -0.04 & 7.68e-03 & 65.3\% & 0.07 & 1.41e-03 & 4.2\% \\ \hline
        3F-Local & 0.38 & 5.81e-04 & 0.0\% & -0.15 & 7.38e-03 & 96.4\% & 0.06 & 2.04e-03 & 9.0\% \\ \hline
        
    \end{tabular}
    \label{tab:red_data_perf_stats}
\end{table}

Direct comparison between the WT-Coarse (blue) and the 3F-Coarse data (red) provides an insight into the usefulness of the lower-fidelity data in improving the overall GP mean and uncertainty estimates. 
The 3F-Coarse database results for the pitch metric are nearly identical to those from the WT database that uses all the wind tunnel data. 

This observation yields that the multi-fidelity fit can perform as well as, if not better than, the single-fidelity fit when there is only sparse high-fidelity data available. 
It is worth reiterating that while we compare results to the single-fidelity WT database, it is not necessarily the best approximation of the actual flight performance.
As seen in the previous section, the uncertainties in the GP model for the WT case can corrupt the expected performance metrics. 
However, in the absence of higher fidelity data than wind tunnel experiments, the wind tunnel data is the primary benchmark. 

The 3F-Local case presents a notable deficiency in the results. 
The addition of localized wind tunnel data improves overall predictions compared to the two-fidelity case shown in the previous section, but it inferior when compared to the other three-fidelity results. 
It cannot get close to the mean metric estimates, nor does it provide a significant reduction in the variance of the performance metrics. 
This deficiency is due to the difference in the low-fidelity and high-fidelity data in areas the high-fidelity data is ignored. 
In these areas, the low-fidelity approximations should capture some of the relevant physics and have correspondingly low uncertainties. 
In reality, RANS CFD simulations only provide some high-quality aerodynamic force and moment coefficient data, while all of the controls databases are built on purely AVL data. 
The high uncertainty and the inadequacy of the AVL data require more than just localized high-fidelity evaluations to improve predictions. 

The data reduction techniques used here are rudimentary. 
Adaptive sampling methodologies leverage the error estimates from the GP models to introduce high-fidelity data at specific locations to minimize the overall uncertainty in the model. 
Simple techniques, such as using high-fidelity data in areas with the highest uncertainty as predicted by the GP, can significantly improve flight performance predictions \cite{wendorff_combining_2016}. 
Since there is a predetermined, limited set of wind tunnel data, adaptive sampling is not explored here.

\section{Certification-based Design Decisions}

The virtual flight testing framework can be taken a step further by making design decisions based on the flight simulation results. 
The current industry standard incorporates Factors of Safety (FoS) to account for possible uncertainties in the design analyses.
The FoS rely on historical experience in designing conventional aircraft. 
While FoS are essential to ensure that the aircraft meets or exceeds the baseline requirements, there is no way to determine \textit{a priori} if they are overly conservative, woefully inadequate, or perfectly sufficient. 
This is further complicated when non-conventional aircraft designs, such as those needed for urban air mobility \cite{silva_vtol_2018} or low-emissions flight \cite{bruner_nasa_2010}, are explored.
Historical experience cannot be relied on in these situations. 

The virtual flight testing framework provides a new, more quantitative method to determine the recommended FoS.
The explicit quantification of the failure rate in completing a certification maneuver can inform design decisions that mitigate risks to an exact level of success.
Since the failure rate is directly related to the aircraft performance predictions and the uncertainty in the design analyses, the methodology is agnostic to the aircraft design. 
Moreover, as simulation techniques improve and their associated uncertainties reduce, the estimation of the failure rates would correspondingly improve. 
These factors result in a versatile framework that evolves alongside the analyses that inform it. 

This new approach to aircraft design is demonstrated by sizing the aileron, at each design stage, based on the virtual flight testing results of the Roll Capability maneuver with the modifications suggested in Section \ref{subsec:sim_mods}.
In addition to quantifying the failure rate of a particular design, the CDF of the output metrics can be inverted to determine the changes required to achieve a prescribed level of success. 
Since the output metrics from the simulations depend on the deflection angle of the control surface, the chosen design variable is the maximum allowable aileron deflection.

Given a desired success rate $x\in[0,1]$, the CDF of the roll metric is queried to find 
\begin{equation} \label{equ:design_roll_metric}
    \rho_{roll}^x:P(\rho_{roll} \leq \rho_{roll}^x) = 1-x.
\end{equation}
Inverting Equation \ref{equ:roll_metric} and using $\rho_{roll}^x$ from Equation \ref{equ:design_roll_metric}, the new maximum allowable aileron deflection is
\begin{equation} \label{equ:design_ail_lim}
    \delta_a^{x} = \delta_a^{lim} \left ( 1 - \rho_{roll}^x \right ).
\end{equation}

The red dashed line is used to demarcate the $\rho_{roll} = 0$.
Recalling that any simulation with $\rho_{roll} < 0$ is considered a failure, the failure rate for a set of simulations is the cumulative probability value at the intersection of the corresponding CDF and the red dashed line.


Current practices would dictate placing an FoS on the deflection limit to account for potential failures due to the uncertainty in the performance predictions.
If the FoS is large, it can result in an overbuilt control surface bigger and bulkier than it needs to be. 
The more critical condition occurs if the FoS is not large enough and the design progresses with insufficient control authority to pass the flight certification tests. 
The later such an issue is recognized, the more expensive a redesign gets. 

The proposed virtual flight testing framework provides success metrics that consider those uncertainties in the design analysis tools. 
Using the CDF output from the simulations and Equations \ref{equ:design_roll_metric}-\ref{equ:design_ail_lim}, engineers can design the control surface to meet a predetermined success rate.
Table \ref{tab:reo_roll_design} lists the current failure rates and the new deflection limits that would be required to achieve a $95\%$ or $100\%$ success rate in the virtual flight testing.

\begin{table}
\centering
    \renewcommand{\arraystretch}{1.2}
    \captionsetup{justification=centering}
    \caption{Certification-based aileron design decisions to achieve prescribed success rate in flight simulations.} 
    \begin{tabular}{|c||c|c|c|}
    \hline
         & Failure Rate & $\delta^{x}_a$ for $x = 95\%$ & $\delta^{x}_a$ for $x = 100\%$ \\ \hline \hline
        AVL & 6.00\% & 15.33 $^\circ$ & 20.91 $^\circ$ \\ \hline 
        AVL+SU2 & 15.10\% & 17.05 $^\circ$ & 22.61 $^\circ$ \\ \hline 
        AVL+SU2+WT & 39.20\% & 16.27 $^\circ$ & 18.21 $^\circ$ \\ \hline 
        WT & 30.80\% & 16.81 $^\circ$ & 19.27 $^\circ$ \\ \hline 
    \end{tabular}
    \label{tab:reo_roll_design}
\end{table}


Section \ref{subsec:sf_vs_mf_cba} discussed the general trends seen when adding fidelity levels to the databases.
Those observations are now placed in the context of making certification-based design decisions.
The low-fidelity databases made using AVL analyses have a low failure rate of $6\%$, but the roll metric covers an extensive range of values.
The AVL analyses over-estimate the effectiveness of the aileron resulting in a lower failure rate. 
However, the significant uncertainties associated with the design analyses propagates through the flight simulation procedure and results in the wide, horizontal spread of the CDF.
Consequently, the $100\%$ design requires a higher deflection limit.

For the AVL+SU2 case, CFD data from SU2 is added to the aerodynamic databases. 
The resulting CDF maintains its shape but shifts towards the three-fidelity (AVL+SU2+WT) and the high-fidelity (WT) case. 
The shift towards a higher failure rate indicates that the baseline aerodynamics databases from AVL were overestimating the aircraft's performance. 
The shape of the CDF remains consistent because the controls databases are identical to the AVL case as there is no CFD-based control derivative information.
These trends are reflected in the higher failure rate and the larger deflection requirements to meet the prescribed success rates. 

Adding high-fidelity wind tunnel data to the aerodynamics and controls databases reduces the uncertainty in the simulation results significantly. 
The relatively small range of values the roll metric achieves, and the steep shape of the CDF for the AVL+SU2+WT case are evidence of this reduction.
The more realistic representation of the flight physics results in a significantly higher failure of $39.2\%$. 
However, the low uncertainty associated with the database results in requiring modest changes to the design to meet the prescribed success rates.
The design requirements are smaller for the three-fidelity case than the high-fidelity case, where the database only relies on only wind tunnel data.

So far, the design variable of choice has been the maximum deflection limit of the control surface. 
However, other design variables, such as aileron size or placement, can also create the same effect.
$\rho_{roll}^x$ acts as a proxy for the supplemental control authority required to achieve the desired success rate. 
The additional control authority can be provided by using larger ailerons or moving the ailerons further outboard on the wing. 
However, in the context of these particular flight simulations, the effect of changing these other design variables is not as straightforward as the effect of changing the deflection limit.
Modifying the simulation to reduce the deflection limits of the aircraft (see Section \ref{subsec:sim_mods}) means that there already existed simulation and experimental data for the effect of the control surface beyond the artificially placed deflection limit. 
If the control surface size or location were changed, it would require rerunning all the design analysis tools to determine the effect of the new aileron design on the aircraft's controls databases. 

Using multi-fidelity modeling and uncertainty quantification to perform virtual flight testing opens new avenues for improving the aircraft design process. 
Flight certification tests can be simulated before building a full-size aircraft prototype, potentially mitigating the enormous costs associated with expensive redesigns late in the aircraft design process.
The probabilistic nature of the simulations also allows for an explicit calculation of failure rates.
This feature can be used to make design decisions to ensure the aircraft can meet the certification requirement with a prescribed success rate. 
\chapter{Conclusions} \label{chap:conclusions}
\section{Summary}
This work aims to introduce a virtual flight testing framework that can be used at any point during the aircraft design process to determine the likelihood that the aircraft passes or fails a particular certification maneuver. 
It is achieved by contributing to the state-of-the-art in three disciplines: uncertainty quantification, multi-fidelity modeling, and certification by analysis. 

\subsection{Uncertainty Quantification}

In the realm of uncertainty quantification, the eigenspace perturbation methodology \cite{iaccarino_eig_pert} was implemented in SU2, validated on a suite of test cases, and applied to full configuration aircraft simulations.
The modular architecture of SU2, an open-source CFD code, allowed for an implementation focused on \textit{versatility}, such that both, experts and non-experts, can use the module.
The validation suite included flow conditions that introduce model-form uncertainties in RANS simulations.
Flow features such as corner flows, shear planes, separation bubbles, and shocks were emphasized. 
Additionally, it was necessary to have high-fidelity data available for the test case to make suitable comparisons.
The methodology performed well across all flow conditions, predicting larger uncertainty bounds in areas where RANS simulations often struggle to capture the flow physics and, conversely, predicting smaller bounds where RANS simulations are usually accurate.
The high-fidelity data points often fell within the uncertainty estimates from the methodology, although this is not mathematically guaranteed. 

This observation is an essential reminder that regardless of the model (in)adequacy of a simulation method, there may be unforeseen uncertainties and errors introduced in the predictions that can cause results to deviate from higher-fidelity information sources. 
Numerical discretization error arising from insufficient grid refinement is one such example that often plagues CFD simulations. 
Its relationship with the turbulence modeling uncertainty was investigated to determine the RANS UQ methodology's grid requirements. The NACA0012 airfoil and ONERA M6 wing were used as test cases. 
It was found that the grid resolution must be sufficient to capture the relevant flow features.
However, given that minimum level of grid resolution,  further refinement did not significantly change the turbulence modeling uncertainty estimate.
Moreover, the uncertainty estimates were significantly larger than the discretization errors. 
These factors allow using coarser meshes to perform the perturbed simulations without losing much accuracy.

The application of the methodology to the NASA Common Research Model (CRM) was presented in detail.
A parameter-sweep in the angle of attack was performed.
Simulations at low angles of attack, where the turbulence model can accurately capture flow features, resulted in smaller uncertainty estimates compared to those performed at high angles of attack, where flow separation is significant, and the turbulence model is unable to provide accurate predictions.
The predicted bounds did not encapsulate the experimental data due to well-known geometric discrepancies between the wind tunnel model and the model used for numerical simulations.
Post-processing of the individual perturbed simulations allowed visualization of the dominant flow features contributing to the uncertainty estimates.
It provides a qualitative use-case to improve design decisions and future high-fidelity data gathering.

\subsection{Multi-Fidelity Modeling}

Multi-fidelity Gaussian processes were used to combine data from different information sources.
Improvements suggested by Gratiet \cite{gratiet_multi-fidelity_nodate}, which significantly reduce the computational cost of learning the Gaussian process, were implemented.
The computational savings compared to the original implementation \cite{kennedy_predicting_2000} were validated using analytical functions. 
The existing equations were extended so that uncertainty could be specified independently for each data point, even when the design sets for each fidelity level are not nested.
This new set of equations were used to create multi-fidelity probabilistic aerodynamic databases for an aircraft configuration, the NASA Common Research Model (CRM).
The multi-fidelity fit outperformed the single-fidelity fit, particularly when high-fidelity data is sparse or localized to areas where low-fidelity functions are inaccurate.
This advantage is more pronounced when using multi-dimensional databases.
As the number of high-fidelity data points is increased, the multi-fidelity and single-fidelity models start performing identically.

\subsection{Probabilistic Aerodynamics and Controls Databases}
A second aircraft configuration, the Generic T-tail Transport (GTT), is used to take the probabilistic aerodynamic databases further.
The wealth of experimental data from wind and water tunnels allows for high-fidelity modeling of the aircraft's lateral dynamics, longitudinal dynamics, and control surface effects.
The generation of multi-fidelity data and the associated uncertainties were discussed, and visualizations of the resulting single-fidelity and multi-fidelity controls databases were compared.
The sparsity of the high-fidelity data in the control surface deflection dimension causes the error estimate from the GP regression to balloon between available data points when using only high-fidelity data to create the databases.
With the multi-fidelity fusion of information sources, the abundance of well-correlated low-fidelity AVL data improves the GP model and results in a smooth, uniformly low error estimate across the control surface deflection dimension. 
This improvement is significant as it is achieved without using any new high-fidelity evaluations, thereby reducing the uncertainty with negligible analysis cost. 

\subsection{Certification by Analysis}
A virtual flight testing framework is the culmination of all the work presented in the previous sections. 
An air-worthiness certification test maneuver formulated by the FAA \cite{romanowski_flight_2018} to ensure sufficient roll capability in engine-out scenarios is used for this purpose.
The aerodynamics and controls databases represent all aspects of an aircraft's dynamics and can be used to perform flight simulations.
To this end, The Boeing Company's existing tools and expertise in flight simulation and control-law mixing \cite{control_law_patent} were leveraged.
A simplified $5$ degree-of-freedom simulator with an open-loop controls configuration is employed, and its details are presented. 
Success in meeting the certification test requirement is defined by completing the flight simulation without over-saturating any control surfaces.

Each deterministic sample of the GP models representing the aircraft databases is a valid instance of the aircraft. 
These instances have slight variations in their performance due to the uncertainty in the underlying data informing the models.
A Monte Carlo analysis of the certification test was performed by taking multiple samples of the same aircraft and running each through the flight simulation. 
Analyzing the percentage of failures provides the likelihood of the aircraft failing the certification test.
One thousand samples were sufficient to create converged sample mean and variance estimates for the simulation metrics. 
The GTT aircraft passes the certification maneuver with $100\%$ certainty.
Derived from a currently certified aircraft (CRJ 700), the GTT should pass the certification tests with a $100\%$  success rate. 

The maneuver is made more challenging by introducing modifications to the control surface functionality.
The modified certification test is simulated with databases built using information representing the aircraft at different stages of the design process.
Early representations use low-fidelity AVL data that has significant uncertainties associated with it.
Successive design stages introduce higher-fidelity information, in the form of RANS CFD simulation results and experimental data from wind and water tunnels, one at a time.
The incremental improvements in the performance predictions propagate through the certification test results. 
Cumulative distribution functions are employed to visualize the resulting simulation metrics. 
The benefits of fusing information sources are evident in the reduced uncertainty in the metrics for simulations using 3-fidelity databases compared to those using single-, high-fidelity information.

Finally, a fundamental shift in the design ethos is discussed.
The current standard is to use conservative safety factors to account for possible uncertainties in the design analyses.
These factors are chosen based on historical experience in designing conventional aircraft. 
While these factors are essential to ensure that the aircraft meets or exceeds the baseline requirements, there is no way to determine \textit{a priori} if they are overly conservative, woefully inadequate, or perfectly sufficient. 
With an explicit quantification of the failure rate, design decisions can be made to mitigate certification failures to a prescribed rate.
These risks, and consequent design decisions, are directly related to the uncertainty in the analyses.
As simulation techniques continue to improve, so will the estimation of these failure rates and the design decisions required to mitigate risks.
While this is an ambitious goal, it is necessary if Certification by Analysis of aircraft is to be achieved.


\section{Contributions} \label{conclusion_contributions}

This thesis established a framework to perform virtual flight testing of an aircraft early in the design process.
This was achieved through contributions in the fields of uncertainty quantification in RANS CFD simulations (Chapter \ref{chap:rans_uq}), multi-fidelity Gaussian Processes (Chapter \ref{chap:mf_gp}), probabilistic aerodynamics and controls databases (Chapter \ref{chap:aero_db}), and certification by analysis (Chapter \ref{chap:cba}).

The eigenspace perturbation methodology was used to quantify epistemic uncertainties injected by turbulence models into RANS CFD simulations.
The methodology, developed by Emory et al. \cite{emory2013modeling} and Mishra et al. \cite{iaccarino_eig_pert}, was implemented in SU2, an open-source CFD solver. 
Test cases featuring a wide variety of complex flow conditions validated the uncertainty estimates from the methodology against high-fidelity experimental data. 
Investigation of the relationship between the epistemic uncertainties and numerical discretization errors involved running two test cases: subsonic flow over a 2D airfoil and transonic flow over a 3D wing.
The eigenspace perturbation methodology was applied to the sequentially refined mesh families used for grid convergence studies \cite{american_society_of_mechanical_engineers_standard_2009}.
The resulting uncertainty estimates were compared to the numerical discretization errors.
This comparison yielded that, given sufficient discretization to capture the relevant flow features, the epistemic uncertainty estimate was independent of the level of mesh refinement.
This observation justified using coarser meshes for UQ purposes, thereby reducing the computational cost of the methodology.   
The UQ methodology was applied to two aircraft, the NASA Common Research Model and the Generic T-tail Transport, to create aerodynamic databases with physics-informed uncertainties. 

This work used multi-fidelity Gaussian processes (GP) to combine different information sources and their associated uncertainties.
The auto-regressive formulation by Gratiet \cite{gratiet_multi-fidelity_nodate} was extended to use noisy data when the design sets of successive fidelity levels are not nested (Equations \ref{equ:mu_Zt} and \ref{equ:sig_Zt}).
Corresponding extensions to the parameter estimation equations were also presented (Equation \ref{equ:param_est_mf}).
These multi-fidelity GP equations were used to represent the aerodynamic predictions, and their associated uncertainties, for the NASA CRM.
AVL simulations, RANS CFD simulations, and wind tunnel experiments informed the low-, medium-, and high-fidelity levels, respectively.
Comparisons between the multi-fidelity and high-fidelity aerodynamic databases for the NASA CRM demonstrated the benefits of using multi-fidelity data fusion, especially when high-fidelity data is sparse or localized to a limited region of the domain. 

The application of the multi-fidelity modeling was extended to the GTT aircraft.
Probabilistic, multi-fidelity, and multi-dimensional aerodynamics and controls databases representing a full-configuration aircraft's lateral and longitudinal dynamics were created.
Low- and medium-fidelity data was generated using AVL and SU2, respectively.
Existing data from wind tunnel experiments informed the highest fidelity.
Uncertainties for the low-fidelity AVL simulations and the high-fidelity wind tunnel experiments were estimated using prior knowledge of the tools and their shortcomings. 
The eigenspace perturbation methodology provided uncertainties for the medium-fidelity CFD simulations. 
A total of $29$ separate Gaussian processes were required to model the databases; a separate GP represented each quantity of interest and its associated uncertainties.

With all aspects of an aircraft's flight dynamics defined, The Boeing Company's $5$ degree of freedom flight simulator was used to perform a flight certification maneuver computationally. 
The roll capability maneuver from the FAA's flight testing guide was chosen for this analysis \cite{romanowski_flight_2018}.
With $1000$ deterministic samples of the databases, a Monte Carlo analysis propogated the uncertainties in the aircraft's design analyses through the maneuver simulation. 
The resulting cumulative distribution functions of the performance metrics for the maneuver were analyzed to explicitly quantify the likelihood of the aircraft succeeding or failing the maneuver.
This certification analysis was performed with databases using varying fidelity levels to simulate different stages in the aircraft design process.
Furthermore, design decisions to mitigate the failure rate to a specific value were suggested.

\section{Future Work}

Certification by Analysis is in the nascent stages of development.
While the goal - completing certain certification tests through simulation alone - is set, there is no defined path to get there, yet.
This work is an an initial exploration of one potential route, through quantification of uncertainties in simulations, their propagation through certification tests, and a statistical, rather than deterministic, analysis of results. 
There are numerous avenues for further research, some of which are outlined in this section. 

\subsection{Uncertainty Quantification}

For this work, subject matter experts (SME) provide the uncertainties for the AVL simulations and the wind tunnel data. 
These were based on historical experience using these analysis techniques for aircraft design. 
Instead, the uncertainties can be handled more rigorously.
Engineers use low-fidelity simulations like AVL to consider significant changes to design parameters in the conceptual design stage.
Conducting parameter sweeps results in a range of performance predictions that can represent the uncertainty in the design parameters at that stage of the process.
This uncertainty is different from the uncertainty due to the modeling simplifications made in the analysis tool itself, but it is a more rigorous methodology than basing uncertainties on SME.

For experimental data, methods for uncertainty quantification have been published \cite{coleman1995engineering} and widely adopted.
Using the raw sensor data for the wind and water tunnel experiments would allow for an explicit quantification of the systematic and precision uncertainties. 
Additional errors due to tunnel-specific phenomena, such as blockage or flow angularity, can be quantified with the help of the tunnel engineers.

A significant focus of this work was the quantification of uncertainties introduced in RANS CFD simulations due to the model inadequacy of turbulence models. 
While the eigenspace perturbation methodology has been effective on a large variety of test cases, there are avenues for improvement. 
Currently, the eigenspace is uniformly perturbed to the limits of physical realizability everywhere in the flow domain. 
This procedure can lead to improbable flow conditions that can overestimate the uncertainty.
More stringent limits on the perturbation magnitude proposed by \cite{mishra_perturbations_2019} can be employed.
Additionally, high fidelity DNS and LES data can be used to learn the ideal perturbation magnitude based on the relevant mean-flow features.
This improvement would allow for spatial variation in the perturbation magnitude across the flow domain.

The principles behind the methodology are valid for any turbulence model that uses the eddy-viscosity hypothesis. 
The current work has used the SST turbulence model for all of the results.
Extensions of the methodology can enable its application to other popular turbulence models, such as the one-equation Spallart-Allmaras \cite{allmaras2012modifications}. 

\subsection{Multi-Fidelity Modeling}

Within the framework of multi-fidelity Gaussian processes, many improvements can be implemented into the existing framework to improve the aerodynamics and controls databases created.
Non-hierarchical information fusion \cite{lam_multifidelity_2015} could be used to combine multiple information sources that are of an unknown or equal fidelity level. 
For example, a single surrogate model could combine the multiple wind tunnel campaigns for the GTT aircraft.
Gradient information, although expensive to generate, can be used to improve the quality of the surrogate models and reduce the error estimates \cite{han_improving_2013,yamazaki_derivative-enhanced_2013}.
This change can be beneficial with CFD data, where adjoint simulations can provide gradient information. 
Similarly, non-linear information fusion \cite{perdikaris_nonlinear_2017} could further enhance the advantages of using low-fidelity data by learning more complex trends between the information sources.
Though care must be taken to ensure the computational cost of processing the GP equations does not balloon to intractable levels.

Outside of Gaussian processes, other multi-fidelity modeling techniques should be explored. 
In particular, multi-fidelity polynomial chaos expansions (PCE) \cite{ng_multifidelity_2014} preserve the essential features of GP models, such as deterministic sampling and providing mean and variance information.
PCE have the added advantage of handling uncertainties with non-Gaussian distributions. 
This avenue would also provide insight into how much the results depend on the multi-fidelity modeling technique used. 

\subsection{Certification by Analysis}

As mentioned earlier, certification by analysis (CbA) is in its nascent stages of development.
For simulations to replace current certification tools, they would have to provide equivalent or, more likely, higher-quality predictions than real-world flight testing. 
The \textit{CFD Vision 2030 Study} \cite{slotnick_cfd_nodate} provides an insight into the advancements in CFD simulations that are essential to achieve in the near future. 
More generally, current low-fidelity techniques are not sufficient to reproduce experimental-level data. 
It is computationally expensive to create a controls database using CFD simulations, and AVL simulations are woefully inadequate in this regard.
They overestimate the control effectiveness and carry significant uncertainties in their predictions.
This shortcoming is evident from the results of the certification simulations in Section \ref{sec:cba_results} where the results from the AVL databases are not reflective of the results from high-fidelity databases. 

Analysis of historical aircraft design data can unearth critical trends between early design analysis and final flight certification performance. 
The design process can be recreated, but with the added virtual flight testing analysis.
Similar to the process for the GTT aircraft, databases can be built using only the information available at each design stage. 
Analyzing the virtual flight testing results with these low- and multi-fidelity databases, design decisions can be suggested and compared to the actual design decisions that were taken. 
This comparison provides a direct assessment of the benefits of virtual flight testing in the design framework. 

Additional maneuvers to stress-test other parts of the aerodynamics and controls databases can be explored.
The chosen maneuver for this work relied heavily on aileron and rudder control authority. 
Similarly, airworthiness testing is not the only part of the certification process that can be made virtual. 
Extensive structural testing is done with prototypes of wings and engines.
The relatively lower cost and relatively high accuracy of structural analysis can lower entry barriers for CbA. 
Choosing tests based on the accuracy of low-fidelity methods in predicting performance is critical. 
If the low-fidelity methods are well developed and accurate in the test-critical parts of the domain, their adoption for CbA is far more likely.
The day that simulations replace real-world testing is far into the future. 
This work paves a potential path to that day with these initial steps.


\end{document}